\documentclass[prr,superscriptaddress,amsmath,amssymb,twocolumn]{revtex4-2}
\usepackage{graphicx}
\usepackage{subfigure}
\usepackage{adjustbox}
\usepackage{bm}
\usepackage{bbm}
\usepackage{color}
\usepackage{xcolor}
\usepackage{braket}
\usepackage{standalone}
\usepackage{multirow}
\usepackage{tikz}
\usepackage{mathrsfs}
\usepackage{dsfont}
\usepackage[colorlinks,bookmarks=true,citecolor=blue,linkcolor=blue,urlcolor=blue]{hyperref}
\usepackage{cleveref}
\usepackage{comment}
\usepackage{mathtools}
\usepackage{soul}

\def\re{\text{Re}}
\def\im{\text{Im}}

\def\sgn{\text{sgn}}
 \Crefname{equation}{Eq.}{Eqs.}
\Crefname{figure}{Fig.}{Figs.}
\Crefname{section}{Sec.}{Secs.}
\def\tot{\text{tot}}

\def\u{\underline}
\def\hc{\text{H.c.}}

\def\diag{\text{diag}}
\numberwithin{equation}{section}
\renewcommand{\theequation}{\arabic{section}.\arabic{equation}}

\begin{document}
\title{Anatomy of higher-order non-Hermitian skin and boundary modes}

\author{Fan Yang}
\email{fan.yang@fysik.su.se}
\affiliation{Department of Physics, Stockholm University, AlbaNova University Center, 10691 Stockholm, Sweden}
\affiliation{Institute of Physics, {\'E}cole Polytechnique F{\'e}d{\'e}rale de Lausanne (EPFL), CH-1015 Lausanne, Switzerland}

\author{Emil J. Bergholtz}
\email{emil.bergholtz@fysik.su.se}
\affiliation{Department of Physics, Stockholm University, AlbaNova University Center, 10691 Stockholm, Sweden}
\date{\today}

\begin{abstract} The anomalous bulk-boundary correspondence in non-Hermitian systems featuring an intricate interplay between skin and boundary modes has attracted enormous theoretical and experimental attention. Still, in dimensions higher than one, this interplay remains much less understood. Here we provide insights from exact analytical solutions of a large class of models in any dimension $d$, with  open boundaries in $d_c \le d$ directions and by tracking their topological origin. Specifically, we show that amoeba theory accounting for the separation gaps of the bulk modes augmented with higher-dimensional generalizations of the biorthogonal polarization and the generalized Brillouin zone approaches accounting for the surface gaps of boundary modes provide a comprehensive understanding of these systems.

\end{abstract}
\maketitle
\section{Introduction}

The non-Hermitian skin effect (NHSE) and the anomalous bulk-boundary correspondence (BBC) of non-Hermitian systems has attracted an enormous amount of theoretical \cite{Lee2016,yao2018,flore2018} as well as experimental \cite{helbig2020generalized,Ghatak2020,photonicNHBBC,Brandenbourger_2019, Neupert_2020,veenstra2023,Ma2024} interest in recent years \cite{emil2021e,Zhang2022,Ma2022,okuma2023,lin2023}. The intense research has amounted to deep insights in one dimension in terms of winding invariants in a generalized Brillouin zone (GBZ) \cite{yao2018,Yokomizo2019}, biorthogonal polarization and (de)localization transitions \cite{flore2018}, spectral winding \cite{Gong18,Lee2019,Herviou2019,Okuma2020}, Green's functions \cite{Zirnstein2021,Borgnia2020}, transfer matrices \cite{flore2019} and spectral sensitivity \cite{Xiong2018,FoaTorres2018,NTOS,McDonald2020,Wanjura2021,elisabet2022}. 

The general case in higher dimensions has been much less explored and open questions remain \cite{zhang2022u,wang2022}. Approximate methods \cite{zhong2018n, liu2019} as well as extensions to hybrid boundary conditions
 \cite{hughes2023,schindler2023,lingfang2024n} have been attempted, yet there are issues including the problem of properly defining the GBZ beyond one dimension (1D) due to geometric obstruction, where the NHSE appears to depend on the lattice geometry \cite{zhang2022u} and the effective dimensionality of the GBZ can also alter \cite{lee2023d}.
Very recently, the novel idea of using the mathematical theory of amoebas \cite{gelfand1994,forsberg2000,passare2004,ronkin1974} to understand this problem was suggested \cite{wang2022}. While very promising, a key limitation in testing this theory lies in the fact that the convergence of both the spectrum under open boundary conditions (OBCs) and the density of states (DOS) predicted by the amoeba requires large system size, which increases computational cost and is hard to check in absence of exact solutions. 

The amoeba formulation also ignores boundary geometries, thus losing track of higher-order skin modes with potential links to NH topology \cite{kawabata2018a,elisabet2019,lee2019b,kohei2020,lee2021o,lee2022e,lee2024a}. In general, these boundary modes may significantly change spectrum properties. To remedy this, it has been suggested to introduce disorder on each boundary site \cite{wang2022} or to consider customized lattice cuts \cite{hu2023} to recover in finite-size systems what has been called the universal spectrum, which corresponds to the bulk spectrum in the thermodynamic limit \cite{wang2022,hu2023}.
\begin{figure}[t]
\centering
\includegraphics[width=1\columnwidth]{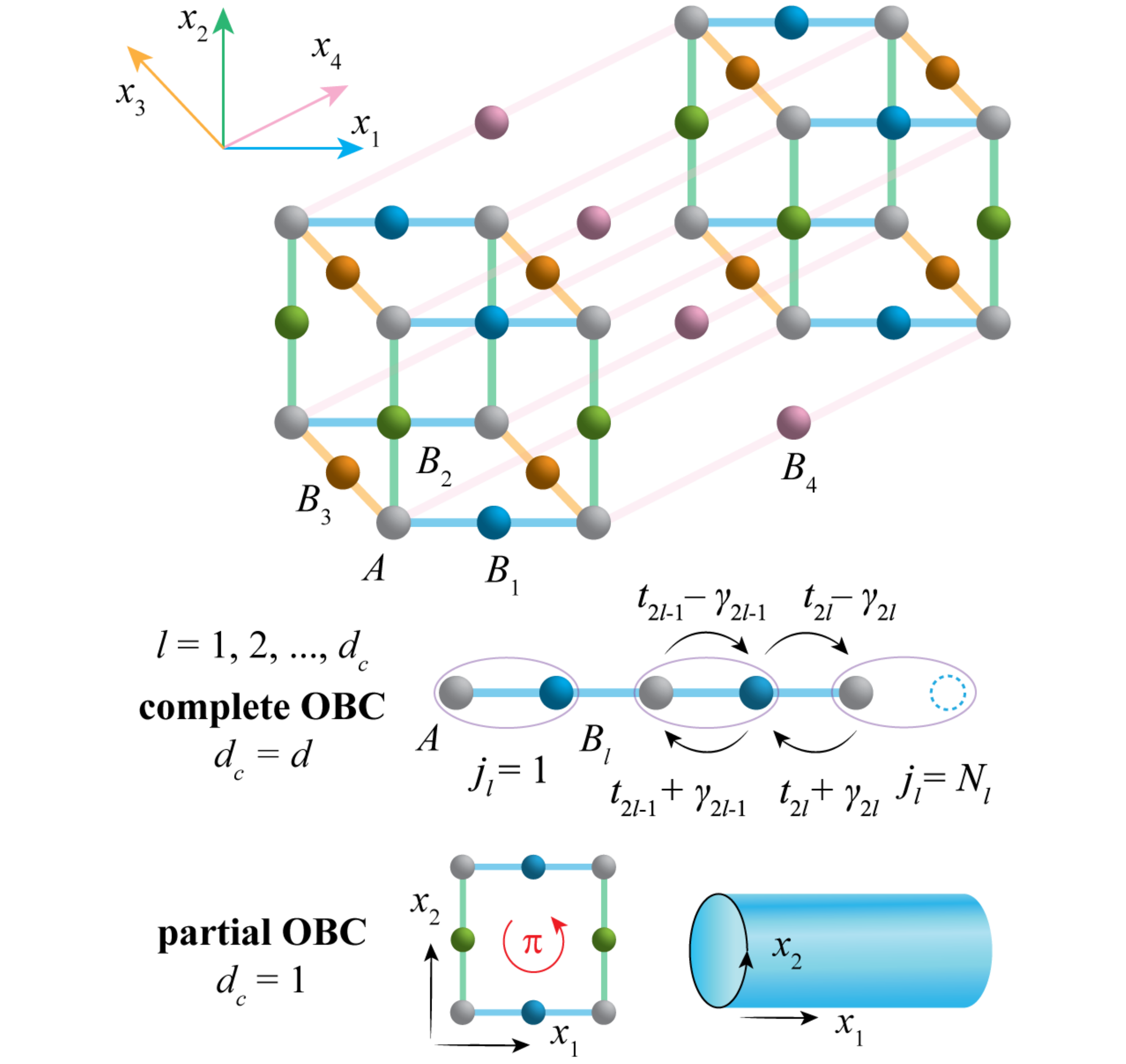}
\caption{Lattice geometry of  exactly solvable $d$-dimensional NH hypercubic models with OBCs in  $d_c \le d$ directions. Denoting unit cells  by index $j_l$ along $x_l$ direction, we assign symmetric and skew-symmetric hopping terms $(t_{2l-1}, \gamma_{2l-1})$ inside each unit cell  and $(t_{2l}, \gamma_{2l})$ between neighboring unit cells. The notation $t_l^\pm = t_l \pm \gamma_l$ is adopted throughout the text.
In the presence of a magnetic field, the NH Lieb lattice with $\pi$ flux per plaquette becomes solvable by projecting to a cylinder. 
}
\label{fig:lattice}
\end{figure}

\textcolor{black}{In this work, we pursue an alternative path to address the outstanding problem of the interplay between higher-dimensional NH skin and boundary modes. We show that the GBZ in 1D  can be naturally extended to generalized surface Brillouin zones (GSBZs) beyond 1D. The SBZs accommodate boundary modes, whose  surface momenta live in lower dimensions than the bulk. We elucidate our GSBZ approach by constructing a class of nonreciprocal NH hypercubic models in \Cref{fig:lattice}, the exact solvability of which in arbitrary spatial dimensions $d$ attached to  $d_c$ open boundaries ($d_c \le d$) and  with arbitrary system size is associated with a spectral mirror symmetry:
    \begin{align}
       \forall k_l, \quad &E^{\text{OBC}}
           ( k_1,\dots,k_l,\dots,k_{d_c}; \vec{k}_{\parallel} )   \notag \\
       = \ &E^{\text{OBC}}(k_1,\dots,-k_l,\dots,k_{d_c}; \vec{k}_{\parallel}). \label{eq:mirror}
    \end{align}
The spectrum remains unchanged when the mirror flips the sign of any element $k_l$ in surface momentum $\vec{k}_{\perp} = (k_1, \dots, k_l, \dots, k_{d_c})$ along the OBC directions.  The quasi momentum $\vec{k}_\parallel$ tracks the elements in the direction of periodic boundary conditions (PBCs). We find that performing hybrid imaginary surface momentum shifts in the Bloch phase factor $e^{i\vec{k}_\perp \cdot \vec{j}_\perp}$ for the right eigenstates:
  \begin{gather}
   \vec{k}'_\perp = \vec{k}_\perp - i \vec{\kappa}(\vec{k}_\perp), \notag \\ 
   \kappa_l (\vec{k}_\perp) = \begin{cases}
     \ln r_l, & k_l \ne 0 \\
     \ln r_{R,l}, & k_l = 0,
  \end{cases}  \label{eq:gsbz}
 \end{gather}
gives rise to the GSBZ as well as the exact OBC spectrum related to the PBC one \cite{yao2018,flore2019,elisabet2020,yang2022}:
    \begin{gather}
E^{\text{OBC}} (\vec{k}_\perp;\vec{k}_\parallel) =   E^{\text{PBC}}(\vec{k}'_\perp; \vec{k}_\parallel). 
\label{eq:opbc}
    \end{gather}
    \textcolor{black}{Since $k_l=0$ in \Cref{eq:gsbz} does not enter the spectrum, $E^\text{PBC}$ is given by the Bloch Hamiltonian living in the subspace expanded by all nonzero surface momentum elements, which becomes the bulk Hamiltonian if $\forall \ k_l \ne 0$ and the boundary Hamiltonian if $\exists \ k_l = 0$.}
Whereas for the left eigenstates, one can make a replacement in the $\kappa_l (\vec{k}_\perp)$ function: $r_l \to (r_l^*)^{-1}$, $r_{R,l} \to r_{L,l}$.
The set of localization parameters 
 \begin{gather}
      r_l  = \sqrt{\frac{t_{2l-1}^-t_{2l}^-}{t_{2l-1}^+t_{2l}^+}}, \quad r_{R,l} = -\frac{t^-_{2l-1}}{t^+_{2l}}, \quad r^*_{L,l} = -\frac{t^+_{2l-1}}{t^-_{2l}},
 \end{gather}
originate from the asymmetric hoppings shown in \Cref{fig:lattice}.}  

\textcolor{black}{The order of skin modes  can be classified  by their total number $O(N^n)$, which scales extensively with the linear system size $N$ \cite{kohei2020}. 
For our models under complete OBC ($d_c = d$),  the power $n$  is given by the number of nonzero surface momentum  elements  in \Cref{eq:gsbz}, considering each $k_l$ takes $\sim N$ discrete values. In this case, $n=0,1, \dots, d$  also encodes the codimension of skin modes in the OBC directions. When $n = d$ ($n < d$), it refers to bulk (boundary) modes. 
In $d=2$, over the NH Lieb lattice, $O(N^2)$, $O(N)$ and $O(1)$ skin modes correspond to the first-order bulk, second-order edge  and third-order corner modes of codimension $n = 2, 1, 0$ respectively. It is noted that our definition is different from the one adopted in NH systems characteristic of the geometry-dependent skin effect \cite{zhang2022u,wan2023}. There, bulk, edge, corner is used to describe the localization profile of first-order $O(N^d)$ skin modes in terms of normal density, which changes in accord with the lattice shape. With a fixed hypercubic lattice geometry, the emergent higher-order skin modes in our models have another physical origin. When the nonreciprocity  is switched on, from normal density, all $O(N^n)$ skin modes are localized at the corners (see \Cref{fig:loc}, upper panel). Yet, after nonunitary gauge transforms, each of them  can be mapped to their reciprocal counterparts (see Figs.~\ref{fig:rs} and \ref{fig:rs0}) which display distinct localization tendencies (bulk, edge, corner), a feature retrieved in biorthogonal density (see \Cref{fig:loc}, lower panel).
Meanwhile, compared with the higher-order NH skin effect induced by spatial symmetry in Ref.~\cite{kohei2020}, our models respect a different symmetry, the
spectral mirror symmetry in \Cref{eq:mirror}. 
 In Ref.~\cite{kohei2020}, the spatial symmetry, in particular, the transposition-associated mirror symmetry plays the role in canceling the first-order skin effect and bringing the appearance of the second-order $O(N)$ skin modes in two dimensions with two OBCs. By contrast, our models always support a complete set of $O(N^n)$ skin modes with $n = 0, \dots, d$ in arbitrary $d$ dimensions under complete OBC.
To properly quantify different orders of NH skin effects they display, it requires an extension of the conventional GBZ to the GSBZ in \Cref{eq:gsbz}, which corresponds to a modified non-Bloch theory exactly formulated in our models through nonunitary gauge transforms (see \Cref{sec:gt}).}

\textcolor{black}{The topology of our models is illuminated through the lens of biorthogonal polarization. Enhanced by a generalized chiral symmetry, the biorthogonal polarization vector of the  
 corner mode is equivalent to an ensemble of non-Bloch winding numbers \cite{yao2018} carried by the edge modes in the GSBZ. Inspired by biorthogonal BBC in 1D \cite{flore2018}, 
a quantized change in polarization signals a gap closing between boundary and bulk modes. In higher dimensions ($d \ge 2$), special attention needs to be paid on the  existence of  two types of energy gaps \cite{fu2018, kang2023}: the surface gap and the separation gap. To clarify, at a given $\vec{k}_\parallel$,
for any $O(N^n)$ skin mode with surface momentum $\vec{k}_{\perp} = (\vec{k}_{n}, \vec{0})$ where $\vec{k}_{n}$ is a collection of $n$ nonzero elements, its surface gap with $O(N^d)$ bulk modes is defined at the same $\vec{k}_{n}$ as
 \begin{gather}
    | \Delta E_{\text{Surf.}} | = \min_{\forall \vec{q},\vec{k}_{n}} \{ | E_{d}^{\text{OBC}} (\vec{k}_{n}, \vec{q}) - E^{\text{OBC}}_{n} (\vec{k}_{n}, \vec{0}) |\}, \label{eq:sfg}
\end{gather}
while the separation gap is relaxed to allowing different nonzero subsets $\vec{p} \ne \vec{k}_{n}$:
\begin{gather}
|\Delta E_{\text{Sep.}} | = \min_{\forall \vec{p},\vec{q},\vec{k}_{n}} \{ |E_{d}^{\text{OBC}} ( \vec{p}, \vec{q}) - E^{\text{OBC}}_{n} (\vec{k}_{n}, \vec{0})| \}. \label{eq:spg}
\end{gather}
To visualize them, the surface gap develops in the multiparameter $\vec{k}_n$ space similar to Hermitian systems, while the separation gap (or its closing) is only discernible on the complex-$E$ plane [see \Cref{fig:gap}(b)].
 For the corner mode ($n=0$) with $\vec{k}_{0} = \varnothing$, one recognizes $ | \Delta E_{\text{Surf.}}| = |\Delta E_{\text{Sep.}}|$. For higher-dimensional boundary modes ($n \ge 1$), $\vec{k}_{n}$ exists and often leads to $|\Delta E_{\text{Surf.}}| \ne |\Delta E_{\text{Sep.}}|$, a general feature in systems beyond 1D.
Remarkably, the polarization jump predicts the surface gap closings in \Cref{eq:sfg} and indicates various types of real-space diffusion among skin modes of different orders. 
The amoeba formulation, by contrast, addresses separation gap closings in \Cref{eq:spg}, and carries information on the first-order NHSE displayed by bulk modes. As will be shown, with a surface gap opening, even if the separation gap closes,  the localization behaviors of bulk and boundary modes remain drastically different. 
Two distinct roles played by the polarization and the amoeba we identify in this work reveal the intricacy of NHSE in higher dimensions. }

\textcolor{black}{The paper is structured as follows: first, we exactly solve a class of mirror symmetric NH hypercubic models in \Cref{fig:lattice} under complete OBC,  and establish thereof the GSBZ approach (\Cref{sec:gsbz}). Specifically, we present complete solutions to NH Lieb lattice, illustrating NHSE in higher dimensions in terms of normal and biorthogonal densities. Second, we compare two approaches, the biorthogonal polarization (\Cref{sec:bp}) and the amoeba (\Cref{sec:af}), by demonstrating different types of gaps and varied aspects of NHSE they disclose. Third, we generalize our models and formulations to more generic cases enriched by hybrid boundary conditions ($d_c < d$) and additional internal degrees of freedoms (\Cref{sec:gen}). In particular, a physical example is given by the NH Lieb model at $\pi$ flux on a cylinder geometry (see \Cref{fig:lattice}, bottom).} 

\section{GSBZ and NHSE beyond 1D}
\label{sec:gsbz}
\subsection{The model}
In \Cref{fig:lattice}, we start by building the NH Hamiltonian on a $d$-dimensional nonreciprocal hypercubic lattice expanded by unit vectors $\vec{e}_l$ and imposing open boundaries in every direction ($d_c = d$, $\vec{k}_\parallel = \varnothing, \vec{k} = \vec{k}_\perp$):
\begin{align}
\mathcal{H} = \sum_{\vec{j}} \sum_{l=1}^{d} & \ t_{2l-1}^+  c^\dagger_{\vec{j}, A} c_{\vec{j}, B_l} + t_{2l-1}^-c^\dagger_{\vec{j}, B_l} c_{\vec{j},A} \notag \\
+ & \  t_{2l}^+  c^\dagger_{\vec{j}, B_l}c_{\vec{j}+\vec{e}_l, A}   + t_{2l}^- c^\dagger_{\vec{j} + \vec{e}_l, A} c_{ \vec{j}, B_l}, \label{eq:hl}
\end{align}
\textcolor{black}{where $t^\pm_l = t_l \pm \gamma_l \in \mathbb{C}$. Given a set of symmetric and skew-symmetric hopping parameters $(t_l, \gamma_l)$, the non-Hermiticity of  the Hamiltonian is manifest in the condition $\im [t_l] \ne \re[\gamma_l]$, and the nonreciprocity is guaranteed in $\gamma_l \ne 0$.} 
 $c^\dagger_{\vec{j}, \lambda}$ ($c_{\vec{j}, \lambda}$) creates (annihilates) a particle on the motif $\lambda \in \{A, B_1, \dots, B_d \}$ inside the $\vec{j}$th unit cell with $\vec{j} = (j_1, j_2, \dots, j_d)$.
 The complete OBC leads to coupled arrays of odd-length NH Su-Schrieffer-Heeger (SSH) chains consisting of $N_l$ unit cells along the direction $x_l$. Each decoupled chain  becomes exactly solvable in 1D \cite{elisabet2020,yang2022}. We endeavour to  extend the exact solutions to higher dimensions. The eigenvalue equations for right and left eigenstates read
 \begin{gather}
   H\  {\u{\psi}}_{Rm} = E_m {\u{\psi}}_{Rm}, \quad
    H^\dagger \  {\u{\psi}}_{Lm}  = E_m^* {\u{\psi}}_{Lm}, 
 \label{eq:rl}
\end{gather}
where $*$ denotes the complex conjugation and $E_m$ the complex energy assigned with band index $m$. We introduce  the notation $\u{x} = (x_1,x_2,...,x_k)^T$ with $T$ being the transpose  to represent a column vector  of scalars or operators \cite{prosen2008,yang2022}, in contrast with a row vector of pure scalars $\vec{x} = (x_1, x_2, \dots, x_k)$ \cite{flore2019e,flore2019b}. 
Biorthogonal relations are satisfied by left and right eigenvectors \cite{brody2013}:
    \begin{gather}
     {\u{\psi}}_{Lm}^* \cdot {\u{\psi}}_{Rm'}  = \delta_{m,m'}. \label{eq:bio}
     \end{gather}

Our real-space multiband Hamiltonian in \Cref{eq:hl} corresponds to a multilevel block Toeplitz matrix \cite{bottcher2005}. After the Fourier transform 
$c_{\vec{j}, \lambda}=\frac{1}{\sqrt{N_1\dots N_d}} \sum_{\vec{k}}  e^{i\vec{k}\cdot \vec{j}} c_{\vec{k},\lambda}$ where $k_l = \frac{2\pi \tilde{n}}{N_l} \in [0, 2\pi)$ and 
           $\tilde{n} = 0, 1, \dots,  N_l -1$, its symbol becomes the Bloch Hamiltonian: $\mathcal{H}= \sum_{\vec{k}} \Psi_{\vec{k}}^\dagger H(\vec{k}) \Psi_{\vec{k}}$, which in the basis $ \Psi_{\vec{k}} = (c_{\vec{k},A}, c_{\vec{k}, B_1}, \dots, c_{\vec{k}, B_d})^T$ shares the form
 	\begin{align}
	  & H(\vec{k}) = \label{eq:hbloch}
      \\ & 
  \begin{pmatrix}
	  		0 & t_1^+ + t_2^- e^{-ik_1}  & \dots & t_{2d-1}^+  + t_{2d}^- e^{-ik_d}  \\
			t_1^- + t_2^+ e^{ik_1} & 0 &  & 0 \\
	   \vdots	 &  & \ddots &  \\
    	 t_{2d-1}^-  + t_{2d}^+ e^{ik_d}  & 0  &  & 0
		    \end{pmatrix}. \notag
	\end{align}
The Bloch Hamiltonian respects a generalized chiral symmetry,  $\mathcal{C}  {H}(\vec{k}) \mathcal{C}^{-1} = -{H}(\vec{k})$ in the representation $\mathcal{C} = \diag \{ 1, -1, -1, \dots, -1\}$. As a result, 
there emerge two dispersive bands with opposite energies (denoted by $\alpha = \pm$) and $(d-1)$ zero-energy flat bands ($\alpha = 0$).


Although spectral mirror symmetry in \Cref{eq:mirror} is not present with PBC, i.e. in the eigenvalues of \Cref{eq:hbloch}, we show later that it is respected in every direction with OBC: $E_\alpha^{\text{OBC}} (k_l) =  E_\alpha^{\text{OBC}} (-k_l)$. Here, the OBC spectrum inherits basic dispersive ($\alpha = \pm$) and flat band ($\alpha = 0$) features from the PBC spectrum. Additionally, spectral mirror symmetry induces a shrink of the BZ from $[0, 2\pi)$ to  $[0, \pi)$, when the other half domain $(-\pi, 0]$ can be obtained by $k_l \to -k_l$. Hence, under OBC the following discrete values are taken for the nonzero momentum element: $k_l = \frac{\pi \tilde{m}}{N_l} \in (0, \pi)$ with $\tilde{m} = 1, 2, \dots, N_l -1$.

 \subsection{Exact solutions from gauge transforms}
 \label{sec:gt}
Next, we set out to derive the exact OBC spectrum for all orders of skin modes by mapping the original nonreciprocal Hamiltonian to its reciprocal counterpart. The diagonalization of the latter is realized by dividing the complete set of eigenmodes to $O (N^n)$ subgroups. These two steps are carried out through a sequence of nonunitary gauge transforms, which can be reinterpreted as the imaginary surface momentum shift in \Cref{eq:gsbz} and naturally gives rise to 
the exact GSBZ in our models. The spectral mirror symmetry in \Cref{eq:mirror} plays a key role in exactly relating the OBC and PBC spectra in 
\Cref{eq:opbc}. 

\subsubsection{Mapping to reciprocal Hamiltonian}
\begin{figure}[t]
\centering
\includegraphics[width=0.85\linewidth]{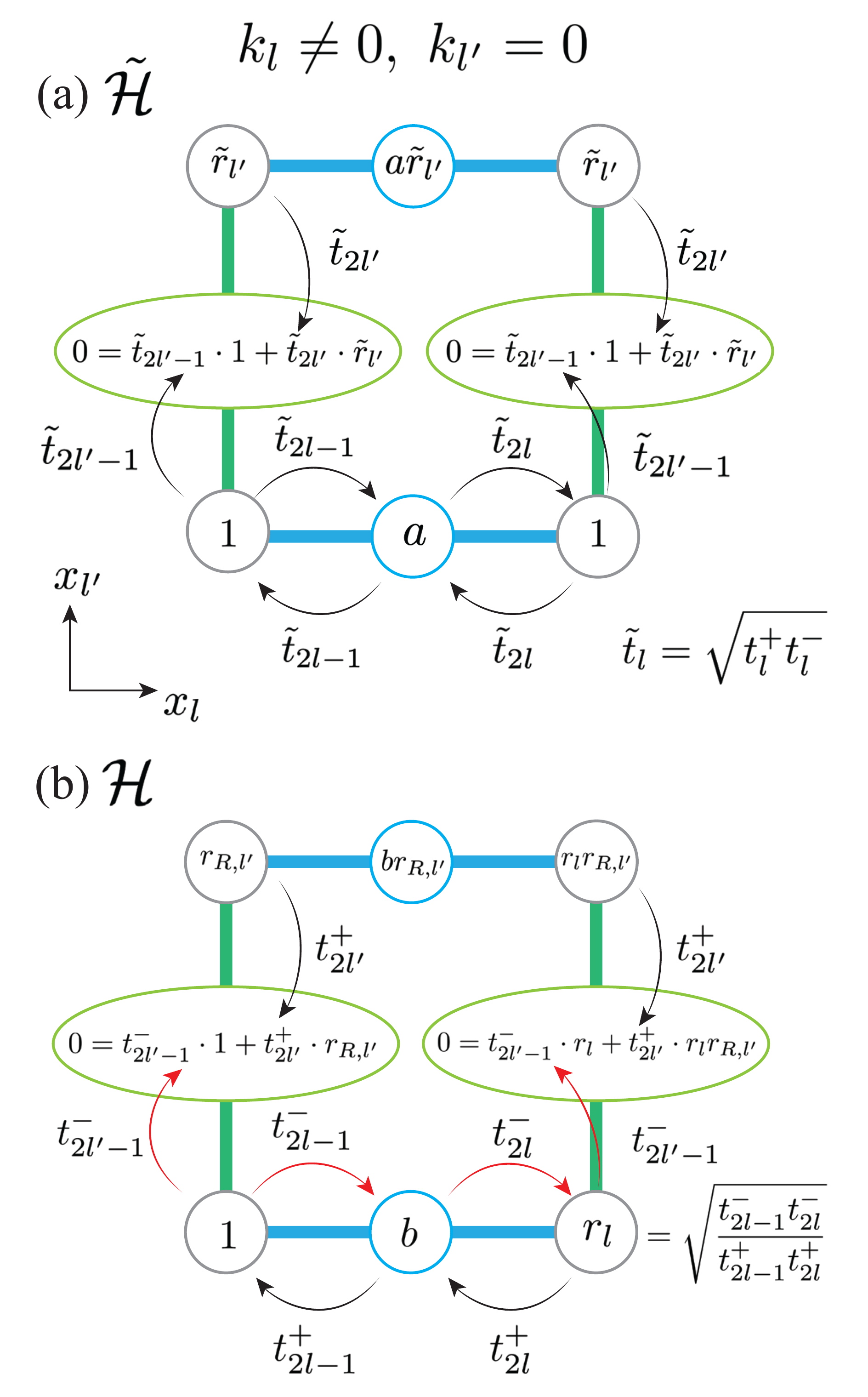} 
\caption{\textcolor{black}{Illustration of the mapping $S$ from (b) the nonreciprocal Hamiltonian $\mathcal{H}$ in \Cref{eq:hl} to (a) its reciprocal counterpart $\tilde{\mathcal{H}}$ in \Cref{eq:map1}. The weight of their wave functions in Eqs.~(\ref{eq:tpsi})  and (\ref{eq:psi_rl})  (right eigenstates) on each site is explicitly shown in the gray ($A$), blue ($B_l$) and green ($B_{l'}$) circles, where the values of $a$ and $b$ can be determined from associated Bloch and non-Bloch Hamiltonians. For boundary modes of both $\tilde{\mathcal{H}}$ and $\mathcal{H}$ with vanishing momentum $k_{l'} = 0$ along the direction $x_{l'}$, the destructive interference effect occurs on $B_{l'}$ (green) sites, giving rise to localization parameters $\tilde{r}_{l'}$ in 
\Cref{eq:map2} and $r_{R,l'}$ in 
\Cref{eq:rRL}. On top of this, in the original  $\mathcal{H}$, NHSE takes place in every direction ($x_l$ and $x_{l'}$) bringing forth localization parameters $r_l$ and  $r_{l'}$, an observation verified by the relation $r_{R,l'} = {r}_{l'}\tilde{r}_{l'}$.}}
\label{fig:rs}
\end{figure}

To remove the nonreciprocity in the original Hamiltonian of \Cref{eq:hl},  we generalize the transformation matrix approach previously developed in one-dimensional NH SSH models \cite{yao2018,yang2022} to higher dimensions.
Since it is more convenient to express the matrix algebra in the operator form, let us define a transformation matrix $S$ as a nonunitary gauge transform on the operators. On each of $A$ and $B_{l=1,\dots,d}$ motifs, the annihilation operators ($\forall l$) are transformed according to 
 \begin{gather}
  c_{\vec{j}, A} \to \prod_{i=1}^{d} r_{i}^{\ j_{i}-1}  c_{\vec{j}, A}, \quad
   c_{\vec{j}, B_{l}} \to  
   {\delta r}_{l}  \prod_{i=1}^d  r_{i}^{\ j_{i} -1}  c_{\vec{j}, B_l}, \label{eq:gauge_s}
 \end{gather}
and simultaneously, the creation operators undergo
 \begin{gather}
   c^\dagger_{\vec{j}, A} \to \prod_{i=1}^d r_{i}^{-(j_{i}-1)}   c^\dagger_{\vec{j}, A}, \quad 
   c^\dagger_{\vec{j}, B_l} \to  {\delta r}^{-1}_{l}  \prod_{i=1}^d  r_{i}^{-(j_{i} -1)}c^\dagger_{\vec{j}, B_l}. \label{eq:gauge_s1}
 \end{gather}
The gauge transform $S$ leaves the commutation (anticommunitation) relations of bosons (fermions) intact: $S^{-1}[c_{\vec{j},\lambda}, c_{\vec{j}',\lambda'}^\dagger]_\mp S = \delta_{\vec{j},\vec{j}'}\delta_{\lambda,\lambda'}$ where $[\hat{C},\hat{D}]_\mp = \hat{C}\hat{D} \mp \hat{D}\hat{C}$.
Introducing a set of localization parameters into $S$,
 \begin{gather}
 r_l  = \sqrt{\frac{t_{2l-1}^-t_{2l}^-}{t_{2l-1}^+t_{2l}^+}}, 
 \quad {\delta r}_{l} = \sqrt{\frac{t_{2l-1}^-}{t_{2l-1}^+}}, \label{eq:rl}
 \end{gather}
one arrives at the reciprocal Hamiltonian,
   \begin{align}
\tilde{\mathcal{H}} &= S^{-1}\mathcal{H}S \label{eq:map1}  \\
&= \sum_{\vec{j}} \sum_{l=1}^{d}  \tilde{t}_{2l-1}  c^\dagger_{\vec{j}, A} c_{\vec{j}, B_l} + \tilde{t}_{2l-1}c^\dagger_{\vec{j}, B_l} c_{\vec{j},A} \notag \\
 &\phantom{= \sum_{\vec{j}} \quad} + \tilde{t}_{2l}  c^\dagger_{\vec{j}, B_l}c_{\vec{j}+\vec{e}_l, A}   + \tilde{t}_{2l} c^\dagger_{\vec{j} + \vec{e}_l, A} c_{ \vec{j}, B_l}, \notag
   \end{align} 
with the symmetric hopping terms
    \begin{gather}
      \tilde{t}_{l} = \sqrt{t_{l}^+t_{l}^-}.
    \end{gather}
In particular, $\tilde{\mathcal{H}}$ can be non-Hermitian if $\im[\tilde{t}_l] \ne 0$. Yet its eigenvectors are devoid of skin effect in absence of nonreciprocity.
Schematically, \Cref{fig:rs} shows the lattice models captured by the reciprocal and nonreciprocal Hamiltonians $\tilde{\mathcal{H}}$ and $\mathcal{H}$ mapped to each other through the transformation matrix $S$.
\subsubsection{Diagonalization via $O(N^n)$ subgroups}

Our goal is now reduced to the diagonalization of the symmetric matrix $\tilde{\mathcal{H}}$, which can be achieved in the subspaces of $O (N^n)$ bulk and boundary modes $| {\tilde{\psi}}_{n}\rangle$ without NHSE:
 \begin{gather}
   \tilde{\mathcal{H}} = \sum_n  \sum_{m} \tilde{E}^{\text{OBC}}_{n,m}  | {\tilde{\psi}}_{nR,m}\rangle \langle {\tilde{\psi}}_{nL,m}|. \label{eq:h_com}
 \end{gather}
Here, $m$ details the band index after an independent diagonalization in each subspace. For the biorthogonal eigenvectors defined in Eqs.~(\ref{eq:rl}) and (\ref{eq:bio}),  $| {\tilde{\psi}}_{nR,m}\rangle 
\ne |{\tilde{\psi}}_{nL,m}\rangle$ if $\tilde{
\mathcal{H}}$ is non-Hermitian.

We take notice that the sum over $n$ is twofold. First, it includes all orders of eigenmodes $n=0, 1, \dots, d$, or more specifically, $O(1)$ corner, $O(N)$ edge, $\dots$, $O(N^d)$ bulk modes. Second, for each $n$, the sum encompasses all possible realizations of $O(N^n)$ eigenmodes covering different motifs $AB_{l_1}\cdots B_{l_n}$, which can be labeled by $\bar{l}_n = \{l_1, \dots, l_n \}$. 
For instance, in 2D, $O(N)$ edge modes can be realized in two ways: $\bar{l}_1 = \{ 1 \}$ and $\{ 2 \}$, one living on $AB_1$ motifs  and the other on $AB_2$ motifs. The twofold summation over $n$ guarantees the decomposition of the reciprocal Hamiltonian  into its $O(N^n)$ subgroups is complete. 

\Cref{eq:h_com} describes a procedure of 
 dimension reduction \cite{flore2019e} from the original space of dimension $d$ to the subspace of codimension $D = n$ supporting the $O(N^n)$ eigenmodes. To systematically achieve this,  we introduce another gauge transform $U_n$ accompanied by an infinitesimal parameter $\epsilon \to 0^+$:
 \begin{gather}
  c_{\vec{j}, A} \to \prod_{i \notin \bar{l}_n} \tilde{r}_{i}^{\ j_{i}}   c_{\vec{j}, A}, \quad
    c_{\vec{j}, B_l} \to  \prod_{i \notin \bar{l}_n} \tilde{r}_{ i}^{\ j_{i}}   c_{\vec{j}, B_l}, \notag \\
   c_{\vec{j}, B_{l'}} \to  \epsilon \ c_{\vec{j}, B_{l'}}, \label{eq:map2_0}
 \end{gather}
where $l \in \bar{l}_n$ ($\ l' \notin \bar{l}_n$) denotes an ensemble of occupied (unoccupied) $B$ motifs of the $O(N^n)$ eigenmodes under consideration. The parameter $\epsilon$ plays the role of formally canceling the occupation on all  $B_{l'}$ motifs.
In the same manner, the creation operators are transformed under $U_n$ according to
  \begin{gather}
    c_{\vec{j}, A}^\dagger \to \prod_{i \notin \bar{l}_n} \tilde{r}_{i}^{\ -j_{i}}  c^\dagger_{\vec{j}, A}, \quad
   c^\dagger_{\vec{j}, B_l} \to  \prod_{i \notin \bar{l}_n} \tilde{r}_{ i}^{\ -j_{i}}  c^\dagger_{\vec{j}, B_l}, \notag \\
    c^\dagger_{\vec{j}, B_{l'}} \to  \epsilon^{-1} \ c^\dagger_{\vec{j}, B_{l'}}. \label{eq:map2_1}
 \end{gather}
Applying $U_n$ to $\tilde{\mathcal{H}}$ in \Cref{eq:map1}, we obtain $  U_n^{-1} \tilde{\mathcal{H}} U_n =  \tilde{\mathcal{H}}_n + \tilde{\mathcal{H}}'_n$, where $\tilde{\mathcal{H}}_n$ acts on the subspace of $O(N^n)$ eigenmodes,
 \begin{gather}
    \tilde{\mathcal{H}}_n  = \sum_{\vec{j}} \sum_{l \in \bar{l}_n} \tilde{t}_{2l-1} c^\dagger_{\vec{j}, A} c_{\vec{j}, B_l}  + \tilde{t}_{2l} c^\dagger_{\vec{j}, B_l} c_{\vec{j}+\vec{e}_l, A}  + \hc, \label{eq:hnh}
\end{gather}
while $\tilde{\mathcal{H}}'_n$ extends beyond this subspace,
\begin{align}
     \left.   \tilde{\mathcal{H}}'_n \right|_{\epsilon \to 0^+}  &= \sum_{\vec{j}} \sum_{l' \notin \bar{l}_n} \epsilon^{-1} \prod_{i \notin \bar{l}_n} \tilde{r}_{i}^{\ j_{i}}
   \\
     &\phantom{=} \left(\tilde{t}_{2l'-1}  c^\dagger_{\vec{j}, B_{l'}} c_{\vec{j}, A}  + \tilde{t}_{2l'} \cdot \tilde{r}_{l'} c^\dagger_{\vec{j}, B_{l'}} c_{\vec{j}+\vec{e}_{l'}, A}\right).  \notag
 \end{align}
Once $\tilde{t}_{2l'-1}  + \tilde{t}_{2l'} \cdot \tilde{r}_{l'} = 0$, the interference from neighboring occupied $A$ sites on $B_{l'}$ sites vanishes. All $B_{l'}$ motifs retain zero occupancy and $\tilde{\mathcal{H}}'_n  = 0$ effectively. Figure~\ref{fig:rs}(a) depicts this destructive interference effect \cite{flore2017a,flore2019e} along the $x_{l'}$ direction, which also renders zero surface momentum $k_{l'} = 0$. 
Hence, choosing localization parameters in $U_n$ as
\begin{gather}
\tilde{r}_{l} = - \frac{\tilde{t}_{2l-1}}{\tilde{t}_{2l}} = - \sqrt{\frac{t_{2l-1}^+t_{2l-1}^-}{t_{2l}^+t_{2l}^-}}, \label{eq:map2}
\end{gather}
and taking the physical limit $\epsilon \to 0^+$,
one  establishes the decomposition  of $\tilde{\mathcal{H}}$ into each $O(N^n)$ subgroup:
  \begin{gather}
  U_n^{-1} \tilde{\mathcal{H}} U_n =  \tilde{\mathcal{H}}_n.
 \end{gather}

\subsubsection{Solvability from spectral mirror symmetry}
\textcolor{black}{It turns out that  the OBC spectrum of $\tilde{\mathcal{H}}_n$ in \Cref{eq:hnh} is exactly solvable due to spectral mirror symmetry. As can be seen in Figs.~\ref{fig:lattice} and \ref{fig:rs}(a), under complete OBC, the last unit cells of the hypercubic lattice are broken in every direction and the boundaries always terminate at the $A$ motif.
 The spectrum is thus invariant by reversing the surface momentum: $\tilde{E}_{n,\alpha}^{\text{OBC}}(k_l) = \tilde{E}_{n,\alpha}^{\text{OBC}}(-k_l), \forall l \in \bar{l}_n$ with $\alpha \in \{0, \pm \} $.
It leads to identical OBC and PBC spectra of $\tilde{\mathcal{H}}_n$.}

Taking into account the nonunitary gauge transform $U_n$ in \Cref{eq:map2_0},
 one resolves  the eigenvalue decomposition problem of the reciprocal Hamiltonian in \Cref{eq:h_com}:
 \begin{align}
   &\tilde{E}^{\text{OBC}}_{n,\alpha} (
\vec{k}) = \tilde{E}_{n,\alpha}^{\text{PBC}}(\vec{k}), \notag \\ 
&\u{\tilde{\psi}}_{nR, (\alpha, \vec{k})}(\vec{j}) \propto  \prod_{l' \notin \bar{l}_n} (\tilde{r}_{ l'})^{j_{l'}}  e^{i\vec{k} \cdot \vec{j}}  \u{\tilde{u}}_{nR,\alpha}(\vec{k}), \label{eq:tpsi} \\
&\u{\tilde{\psi}}_{nL, (\alpha, \vec{k})}(\vec{j}) \propto  \prod_{l' \notin \bar{l}_n} (\tilde{r}_{l'}^*)^{j_{l'}}  e^{i\vec{k} \cdot \vec{j}}  \u{\tilde{u}}_{nL,\alpha}(\vec{k}),
\notag
 \end{align}
with the band index $m = (\alpha, \vec{k})$.
 The localization parameter of the left eigenmode changing from $\tilde{r}_{l'}$ to  $\tilde{r}^*_{l'}$ is determined from the relation
  \begin{gather}
  \u{\tilde{\psi}}_{nL, (\alpha, \vec{k})} (\vec{\gamma}) =  \u{\tilde{\psi}}^*_{nR, (\alpha, \vec{k})} (\vec{\gamma}), \label{eq:rhsym}
  \end{gather} 
  with $\vec{\gamma} = (\gamma_1, 
\gamma_2, \dots, \gamma_{2d})$, on account of the symmetric reciprocal matrix  $\tilde{\mathcal{H}}^T(\vec{\gamma}) =  \tilde{\mathcal{H}}(\vec{\gamma})$ and $\tilde{\mathcal{H}}^\dagger(\vec{\gamma}) =  \tilde{\mathcal{H}}^*(\vec{\gamma})$.
Besides, $\u{\tilde{u}}_{nR(L),\alpha}(\vec{k})$ denotes the right (left) biorthogonal column eigenvector of the associated  Bloch Hamiltonian: 
\begin{gather} 
\tilde{H}_n(\vec{k}) = \sum_{\alpha} \tilde{E}^{\text{PBC}}_{n,\alpha} (\vec{k}) | {\tilde{u}}_{nR,\alpha}(\vec{k})\rangle \langle {\tilde{u}}_{nL,\alpha}(\vec{k})|,
\end{gather}
where  $\u{\tilde{u}}_{Ln,\alpha}^*(\vec{k}) \cdot \u{\tilde{u}}_{Rn, 
\alpha'}(\vec{k}) = \delta_{\alpha, \alpha'}$.
In the basis $ \Psi_{n,\vec{k}} = (c_{\vec{k},A}, c_{\vec{k}, B_{l_1}}, \dots, c_{\vec{k}, B_{l_n}})^T$, $\tilde{\mathcal{H}}_n (\vec{k})$ takes the form
  	\begin{align}
	   \tilde{H}_n(\vec{k}) &= \begin{pmatrix}
	       0 & \tilde{X}_{n,-} \\
          \tilde{X}_{n,+}^T & 0
	   \end{pmatrix},  \\
      \tilde{X}_{n, \pm} &= \begin{pmatrix}
          \tilde{t}_{2l_1-1} + \tilde{t}_{2l_1} e^{\pm ik_{l_1}},  & \dots, & \tilde{t}_{2l_n-1} + \tilde{t}_{2l_n} e^{\pm ik_{l_n}} \notag
      \end{pmatrix}.
	\end{align}
It is easy to recognize that $\tilde{\mathcal{H}}_n (\vec{k})$ also respects generalized chiral symmetry, and hosts a pair of opposite energy bands: $ \tilde{E}^{\text{PBC}}_{n, +} (\vec{k})  = - \tilde{E}^{\text{PBC}}_{n, -} (\vec{k})$, together with  $(n-1)$ degenerate zero-energy boundary (bulk) flat bands for $1 < n < d$ ($n = d$): $\tilde{E}_{n, 0}^{\text{PBC}}(\vec{k}) = 0$. 

\subsubsection{Momentum shifts in nonreciprocal Hamiltonian}

We are now ready to obtain the analytical solutions to the original nonreciprocal NH Hamiltonian by taking the inverse of the first mapping $S$ in \Cref{eq:map1},
\begin{align}
  \mathcal{H} = S \tilde{\mathcal{H}}S^{-1} &= \sum_n  \sum_{\alpha, \vec{k}} \tilde{E}^{\text{OBC}}_{n, \alpha} (\vec{k}) S \tilde{P}_{n,\alpha}(\vec{k}) S^{-1}, \notag \\ \tilde{P}_{n,\alpha}(\vec{k}) &= |{\tilde{\psi}}_{nR,(\alpha,\vec{k})}\rangle \langle {\tilde{\psi}}_{nL,(\alpha,\vec{k})}|. \label{eq:map3}
\end{align}
Thanks to its reciprocal counterpart in 
\Cref{eq:h_com}, $\mathcal{H}$ is already formally decomposed into $O (N^n)$ skin modes.  The projection operator $\tilde{P}_{n,\alpha}(\vec{k})$ enables us to focus on the subspace of occupied motifs where an effective basis ${\Psi}_{n,\vec{j}} = (c_{\vec{j},A}, c_{\vec{j},B_{l_1}}, \dots, c_{\vec{j},B_{l_n}})^T$ has been constructed for each unit cell. In this subspace,  the transformation matrix $S$ of \Cref{eq:gauge_s} finds expression as
 \begin{gather}
  S\tilde{P}_nS^{-1} = S_n\tilde{P}_nS_n^{-1}, \quad   S_n = \bigoplus_{\vec{j}} \prod_{i \in \bar{l}_n} (r_{i})^{j_i}   S_{n, \vec{j}}, \notag \\ S_{n,\vec{j}} =   \diag \{1, 
  \delta r_{l_1}, \dots, \delta r_{{l_n}} \} \times \prod_{i \in \bar{l}_n} (r_{i})^{-1}. \label{eq:sn}
 \end{gather}
Combining Eqs.~(\ref{eq:tpsi})-(\ref{eq:sn}), we identify in the $\vec{j}$th unit cell the essential building block of a non-Bloch Hamiltonian:
  \begin{gather}
      S_{n,\vec{j}} \tilde{H}_n(\vec{k}) S_{n,\vec{j}}^{-1} = \begin{pmatrix}
	       0 & {Y}_{n,-} \\
           {Y}_{n,+}^T & 0
	   \end{pmatrix}, \label{eq:hnb} \\
    \begin{split}
    & {Y}_{n,\pm} = \notag \\
    & \left[t_{2l_1-1}^\mp + t_{2l_1}^\pm (r_{l_1})^{\pm 1} e^{\pm ik_{l_1}},  \dots,  t_{2l_n-1}^\mp + t_{2l_n}^\pm (r_{l_n})^{\pm 1} e^{\pm ik_{l_n}} \right].  \notag 
     \end{split}
  \end{gather}
Comparing with the submatrix of the original Bloch Hamiltonian in \Cref{eq:hbloch} generated by the same motifs $AB_{l_1}\cdots B_{l_n}$, an equality can be established through imaginary momentum shifts: 
$\forall l \in \bar{l}_n$,
\begin{gather}
     S_{n,\vec{j}} \tilde{H}_n(k_l)  S_{n,\vec{j}}^{-1} =  H_{n}(k_l - i\ln r_l). 
     \label{eq:ms1}
  \end{gather}
Finally, by making a change of variables $k'_l = k_l - i\ln r_l$ and taking the decomposition 
\begin{gather}
{H}_n(\vec{k}') = \sum_{\alpha, \vec{k}} {E}^{\text{PBC}}_{n,\alpha} (\vec{k}') | {{u}}_{Rn, 
\alpha}(\vec{k}')\rangle \langle {{u}}_{Ln,\alpha}(\vec{k}')|,
\end{gather} we obtain the eigenvalue solutions to the original nonreciprocal Hamiltonian:
    \begin{gather}
       \mathcal{H} =  \sum_n  \sum_{\alpha, k} E_{n, \alpha}^{\text{OBC}} (\vec{k})| {\psi}_{nR,\alpha}(\vec{k})\rangle \langle {\psi}_{nL,\alpha}(\vec{k})|,\label{eq:dc_nh} 
       \\  E_{n, \pm}^{\text{OBC}} (k_l) =   E^{\text{PBC}}_{n,\pm} (k_l -i \ln r_l), \quad E_{n, 0}^{\text{OBC}} (\vec{k}) = 0 \ (n > 1), \notag 
    \end{gather}
with the analytical functions 
 \begin{align}
   E^{\text{PBC}}_{n,\pm} (\vec{k}) &=\pm \sqrt{\sum_{l=l_1}^{l_n} f_{2l-1,2l}(k_l)},  \notag \\    f_{i,j} (q) &= t_i^+t_i^- + t_j^+t_j^- + 2(t_it_j + \gamma_i \gamma_j) \cos q \notag \\
   &\phantom{=} +  2i(t_i \gamma_j + t_j \gamma_i) \sin q, \notag \\
        E_{n, \pm}^{\text{OBC}} (\vec{k}) &=\pm \sqrt{\sum_{l=l_1}^{l_n} g_{2l-1,2l}(k_l)}, \label{eq:gf} \\
     g_{i,j} (q) &= t_i^+t_i^- + t_j^+t_j^-+  \textcolor{black}{\sgn{(t_i^+t_j^+)} 2 \sqrt{t_i^+t_j^+t_i^-t_j^-}} \cos q. \notag
 \end{align}
While the spectrum under OBC inherits the spectral mirror symmetry reflected in the even-$g$ functions, the PBC spectrum is not invariant under $k_l \to -k_l$.  
 \begin{figure}[t]
\centering
\includegraphics[width=0.7\linewidth]{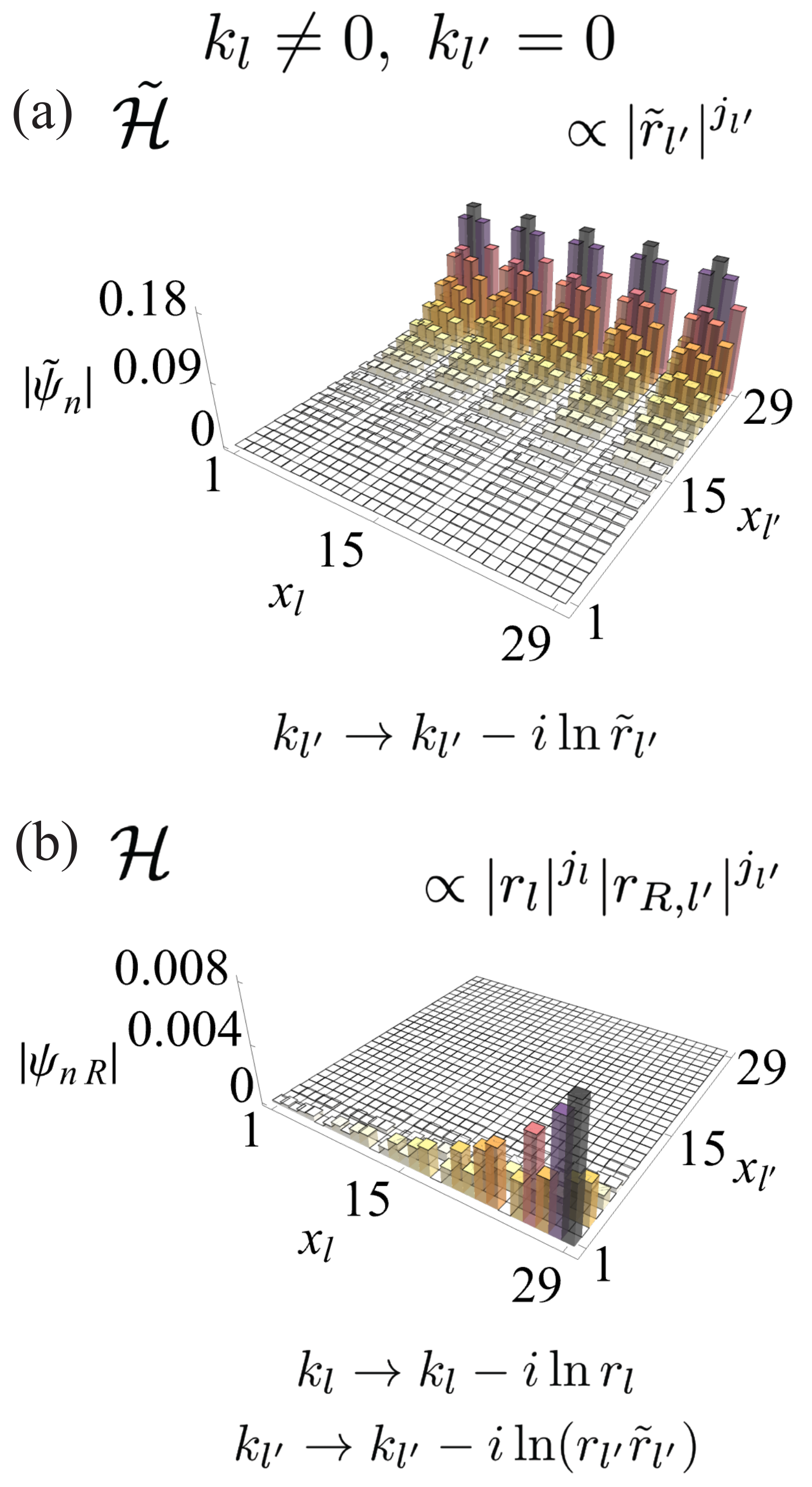} 
\caption{\textcolor{black}{To reveal the origins of the GSBZ, we examine in the NH Lieb model the right eigenvector of the edge $AB$ modes. It can also be taken as a growth of the microstructure in \Cref{fig:rs} over the two-dimensional (2D) lattice and as a visualization of the analytical solution in \Cref{eq:vec_edge}.  We choose a system  size $N_{1,2} = 15$ and select the band index $m=(+, \vec{k})$ at $\vec{k} = (k_1, k_2) = (\pi/3, 0)$. Hopping parameters are chosen as $t_1 = -1, t_2 = 2, \gamma_1 = 0.25,  \gamma_2 = 0$ ($t_3 = 1, t_4 = 0.3, \gamma_3 = 0.9,\gamma_4=0$) along the $x_l = x_1$ ($x_{l'} = x_2$) axis indexed by site number. It sets the localization parameters: (a) $\tilde{r}_2 = -1.45$; (b) $ r_1 =  1.29$ and $r_{R,2}=  r_2 \tilde{r}_2=  -0.33$. The localization profile of the wave function changes drastically after two successive imaginary momentum shifts performed on the Bloch phase factor $e^{i\vec{k}\cdot \vec{j}}$. The first shift, brought forth by the gauge transform $U_n$, produces boundary modes in $\tilde{\mathcal{H}}$ and the second one, arising from $S$, 
 generates higher-order NH skin modes in $\mathcal{H}$.}}
\label{fig:rs0}
\end{figure}

The inverse map  from the reciprocal $\tilde{\mathcal{H}}$ to the nonreciprocal $\mathcal{H}$ in \Cref{eq:map3} not only introduces 
a non-Bloch Hamiltonian with a momentum shift in \Cref{eq:ms1} but also imposes additional localization parameters $r_l$ in  \Cref{eq:rl} to all the motifs, both occupied ($\in \bar{l}_n$) and unoccupied  ($\notin \bar{l}_n$), through the gauge transform $S$ in \Cref{eq:gauge_s}. NHSE occurs 
if there exists at least one $\gamma_{2l-1} \ne 0$ or $\gamma_{2l} \ne 0$ with $l \in \{1, 2, \dots, d\}$ such that 
\begin{gather}
 |r_l| \ne 1, \quad \mathcal{H} \ne \tilde{\mathcal{H}}.
 \end{gather}
For instance, the skin effect on the occupied motifs is directly encoded in the factor  $\prod_{i \in \bar{l}_n} (r_{i})^{j_i}$ in $S_n$ of \Cref{eq:sn}. Combining two nonunitary gauge transforms,
the left and right $O (N^n)$ skin modes of $\mathcal{H}$ 
exhibit hybrid localization behaviors:
   \begin{align}
       &\u{\psi}_{nR, (\alpha, \vec{k})}(\vec{j}) \label{eq:psi_rl} \\
    &\propto    \prod_{l \in \bar{l}_n} (r_{l})^{j_{l}}  \prod_{l' \notin \bar{l}_n} (r_{R,l'})^{j_{l'}}  e^{i \vec{k} \cdot \vec{j}}  \u{u}_{nR,\alpha}(k_l -i \ln r_l), \notag \\
       & \u{\psi}_{nL, (\alpha, \vec{k})}(\vec{j}) \notag \\
      & \propto  \prod_{l \in \bar{l}_n} (r_{l}^*)^{-j_{l}}  \prod_{l' \notin \bar{l}_n} (r_{L,l'})^{j_{l'}}  e^{i \vec{k} \cdot \vec{j}}  \u{u}_{nL,\alpha}(k_l -i \ln r_l), \notag
    \end{align}
where the localization parameters read 
  \begin{align}
      r_{R,l} &= r_l\tilde{r}_{l} = -\frac{t^-_{2l-1}}{t^+_{2l}}, \label{eq:rRL} \\
      r^*_{L,l} &=  r^{-1}_{l}\tilde{r}_{l} = -\frac{t^+_{2l-1}}{t^-_{2l}}. \notag    
  \end{align}
In contrast with the nonskin eigenmodes in \Cref{eq:rhsym}, we identify the relation between right and left skin modes as
    \begin{gather}
  \u{{\psi}}_{nL,(\alpha, \vec{k})} (\vec{\gamma}
) = \u{{\psi}}^*_{nR,(\alpha, \vec{k})} (-\vec{\gamma}),  \label{eq:psirl}
    \end{gather}
 from the symmetries of the nonreciprocal Hamiltonian: $\mathcal{H}^T (\vec{\gamma}) = \mathcal{H} (-\vec{\gamma})$ and $ E_{n, \alpha}^{\text{OBC}}(\vec{k},\vec{\gamma
}) = E_{n, \alpha}^{\text{OBC}}(\vec{k}, -\vec{\gamma
})$. It brings about a spontaneous change in localization parameters $r_l$ defined in \Cref{eq:rl} when going from the right eigenmode to the left one: $r_l \to (r_l^*)^{-1}, \forall l$. Moreover, as illustrated in \Cref{fig:rs}(b), while $r$ is responsible for the exponential localization behavior of skin modes on occupied $A$ and $B_l$ motifs along the $x_l$ direction, 
$r_{R} = r \tilde{r}$ (or $r^*_{L} = r^{-1} \tilde{r}$) 
plays the role of canceling the contributions from the neighboring $A$ motifs to the unoccupied $B_{l'}$ motifs along the $x_{l'}$ direction. In terms of nonunitary gauge transforms, two cumulative effects give birth to $r_R$ ($r_L^*$), the NHSE characterized by $r$ in $S$ producing skin modes and the destructive interference effect  determined by $\tilde{r}$ in $U_n$ generating boundary modes.

Alternatively, $\u{\tilde{\psi}}_{nR}$ and $\u{\psi}_{nR}$ in Eqs.~(\ref{eq:tpsi}) and (\ref{eq:psi_rl}), belonging to the transformed reciprocal $\tilde{\mathcal{H}}$ and original nonreciprocal $\mathcal{H}$, can be obtained by performing imaginary momentum shifts on their respective Bloch wavefuctions: $\forall l \in \bar{l}_n, l' \notin  \bar{l}_n$, or equivalently, $ \forall k_l \ne 0, k_{l'} = 0$,
  \begin{align}
  \u{\tilde{\psi}}_{nR,(\alpha, \vec{k})}: \quad 
  &k_{l'} \to k_{l'} - i \ln \tilde{r}_{l'},  \notag \\
   \u{\psi}_{nR, (\alpha, \vec{k})}: \quad  &k_l \to k_l - i \ln r_{l}, \notag \\
   &k_{l'} \to k_{l'} - i \ln (r_{l'} \tilde{r}_{l'}), \label{eq:ms}
\end{align}
with $r_{R,l'} = r_{l'} \tilde{r}_{l'}$. As illustrated in \Cref{fig:rs0}, a boundary mode is first developed in $\tilde{\mathcal{H}}$, and then evolves to a higher-order NH skin mode in $\mathcal{H}$. We complete the construction of the exact GSBZ hosting right $O (N^n)$ skin modes in \Cref{eq:gsbz}.
Extended from the normal GBZ which describes bulk modes ($n=d$), the GSBZ is able to capture all orders of skin modes: $n = 0, 1, \dots, d$.

Starting from \Cref{eq:psi_rl},  the exact $O (N^n)$ skin modes can be further built by taking a superposition of these non-Bloch waves at opposite momenta $\pm k_l$, such that the total wave function meets the boundary condition, which requires all $B_l$ sites in the last broken unit cells to be empty (see \Cref{fig:lattice}). 
 It is also interesting to comment on the structures of biorthogonal eigenvectors $\u{u}_{nR(L)}$ of the non-Bloch Hamiltonian in \Cref{eq:hnb}.
 It is found that while the eigenvectors of the two dispersive bands in \Cref{eq:dc_nh} live on the $AB_{l_1}\cdots B_{l_n}$ motifs, the ones for those degenerate zero-energy flat bands have zero occupancy on the $A$ motif. By satisfying one more condition,  i.e., the eigenvalue equation, flatband eigenvectors can be constructed as $(n-1)$ linearly independent states covering $B_{l_1} \cdots B_{l_n}$ motifs, thus fulfilling the requirement of $O (N^n)$ skin modes of a codimension $D = n$.

\subsection{Example in $d=2$: NH Lieb lattice}
\label{sec:lieb}
\begin{figure}[t]
\centering
\includegraphics[width=1\linewidth]{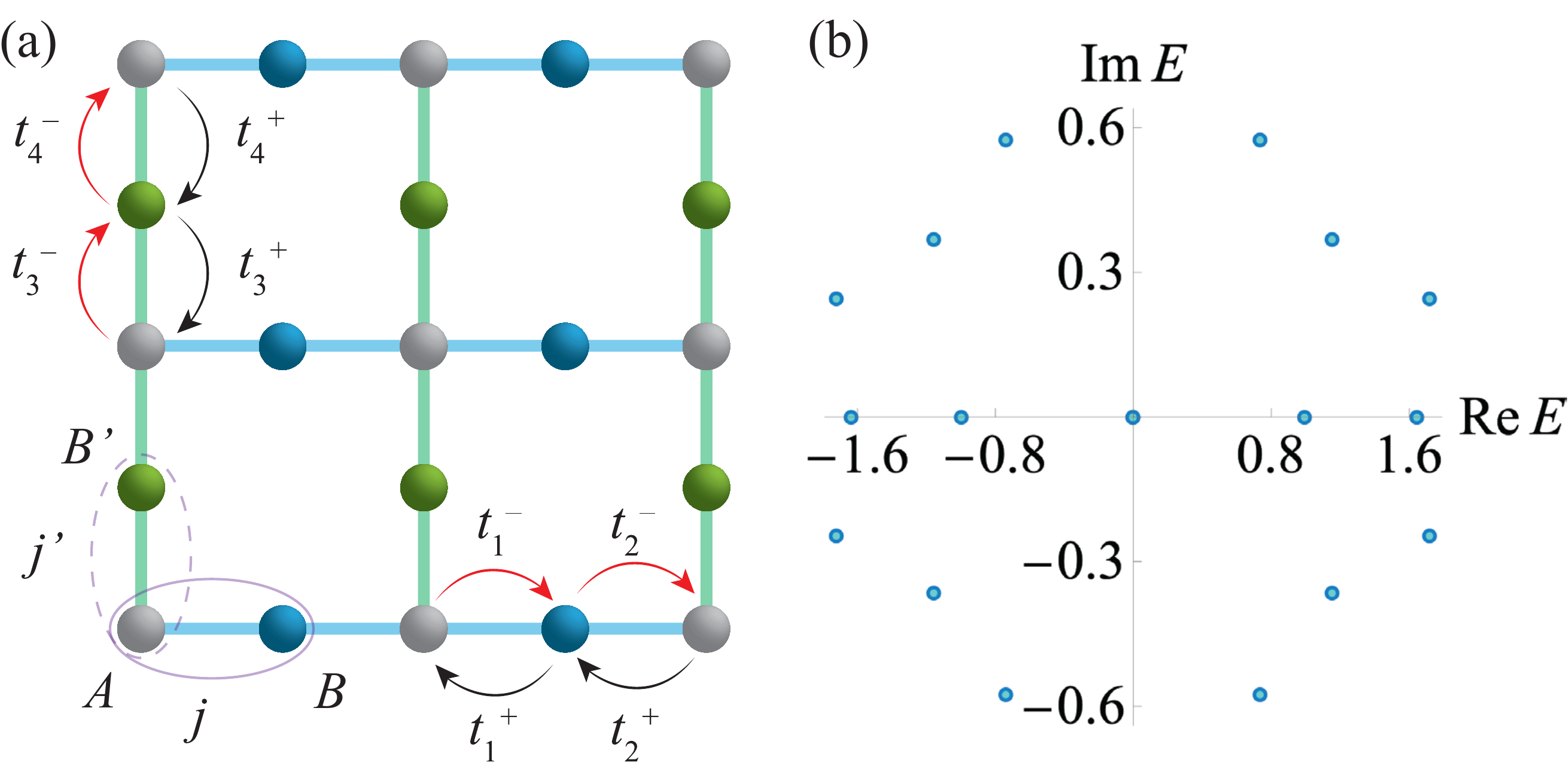}
\caption{(a) Lattice geometry of the NH Lieb lattice; (b) Comparison of complex eigenvalues between the numerical (dark blue) and analytical (light blue) results for a finite-size Lieb lattice with $N_{1,2} =3$. We choose distinct values for all hopping parameters in $t_l^\pm = t_l \pm \gamma_l$: $\{ t_1, t_2, t_3, t_4 \} = \{ 1.5, 1, 1.6, 1.2 \}$, $\{ \gamma_1, \gamma_2, \gamma_3, \gamma_4 \} = \{ \sqrt{3}, 0.2, \sqrt{2}, 0.4 \}$.}
\label{fig:lieb}
\end{figure}

In this section, we present the exact solutions to our NH hypercubic lattice in 2D, the NH Lieb model. Based on the results of gauge transforms in \Cref{eq:psi_rl}, we finalize the construction of  $O (N^n)$ skin modes by taking into account spectral mirror symmetry and boundary constraints under complete OBC.

\subsubsection{Complete spectrum}

Figure~\ref{fig:lieb}(a) illustrates the geometry of the NH Lieb lattice of size $(2N_1-1) \times  (2N_2-1)$, where we designate three motifs $ABB'$($=AB_1B_2$)  in each unit cell at the position $\vec{j} = (j, j')$ with $j=1, \dots, N_1$ and $j'=1, \dots, N_2$. There are $(3N_1N_2 - N_1 - N_2)$ sites, equal to the total number of $O (N^n)$ skin modes ($n=0,1,2$). Among them, we recognize one corner mode  at $k_0 = (0,0)$, $2(N_1-1) + 2(N_2-1)$ edge  modes at $k_{\text{edge},AB} = (k_1, 0)$ and $k_{\text{edge},AB'} = (0, k_2)$ as well as $3(N_1-1)(N_2-1)$ bulk  modes  at $k_{\text{bulk}} = (k_1, k_2)$, manifesting the third-order, second-order and first-order NHSEs in two dimensions respectively. 

Let us first test the exact spectrum of $O (N^n)$ skin modes  given by \Cref{eq:dc_nh} in the GSBZ,
  \begin{align}
     {O(1)}:   \ &E^{\text{OBC}}_{\text{corner}}(0,0) = 0,  \label{eq:on} \\
     {O(N)}:   \  &E^{\text{OBC}}_{\text{edge},AB,\pm}(k_1,0)  = \pm \sqrt{g_{1,2}(k_1)}, \notag \\
     &E^{\text{OBC}}_{\text{edge},AB',\pm}(0, k_2) = \pm \sqrt{g_{3,4}(k_2)}, \notag \\
     {O(N^2)}:\   &E^{\text{OBC}}_{\text{bulk}, 0} (k_1,k_2) = 0,  \notag \\
     &E^{\text{OBC}}_{\text{bulk}, \pm} (k_1,k_2) = \pm \sqrt{g_{1,2}(k_1) + g_{3,4}(k_2)}. \notag
  \end{align} 
Recalling our convention for nonzero surface momentum under OBC: $k_l = \frac{\pi \tilde{m}}{N_l} 
\in (0, \pi)$, with $\tilde{m} = 1, 2, \dots, N_l-1$ and the form of $g$ function in \Cref{eq:gf}, we find our analytical solutions are consistent with the results from numerical diagonalization of the real-space NH Hamiltonian allowing arbitrary system size and arbitrary hopping parameters. Figure~\ref{fig:lieb}(b) shows the precision of complex eigenvalues on a finite $5 \times 5$ NH Lieb lattice.

\begin{widetext}

\subsubsection{Exact corner modes}
We proceed to build NH skin modes on the Lieb lattice.  First, the zero-energy corner mode lives only on the $A$ motif in \Cref{fig:lieb}(a), and by setting correspondingly $\bar{l}_0 = \varnothing$ in  \Cref{eq:psi_rl}, is given by
 \begin{gather}
  |{\psi}_{R/L,0} \rangle = \mathcal{N}_{R/L} \sum_{j=1}^{N_1} \sum_{j'=1}^{N_2} r_{R/L,1}^j r_{R/L,2}^{j'} c^\dagger_{A, (j,j')} |0\rangle, \label{eq:cor2}
 \end{gather}
with localization parameters defined in \Cref{eq:rRL},
\begin{gather}
    	 r_{R,1} = -\frac{t_1^- }{t_2^+}, \quad r^*_{L,1} = -\frac{t_1^+}{t_2^-}, \quad
      r_{R,2} = -\frac{t_3^-}{t_4^+}, \quad r^*_{L,2} = -\frac{t_3^+}{t_4^-}.
    \end{gather}
The normalization factors $\mathcal{N}_{R/L}$ are introduced to meet the biorthogonal relation $ \langle {\psi}_{L0}  |{\psi}_{R0} \rangle = 1$, rendering
    \begin{gather}
    \mathcal{N}^*_L\mathcal{N}_R = \frac{(r^*_{L,1}r_{R,1}r^*_{L,2}r_{R,2})^{-1} (r^*_{L,1}r_{R,1}-1)(r^*_{L,2}r_{R,2}-1)}{ [(r^*_{L,1}r_{R,1})^{N_1}-1][(r^*_{L,2}r_{R,2})^{N_2}-1]}. 
   \end{gather}
Equation~(\ref{eq:cor2}) indicates that the localization of the left and right corner modes can be different depending on the choice of parameters $\{|r_{R,l}|, |r_{L,l}|\}$ with $l=1,2$.  It turns out that 
once the spectral winding number of the associated Bloch edge Hamitonian defined by
  \begin{gather}
   H_{AB_l}(k_l) = \begin{pmatrix}
	  		0 & t_{2l-1}^+  + t_{2l}^- e^{-ik_l} \\
			t_{2l-1}^-  + t_{2l}^+ e^{-ik_l} & 0 
		    \end{pmatrix}, \quad \omega_{l} = \frac{1}{2\pi i } \int_{-\pi}^{\pi} d k_l \partial_{k_l}
 \log \{ \det [ H_{AB_l}(k_l) ] \} \label{eq:ebh}
 \end{gather}
 becomes nontrivial, the left and right corner modes spontaneously separate in real space, each exponentially localized towards different corners \cite{NTOS,yang2022},
    \begin{gather}
         |\omega_{l} | = 1 \  \Leftrightarrow \ 
\sgn [\log (|r_{R,l}|)] \ne \sgn [\log (|r_{L,l}|)].
    \end{gather}

\subsubsection{Exact edge modes}
The NH Lieb model host two types of edge modes: one resides on the $AB$ motifs and the other on the $AB'$ motifs in \Cref{fig:lieb}(a). The spectral mirror symmetry in their OBC spectra,
   \begin{gather}
     E^{\text{OBC}}_{AB, \pm}(k_1) = E^{\text{OBC}}_{AB, \pm}(-k_1), \quad
     E^{\text{OBC}}_{AB', \pm}(k_2) = E^{\text{OBC}}_{AB', \pm}(-k_2),  \label{eq:mirror2}
   \end{gather} 
enables us to build their eigenvectors as a superposition of non-Bloch waves from \Cref{eq:psi_rl} at opposite momenta $\pm k_l$. Identifying $\bar{l}_1 = \{1\}$ and $\{2\}$, the right edge modes share the form, 
\begin{align}
  \u{{\psi}}^{AB}_{R,(\pm,k_1)} (j, j')  
  &=  \mathcal{N}_{R,AB} \frac{ r_1^j r_{R,2}^{j'}}{\sqrt{2N_1}} [ e^{ik_1j} \u{u}^{\text{OBC}}_{R,\pm} (k_1) - e^{-ik_1j} \u{u}^{\text{OBC}}_{R,\pm} (-k_1) ], \notag \\
     \u{{\psi}}^{AB'}_{R,(\pm,k_2)} (j,j')
     &= \mathcal{N}_{R,AB'} \frac{r_{R,1}^j r_2^{j'} }{\sqrt{2N_2}} [e^{ik_2j'} \u{u}^{\text{OBC}}_{R,\pm} (k_2) - e^{-ik_2j'} \u{u}^{\text{OBC}}_{R,\pm} (-k_2) ], \label{eq:vec_edge} 
\end{align}
with localization parameters chosen in \Cref{eq:rl},
  \begin{gather}
    r_1 = \sqrt{\frac{t_1^-t_2^-}{t_1^+t_2^+}}, \quad r_2 = \sqrt{\frac{t_3^-t_4^-}{t_3^+t_4^+}}. \label{eq:liebr}
  \end{gather}
In the above, $\u{u}^{\text{OBC}}_{R,\pm} (k_{l}) =\u{u}_{R,\pm}^{\text{PBC}} (k_l-i\ln r_l) $ denote two right eigenvectors  of non-Bloch edge Hamiltonians, which appear after the imaginary momentum shift: $k_l 
\to k_l -i \ln r_l$ in the Bloch Hamiltonians of \Cref{eq:ebh}. In the basis $\u{\psi}_{AB} (k_1) = (c_A (k_1), c_B (k_1))^T$ and $\u{\psi}_{AB'} (k_2) = (c_A (k_2), c_{B'} (k_2))^T$, they take the explicit form
  \begin{gather}
    H^{\text{non-Bloch}
    }_{\text{edge}, AB}(k_1) = \begin{pmatrix}
	  		0 & t_1^+ + t_2^- r_1^{-1}e^{-ik_1} \\
			t_1^- + t_2^+r_1 e^{ik_1} & 0 
		    \end{pmatrix}, \quad
 H^{\text{non-Bloch}
    }_{\text{edge}, AB'}(k_2) = \begin{pmatrix}
	  		0 & t_3^+ + t_4^- r_2^{-1}e^{-ik_2} \\
			t_3^- + t_4^+r_2 e^{ik_2} & 0 
		    \end{pmatrix},
  \end{gather}
which in turn gives
  \begin{gather}
   \u{u}^{\text{OBC}}_{R,\pm} (k_1)  
   = \frac{1}{\sqrt{2}} \begin{pmatrix}
       (t_1^+ + t_2^- {r_1}^{-1}e^{-ik_1})/E^{\text{OBC}}_{AB,\pm} (k_1)\\
       1
   \end{pmatrix}, \quad
 \u{u}^{\text{OBC}}_{R,\pm} (k_2)
 =
 \frac{1}{\sqrt{2}} \begin{pmatrix}
        (t_3^+ + t_4^- r_2^{-1}e^{-ik_2})/E^{\text{OBC}}_{AB',\pm} (k_2) \\
       1 
   \end{pmatrix}. \label{eq:ur_edge} 
  \end{gather}
In \Cref{eq:vec_edge}, the relative minus sign between two non-Bloch waves at $\pm k_l$ comes from the boundary condition: the total wave function must vanish on empty $B$ and $B'$ sites in last unit cells on both directions [see Figs.~\ref{fig:lieb}(a) and \ref{fig:lattice}]:
    \begin{gather}
 \left. \u{{\psi}}^{AB}_{R,(\pm,k_1)} (N_1, j')\right|_B = 0, \quad  \forall \ k_1, \forall \  j'; \quad
 \left. \u{{\psi}}^{AB'}_{R,(\pm,k_2)} (j, N_2)\right|_{B'} = 0, \quad   \forall \ k_2, \forall \ j.
    \end{gather}
    
The left edge modes can be obtained  directly  from the right ones according to their relation in \Cref{eq:psirl}:
    \begin{align}
  \u{{\psi}}^{AB}_{L,(\pm,k_1)} (j,j') &=  \mathcal{N}_{L,AB}  \frac{(r_1^*)^{-j} (r_{L,2})^{j'}}{\sqrt{2N_1}}   [ e^{ik_1j} \u{u}_{L,\pm} (k_1) - e^{-ik_1j} \u{u}_{L,\pm} (-k_1) ], \notag \\
     \u{{\psi}}^{AB'}_{L,(\pm,k_2)} (j,j') &=  \mathcal{N}_{L,AB'}  \frac{(r_{L,1})^j (r_2^*)^{-j'}}{\sqrt{2N_2}} [ e^{ik_2j'} \u{u}_{L,\pm} (k_2) - e^{-ik_2j'} \u{u}_{L,\pm} (-k_2) ],  \label{eq:vec_edge1}
\end{align}
where
  \begin{gather}
   \u{u}_{L,\pm} (k_1) = \frac{1}{\sqrt{2}} \begin{pmatrix}
      [ (t_1^- + t_2^+ r_1 e^{ik_1})/E^{\text{OBC}}_{AB,\pm} (k_1)]^*\\
       1
   \end{pmatrix}, \quad
 \u{u}_{L,\pm} (k_2)  =  
 \frac{1}{\sqrt{2}} \begin{pmatrix}
        [(t_3^- + t_4^+ r_2 e^{ik_2})/E^{\text{OBC}}_{AB',\pm} (k_2)]^* \\
       1
   \end{pmatrix}. \label{eq:ul_edge} 
  \end{gather}
  
It can be checked that the right and left eigenvectors of non-Bloch edge Hamiltonians in Eqs.~(\ref{eq:ur_edge}) and (\ref{eq:ul_edge}) are already biorthogonal to each other: $\u{u}^*_{L,\alpha}(k_1) \cdot \u{u}_{R,\alpha'}(k_1) = \u{u}^*_{L,\alpha}(k_2) \cdot \u{u}_{R,\alpha'}(k_2)  = \delta_{\alpha, \alpha'}$. To satisfy  ${\u{\psi}}_{Lm}^{AB *} \cdot {\u{\psi}}_{Rm'}^{AB} = {\u{\psi}}_{Lm}^{AB' *} \cdot {\u{\psi}}_{Rm'}^{AB'}  = \delta_{m,m'}$ with band index $m = (\pm, k_l)$, 
we can fix the normalization factors: 
 \begin{gather}
   \mathcal{N}_{L,AB}^* \mathcal{N}_{R,AB} = \frac{(r^*_{L,2}r_{R,2})^{-1} (r^*_{L,2}r_{R,2}-1)}{(r^*_{L,2}r_{R,2})^{N_2}-1}, \quad  \mathcal{N}_{L,AB'}^* \mathcal{N}_{R,AB'} = \frac{(r^*_{L,1}r_{R,1})^{-1} (r^*_{L,1}r_{R,1}-1)}{(r^*_{L,1}r_{R,1})^{N_1}-1}.
 \end{gather}

 \subsubsection{Exact bulk  modes}
The construction of bulk modes on the NH Lieb lattice is more involved because, given $\vec{k} = (k_1, k_2)$ and $\alpha \in \{ 0, \pm\}$, there are four possible combinations of momentum allowed by spectral mirror symmetry:
   \begin{align}
     E_{\text{bulk},\alpha}^{\text{OBC}} (k_1, k_2) &=  E_{\text{bulk},\alpha}^{\text{OBC}} (-k_1, k_2)  \notag \\ = E_{\text{bulk},\alpha}^{\text{OBC}} (k_1, -k_2) &= E_{\text{bulk},\alpha}^{\text{OBC}} (-k_1, -k_2). 
   \end{align}  
Let us start from an arbitrary superposition of these non-Bloch waves covering the $ABB'$ motifs and holding the form of \Cref{eq:psi_rl} with $\bar{l} = \{1, 2\}$:
   \begin{align}
\u{{\psi}}_{R,(\alpha, \vec{k})} (j,j') &= \frac{r_1^j{r_2}^{j'}}{2\sqrt{N_1N_2}} \left[ C_1 e^{i(k_1j + k_2j')} \u{u}_{R,\alpha}(k_1,k_2) \right.  \label{eq:vec_bulk0}\\
&  \quad \quad \left. +C_2 e^{i(-k_1j + k_2j')} \u{u}_{R,\alpha} (-k_1,k_2) +C_3 e^{i(k_1j - k_2j')} \u{u}_{R,\alpha} (k_1,-k_2) +C_4 e^{-i(k_1j + k_2j')} \u{u}_{R,\alpha}(-k_1,-k_2) \right]. \notag
       \end{align}
Here, $\u{u}_{R, \alpha} (\vec{k})$ denote three right eigenvectors  of the non-Bloch bulk Hamiltonian, which is transformed from the Bloch one in \Cref{eq:hbloch} according to momentum shifts: $k_1 
\to k_1-i\ln r_1, k_2 \to  k_2-i\ln r_2$. In the basis $\u{\psi} (\vec{k}) = (c_A (\vec{k}), c_B (\vec{k}),  c_{B'} (\vec{k}))^T$,
 \begin{gather}
	  H^{
\text{non-Bloch}}_{\text{bulk}}(k_1, k_2) = \begin{pmatrix}
	  		0 & t_1^+ + t_2^- r_1^{-1}e^{-ik_1} &    t_3^+ + t_4^-r_2^{-1} e^{-ik_2} \\
			t_1^- + t_2^+r_1 e^{ik_1} & 0 & 0 \\
			 t_3^- + t_4^+ r_2e^{ik_2} & 0 & 0
		    \end{pmatrix}. 
	\end{gather}
A convenient choice for $\u{u}_{R, \alpha} (\vec{k})$ turns out to be
    \begin{align}
    &\u{u}_{R,0}(k_1, k_2) = \mathcal{N}_{R,0} (k_1, k_2) \begin{pmatrix}
        0 \\
        t_3^+ + t_4^- r'^{-1} e^{-ik_2} \\
        -(t_1^+ + t_2^- r^{-1} e^{-ik_1})
    \end{pmatrix}, \notag \\
    &\u{u}_{R,\pm} (k_1, k_2) = \mathcal{N}_{R, \pm} (k_1, k_2)  \begin{pmatrix}
     E_{\text{bulk},\pm}^{\text{OBC}}(k_1, k_2) /[ (t_1^- + t_2^+ r e^{ik_1})   (t_3^- + t_4^+ r' e^{ik_2})] \\
     (t_3^- + t_4^+ r' e^{ik_2})^{-1}
    \\
         (t_1^- + t_2^+ r e^{ik_1})^{-1}    
    \end{pmatrix}, \label{eq:ubk}
    \end{align}
with  mirror-symmetric normalization factors,
  \begin{gather}
  \mathcal{N}_{R,0} (k_1, k_2) = \frac{1}{\sqrt{g_{1,2}(k_1)+ g_{3,4}(k_2)}}, \quad
  \mathcal{N}_{R, \pm}(k_1, k_2) = \sqrt{\frac{g_{1,2}(k_1)  g_{3,4}(k_2)}{2[g_{1,2}(k_1)+ g_{3,4}(k_2)]}},
 \end{gather}
 which satisfy
   \begin{gather}
 \mathcal{N}_{R, \alpha}(k_1, k_2) = \mathcal{N}_{R, \alpha}(-k_1, k_2) 
 = \mathcal{N}_{R, \alpha} (k_1, -k_2) = \mathcal{N}_{R, \alpha} (-k_1, -k_2).
  \end{gather}
In the same way, the coefficients $C_i$ in \Cref{eq:vec_bulk0} can be determined by ensuring that  the total wave function vanishes on all the $B$ and $B'$ sites belonging to the last unit cells in \Cref{fig:lieb}(a). With our choice of $\u{u}_{R, \alpha} (\vec{k})$ in \Cref{eq:ubk}, both the normalization factors and the  weight on the $B$ ($B'$) motif are mirror-symmetric with respect to $k_1 \to -k_1$ ($k_2 \to -k_2$). This feature, also present in eigenvectors for  non-Bloch edge Hamiltonians in \Cref{eq:ur_edge},
greatly simplifies meeting the boundary constraint.
$\forall \alpha \in \{ 0, \pm \}, \forall \vec{k}$, it is easy to verify that 
    \begin{gather}
  \left. \u{{\psi}}_{R,(\alpha, \vec{k})} (N_1,j')\right|_B = 0,  \forall \ j': \quad C_1 = -C_2, \quad C_3 = -C_4, \notag \\
    \left. \u{{\psi}}_{R,(\alpha, \vec{k})} (j,N_2)\right|_{B'} = 0, \forall \  j: \quad C_1 = -C_3, \quad C_2 = -C_4.
    \end{gather}
Choosing $(C_1,C_2,C_3,C_4) = (1,-1,-1,1)$, we arrive at a closed form for the right bulk modes:
   \begin{align}
\u{{\psi}}_{R,(\alpha, \vec{k})} (j,j') = &\frac{r_1^j{r_2}^{j'}}{2\sqrt{N_1N_2}} \left[ e^{i(k_1j + k_2j')} \u{u}_{R,\alpha} (k_1,k_2) 
- e^{i(-k_1j + k_2j')} \u{u}_{R,\alpha} (-k_1,k_2) - e^{i(k_1j - k_2j')} \u{u}_{R,\alpha} (k_1,-k_2) \right.  \notag \\
& \left.+ e^{-i(k_1j + k_2j')} \u{u}_{R,\alpha} (-k_1,-k_2) \right].
 \label{eq:vec_bulk}
       \end{align}

The left bulk modes follow from the relation in \Cref{eq:psirl}: 
    \begin{align}
\u{{\psi}}_{L,(\alpha, \vec{k})} (j,j') = &\frac{(r_1^*)^{-j}{(r_2^*)}^{-j'}}{2\sqrt{N_1N_2}} \left[ e^{i(k_1j + k_2j')} \u{u}_{L,\alpha} (k_1,k_2)  - e^{i(-k_1j + k_2j')} \u{u}_{L,\alpha} (-k_1,k_2)- e^{i(k_1j - k_2j')} \u{u}_{L,\alpha} (k_1,-k_2) \right. \notag \\
& \left. + e^{-i(k_1j + k_2j')} \u{u}_{L,\alpha} (-k_1,-k_2) \right],
\label{eq:vec_bulk1}
       \end{align}
where
 \begin{align}
   &\u{u}_{L,0} (k_1, k_2) = \mathcal{N}_{L,0}(k_1, k_2)  \begin{pmatrix}
        0 \\
        (t_3^- + t_4^+ r' e^{ik_2})^* \\
        -(t_1^- + t_2^+ r e^{ik_1})^*
    \end{pmatrix}, \notag  \\
 &\u{u}_{L,\pm} (k_1, k_2) = \mathcal{N}_{L,\pm}(k_1, k_2)\left[E_{\text{bulk},\pm}^{\text{OBC}}(k_1, k_2)\right]^*  \begin{pmatrix}
    1/[ (t_1^+ + t_2^- r^{-1} e^{-ik_1})   (t_3^+ + t_4^- {r'}^{-1} e^{-ik_2})]^* \\
     1/[E_{\text{bulk},\pm}^{\text{OBC}}(k_1, k_2)(t_3^+ + t_4^- {r'}^{-1} e^{-ik_2})]^*
    \\
1/[E_{\text{bulk},\pm}^{\text{OBC}}(k_1, k_2)(t_1^+ + t_2^- r^{-1} e^{-ik_1})]^*    
    \end{pmatrix}, 
   \end{align}
and $\mathcal{N}_L(k_1, k_2) = \mathcal{N}_R^*(k_1, k_2)$. Meanwhile, $\u{u}_{R/L, \alpha}(\vec{k})$ are biorthogonal to each other:  $\u{u}^*_{L,\alpha}(k_1,k_2) \cdot \u{u}_{R,\alpha'}(k_1,k_2) = \delta_{\alpha, \alpha'}$.

In the end, we verify that the analytical wave functions built for the corner mode in \Cref{eq:cor2}, the edge modes in Eqs.~(\ref{eq:vec_edge}) and (\ref{eq:vec_edge1}), and the bulk modes in Eqs.~(\ref{eq:vec_bulk}) and (\ref{eq:vec_bulk1}) mutually satisfy biorthogonal relations. We therefore obtain the entire set of  $O(N^n)$ skin eigenmodes of the NH Lieb model under complete OBC:
 \begin{gather}
  {\u{\psi}}_{nL,m}^* \cdot {\u{\psi}}_{n'R,m'}  = \delta_{n,n'} \delta_{m,m'}, \quad
\mathcal{H}_{\text{Lieb}}^{\text{NH}} = \sum_n \sum_{m} E^{\text{OBC}}_{n,m} |{\u{\psi}}_{nR,m} \rangle  \langle {\u{\psi}}_{nL,m}|, \label{eq:bor}
 \end{gather}
   \end{widetext}
with $n = 0,1,2$ and $m = (\alpha, \vec{k})$ indexing their complex energy bands in  \Cref{eq:on}.
Notably, our exact solutions are valid for the NH Lieb lattice of arbitrary system sizes $2(N_1-1) \times 2(N_2-1)$.

 \begin{figure*}[ht]
\centering
\includegraphics[width=2\columnwidth]{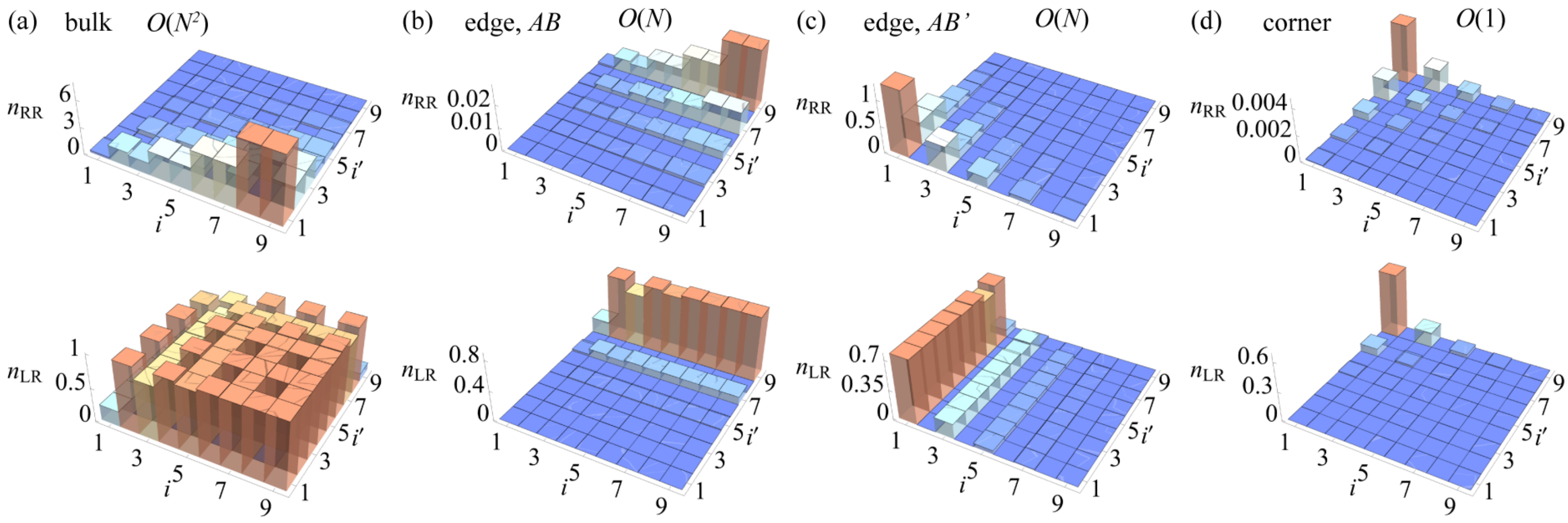}
\caption{Total density comparison for the NH Lieb lattice: $n_{RR}$ vs $n_{LR}$. While $n_{RR}$ of (a) the bulk , (b) edge $AB$ ,   (c) edge $AB'$ and  (d) corner  modes are centered towards different corners of the Lieb lattice, $n_{LR}$ shows whether the modes are of bulk, edge, or corner nature. $n_{LR}$ of the bulk modes becomes almost uniform (excluding holes, the empty sites on the Lieb lattice in \Cref{fig:lattice}). Summing over all types of skin modes, $n_{LR,\text{tot}}$ is normalized to $1$ on each available site.   We take small size $N_{1,2} = 5$ and choose hopping parameters $t_1 = -1, t_2 = 2, \gamma_1 = 0.25,  \gamma_2 = 0$ ($t_3 = 1, t_4 = 0.3, \gamma_3 = 0.5,\gamma_4=0$) along the $x_1$ ($x_2$) direction denoted by site index $i$ ($i'$). The localization parameters in Eqs.~(\ref{eq:loc1}) and (\ref{eq:loc2}) follow: $\{r_1, r_2 \} = \{1.291, 0.577 \}$, $\{r_{R,1}, r_{R,2} \} = \{0.625, -1.667 \}$ and $\{r_{L,1}, r_{L,2} \} = \{0.375, -5\}$.}
\label{fig:loc}
\end{figure*}
\subsection{Normal and biorthogonal densities}

\textcolor{black}{Based on the exact solutions, we can now study localization properties of different orders of skin modes in terms of total normal  ($n_{RR}$) and biorthogonal ($n_{LR}$) densities,
  \begin{gather}
    n_{RR}(\vec{i}) = \sum_m   \langle \psi_{Rm}|  \Pi_{\vec{i}} | \psi_{Rm} \rangle, \notag \\
    n_{LR}(\vec{i}) = \sum_m   \langle \psi_{Lm}|  \Pi_{\vec{i}} | \psi_{Rm} \rangle. \label{eq:den}
  \end{gather}
The summation goes over all bands $m = (\alpha, \vec{k})$ held by the type of skin modes of interest. The density operator at the site position $\vec{i} = (\vec{j}, \lambda)$ is defined by $\Pi_{\vec{i}} =  |e_{\vec{j}, \lambda} \rangle \langle e_{\vec{j},\lambda}|$ and $|e_{\vec{j}, \lambda} \rangle = c^\dagger_{\vec{j},\lambda} |0\rangle$ with $\vec{j}$ locating the unit cell and $\lambda \in \{ A, B_1, \dots, B_d \}$ the motif.}

\textcolor{black}{Let us compare the two densities in the NH Lieb model as an example. On one hand, the total normal density $n_{RR}$ collects the impact of localization factors $r$ and $r_R$ on each skin mode  $\u{\psi}_{Rm}$ from Eqs.~(\ref{eq:vec_bulk}), (\ref{eq:vec_edge}) and (\ref{eq:cor2}), such that in the unit cell $\vec{j} = (j, j')$,
    \begin{align}
       n_{RR}^{\text{bulk}
      } &\propto |r_1|^{2j}|r_2|^{2j'}, \label{eq:loc1}\\
       n_{RR}^{\text{edge}, AB
      } &\propto |r_1|^{2j}|r_{R,2}|^{2j'}, \notag \\
       n_{RR}^{\text{edge}, AB'
      } &\propto |r_{R,1}|^{2j}|r_{2}|^{2j'}, \notag \\
       n_{RR}^{\text{corner}
      }&\propto |r_{R,1}|^{2j}|r_{R,2}|^{2j'}.  \notag
    \end{align}
As a result,
when cast in $n_{RR}$, all types of
skin modes are centered at corners of the Lieb lattice. By fine-tuning  the parameters, one is able to completely separate them in real space in large $N$ limit, as indicated by \Cref{fig:loc}~(upper panel). Since the normal densities ($n_{RR}$ or $n_{LL}$) can be indirectly measured in open quantum systems during the relaxation process in the form of chiral damping \cite{fei2019,yang2022}, their distinct distribution of boundary and bulk modes has potential application in cold-atom setups, where any boundary mode of well-controlled spatial structure can be dynamically prepared through engineered dissipation \cite{emil2023, meng2023}.} 

\textcolor{black}{The total biorthogonal density $n_{LR}$, on the other hand, is independent of the localization parameter $r$ after  an overlap of $\u{\psi}_{Rm}$ with $\u{\psi}_{Lm}$ in Eqs.~(\ref{eq:vec_bulk1}), (\ref{eq:vec_edge1})
 and (\ref{eq:cor2}). Shown in \Cref{fig:loc}~(lower panel), $n_{LR}$ exhibits localization properties analogous to Hermitian systems and thereby motivates their categorization in terms of bulk, edge and corner modes:
    \begin{align}
       n_{LR}^{\text{bulk}
      } &\propto 1, \label{eq:loc2} \\
       n_{LR}^{\text{edge}, AB
      } &\propto (r_{L,2}^*r_{R,2})^{j'}, \notag \\
       n_{LR}^{\text{edge}, AB'
      } &\propto (r_{L,1}^*r_{R,1})^{j}, \notag \\
       n_{LR}^{\text{corner}
      }&\propto (r_{L,1}^*r_{R,1})^{j}(r_{L,2}^*r_{R,2})^{j'}.  \notag
    \end{align}
Supported also by our nonunitary gauge transforms, the emerging factors above restore the localization parameters displayed by  boundary modes of the reciprocal Hamiltonian in \Cref{eq:tpsi}:
    \begin{gather}
        r^*_{L,l} r_{R,l}  = \tilde{r}_l^2,
    \end{gather}
thus revealing their correct order.
One can probe the nature of $O(N^n)$  skin modes directly through $n_{LR}$.} 

\textcolor{black}{Besides, the biorthogonal relation in \Cref{eq:bor} ensures that $\mathbbm{1}_{N_{\tot} \times N_{\tot}} = \sum_n \sum_m |\psi_{nR,m}\rangle \langle \psi_{nL,m}|$. When summing over all types of skin modes $\sum_n$, only $n_{LR}$  gives back the identity $1$ on each lattice site, a normalization property lacking in $n_{RR}$ from \Cref{fig:loc}.}

\section{Biorthogonal polarization}
\label{sec:bp}
In this section, we track the topological origin of our exactly solvable NH hybercubic models through biorthorgonal polarization, which is generalized to a vector form in higher dimensions and equivalent to an ensemble of non-Bloch winding numbers of higher-order edge modes. Quantized jump in polarization indicates the occurrence of surface gap closings between boundary and bulk modes.  
It also accurately predicts a real-space diffusion among these skin modes in terms of biorthogonal density. 

\subsection{Relation to non-Bloch winding number}

\begin{figure}[b]
\centering
\includegraphics[width=0.75\linewidth]{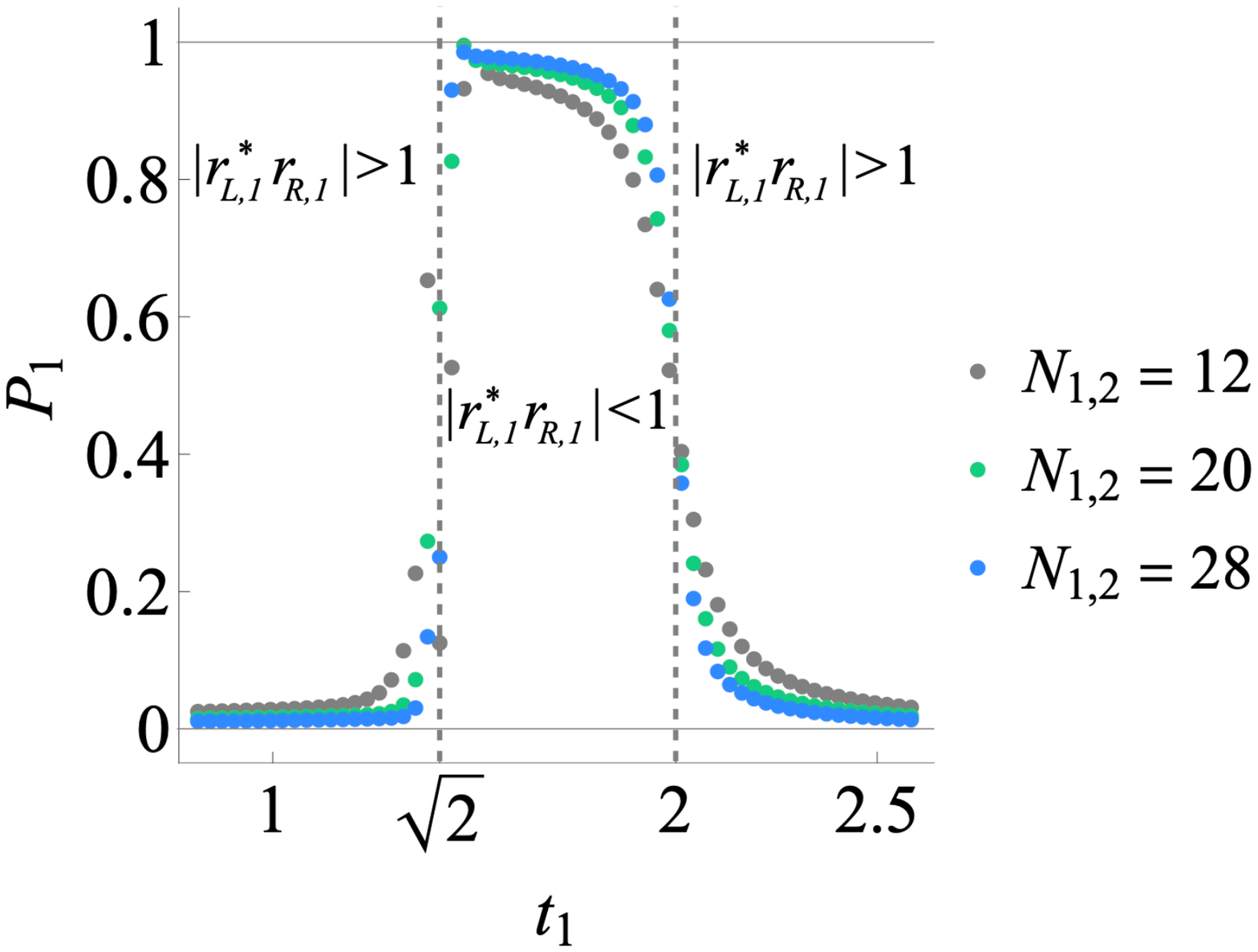}
\caption{Quantization of polarization $P_1$ as a function of $t_1$ on the NH Lieb lattice with different system sizes $(2N_1-1) \times (2N_2-1)$. We keep $t_1 = t_3$ and fix $t_{2,4} = 1, \gamma_{1,3} =  \sqrt{3}, \gamma_{2,4} =  0$. $P_1$ jumps at $|r^*_{L,1}r_{R,1}|=1$ marked by two dashed gray lines.}
\label{fig:lieb-p}
\end{figure}

The biorthogonal polarization vector can be built on the zero-energy corner mode, of which the associated right and left eigenvectors normalized by $\langle \u{\psi}_{L0}  |\u{\psi}_{R0} \rangle = 1$ live on the motif $A$ only: $E^{\text{OBC}}_{n=0} = 0$,
	\begin{gather}
    |{\psi}_{R/L, 0} \rangle = \mathcal{N}_{R/L} \sum_{\vec{j}} \prod_{l=1}^d (r_{R/L,l})^{j_l} c^\dagger_{ \vec{j}, A} |0\rangle.
	  \label{eq:cor}
	\end{gather}
The product gathering localization factors can be viewed as a special case of \Cref{eq:psi_rl} with $\bar{l_0} = \varnothing$, and in $d=2$ it recovers the form in \Cref{eq:cor2}.
The polarization vector $\vec{P} = (P_1, P_2, \dots, P_d)$ of the corner mode is quantized when its component in the $x_l$ direction is defined as \cite{flore2018,elisabet2019}
 \begin{gather}
    P_l =  1- \lim_{N_l \to \infty} \frac{1}{N_l} | \langle \psi_{L0}| \sum_{\vec{j}} j_l \Pi_{\vec{j}} | \psi_{R0}\rangle |. \label{eq:pol}
 \end{gather}
 Here, the density operator acts on $\vec{j}$th unit cell: $\Pi_{\vec{j}} = \sum_{\lambda} |e_{\vec{j}, \lambda} \rangle \langle e_{\vec{j}, \lambda}|$ with $\lambda \in  \{A, B_1, \dots, B_d\}$.
 $P_l = 1 \ (0)$ when $|r^*_{L,l} r_{R,l}| < 1 \ (> 1)$.
  In \Cref{fig:lieb-p}, we show the value of $P_1$ on the finite-size NH Lieb lattices. By changing $t_1$, $P_1$ jumps at the transition lines (dashed gray): $|r_{L,1}^*r_{R,1}|=1$, the quantization of which becomes more and more ideal when the system size increases.
  
In the meantime, from \Cref{eq:hnb} the non-Bloch Hamiltonian of the edge modes on  motifs $AB_l$, 
  \begin{gather}
        H^{\text{non-Bloch}
    }_{\text{edge}, AB_l}(\vec{k}) = \begin{pmatrix}
	  		0 & t_{2l-1}^+ + t_{2l}^- r_l^{-1}e^{-ik_l} \\
			t_{2l-1}^- + t_{2l}^+r_l e^{ik_l} & 0 
		    \end{pmatrix} 
  \end{gather}
exhibits chiral symmetry as well,  $\mathcal{C}  {H}^{\text{non-Bloch}}_{\text{edge}, AB_l}(\vec{k}) \mathcal{C}^{-1} = -{H}^{\text{non-Bloch}}_{\text{edge}, AB_l}(\vec{k})$ where $\mathcal{C} = \diag \{ 1, -1\}$.  It allows to define another topological invariant, i.e. the non-Bloch winding number \cite{yao2018}, for the edge modes in the GSBZ:
  \begin{gather}
    W_l = \frac{i}{2\pi} \int q_l^{-1}dq_{l}, \label{eq:wn}
  \end{gather}
with $q_l = -(t_{2l-1}^+ + t_{2l}^- r_l^{-1}e^{-ik_l})/E_{AB_l,+}^{\text{OBC}}(k_l)$. 
By analogy to 1D NH SSH chain \cite{yao2018,flore2018}, in our models, these two invariants also turn out to be equivalent: 
  \begin{gather}
       P_l = |W_l|. 
  \end{gather}
Hence, by reproducing the complete set of non-Bloch winding numbers of the edge modes, the polarization vector we construct contains all the topology of the NH hypercubic models. 

\textcolor{black}{We briefly comment on the implications from a nontrivial non-Bloch topological invariant. When $P_l = |W_l| = 1$, one may interpret the associated  edge $AB_l$ mode to be topological in the sense that if the underlying NH SSH chain capturing its edge Hamiltonian is switched to even lengths, there emerges a pair of zero-energy corner modes in correspondence to a nonzero bulk invariant, whereas no corner mode if $P_l = |W_l| = 0$.
In our hypercubic models, the difference lies in that to acquire spectral mirror symmetry, the coupled NH SSH chains are constructed with odd lengths. Our models support one corner mode for both topological $(
|r^*_{L,l}r_{R,l}| < 1)$ and nontopological  $(|r^*_{L,l}r_{R,l}| > 1)$ regions. Although the number of the corner mode will not change for specific parameters, in the topological region the wave function  of our corner mode in \Cref{eq:cor} will match one of the two topologically protected corner modes (the one localized at the $A$-site corner) on an even-site lattice in the large-$N_l$ limit.}

\subsection{Polarization jump and surface  gap closing}
 \begin{figure}[t]
\centering
\includegraphics[width=1\columnwidth]{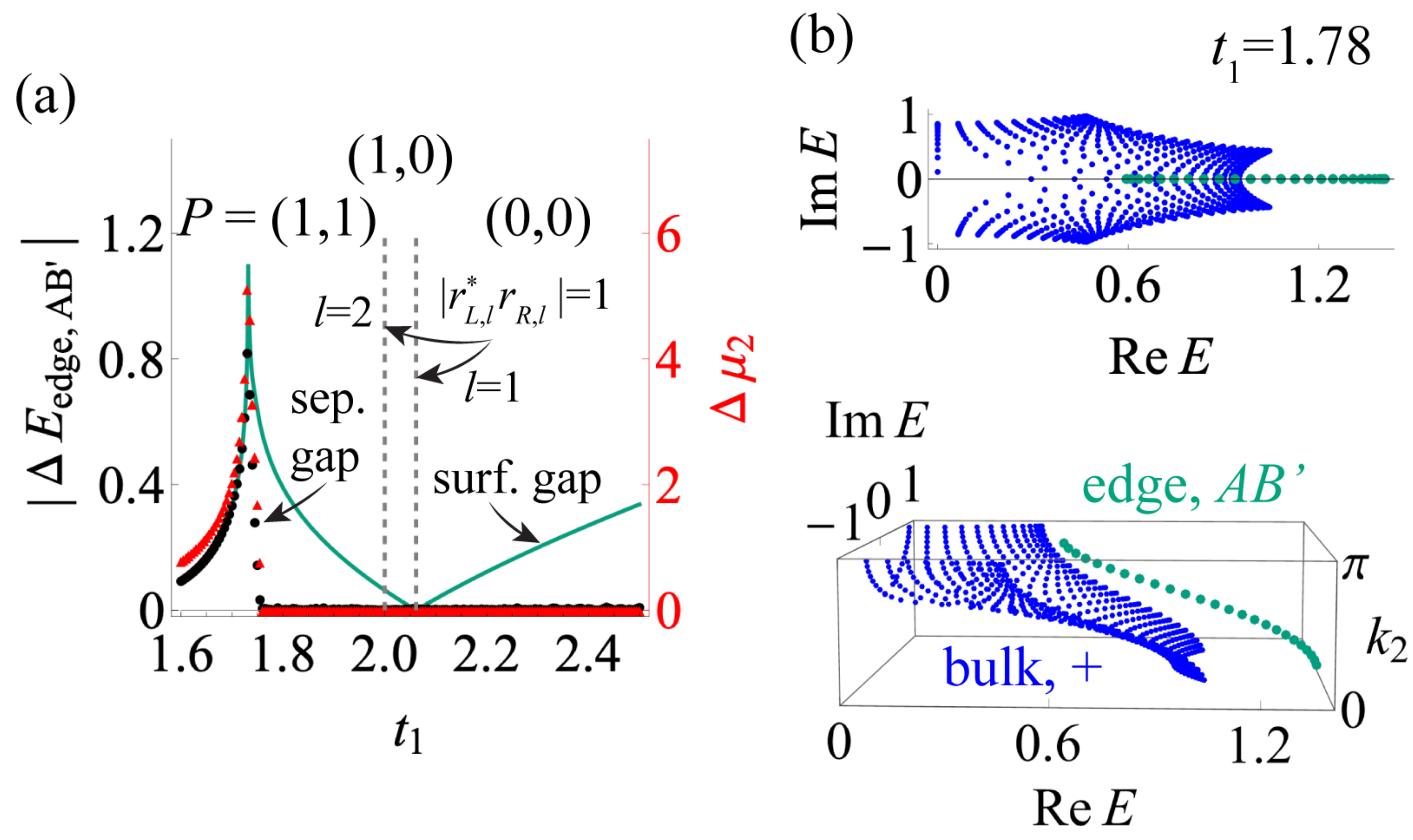}
\caption{Gap comparison in the NH Lieb model: surface gap  vs separation gap formed by the edge mode $AB'$. Panel~(a) illustrates while
its surface gap (green curve) closes as soon as the biorthogonal polarization vector $\vec{P}$ jumps with $|\Delta P_1| = 1$ at $|r^*_{L,1}r_{R,1}|=1$ (gray dashed line), its separation gap (black dot) closing is signaled by a vanishing amoeba hole (red triangle). $\Delta 
\mu_2$ denotes the longest distance the hole extends inside the amoeba along the $\mu_2$ direction [see Figs.~\ref{fig:rk}(a) and \ref{fig:rk}(b)].  We take size $N_{1,2} = 26$ and choose a parameter path with $t_1 = t_3$ and fixed $t_{2,4}=1$, $\gamma_{1,3}=\sqrt{3}$, $\gamma_{2}=1.5$, $\gamma_4=0$. Along this path, $\vec{P}$ jumps twice across two vertical transition lines at $|r^*_{L,l}r_{R,l}|=1$. Panel~(b) shows the distribution of energies of bulk ($+$, in blue) and edge $AB'$ (in green)  modes at $t_1 = 1.78$, displaying a closed separation gap (upper panel) in the complex-energy plane and a finite surface gap measured at the same nonzero surface momentum $k_2$ (lower panel).}
\label{fig:gap}
\end{figure}

\textcolor{black}{Next, we extend the notion of the biorthogonal BBC  to higher dimensions by relating surface gap closings to the polarization jump.}

\textcolor{black}{Applying the analytical structure of the exact OBC spectrum in \Cref{eq:dc_nh} to the surface gap of $O(N^n)$ skin modes defined in \Cref{eq:sfg}, we arrive at
 \begin{align}
    | \Delta E_{\text{Surf.}} | 
    = &\min_{\forall \vec{q},\vec{k}_{n}, \alpha} \left|   \sqrt{\sum_{l \in \bar{l}_n} g_{2l-1,2l}(k_l) + \sum_{l' \notin \bar{l}_n} g_{2l'-1,2l'}(q_{l'})}  \right. \notag \\
    &- \alpha  \left. \sqrt{\sum_{l \in \bar{l}_n} g_{2l-1,2l}(k_l)} \right| , \label{eq:surg}
\end{align}
where $\bar{l}_n = \{l_1, \dots, l_n \}$ collects all occupied $B$ motifs  supporting nonzero surface momentum $\vec{k}_n$ and $\alpha = \pm$ denotes the pair of opposite energies $(E, -E)$. 
One immediately sees that the surface gap closes at the zeros of $g$ functions: 
\begin{gather}
\sum_{l' \notin \bar{l}_n} g_{2l'-1,2l'}(q_{l'}) = 0,
\end{gather} which allows the bulk spectrum  to collapse into the boundary spectrum.}

\textcolor{black}{Individually, each $g$ function reconstructs a spectrum of one edge mode:
 \begin{align}
  E_{AB_l,\pm}^{\text{OBC}}(k_l) &= \pm \sqrt{g_{2l-1,2l}(k_l)}, 
\\
  g_{2l-1,2l}(k_l) &= t_{2l-1}^+t_{2l-1}^- + t_{2l}^+t_{2l}^-
  \notag \\
  &\phantom{=} 
  + \sgn (t_{2l-1}^+t_{2l}^+) 2 \sqrt{t_{2l-1}^+t_{2l}^+ t_{2l-1}^-t_{2l}^-} \cos k_l, \notag
  \end{align}
 the gap closing   of which can be retrieved from the gauge transform in \Cref{eq:map1}.
On motifs $AB_l$, the associated edge mode has zero energy as soon as its reciprocal Hamiltonian $\tilde{\mathcal H}_{AB_l} = \sum_{\vec{j}} \tilde{t}_{2l-1} c^\dagger_{\vec{j}, A} c_{\vec{j}, B_l} + \tilde{t}_{2l} c^\dagger_{\vec{j}, B_l}c_{\vec{j}+\vec{e}_l, A}  + \hc$ becomes gapless:
    \begin{gather}
       |\tilde{t}_{2l-1}|  = |\tilde{t}_{2l}|
       \ \Longleftrightarrow \   |r^*_{L, l} r_{R, l}| = 1, \label{eq:gc}
    \end{gather}
where $r^*_{L, l} r_{R, l} =   (\tilde{t}_{2l-1} / \tilde{t}_{2l})^2 = (t_{2l-1}^+  t_{2l-1}^-)/( t_{2l}^+  t_{2l}^-)$. Indeed, when $r^*_{L, l} r_{R, l} = e^{i\theta}$,  one verifies $g_{2l-1,2l}(k_l) = 0$ at $k_l = \arccos[-\cos(\theta/2)]$.}

\textcolor{black}{On the hypercubic lattice, these gapless lines of edge modes signal both surface gap closings and a quantized change in the polarization vector in \Cref{eq:pol}. It enables us to use the latter as a diagnostic of the biorthogonal BBC in higher dimensions:
\begin{gather}
 \left. \Delta E_{\text{Surf.}} \right|_{N \to \infty} = 0 \quad \text{if} \quad |\Delta P_{l'}| = 1 \  \forall \  l' \notin \bar{l}_n. \label{eq:ptot}
 \end{gather}
Shown in \Cref{fig:gap}(a), on the NH Lieb lattice, we plot the surface gap (in green) of the second-order edge mode $AB'$  against the value of $\vec{P}$ ($\vec{W}$) along selected parameter paths.
 If $|\Delta E_\text{Surf.}| \ne 0$, the gap between its spectrum and the bulk remains open at any nonzero surface momentum, e.g. $\vec{k}_n = k_2$ in  \Cref{fig:gap}(b) (lower panel).  To reach $|\Delta E_\text{Surf.}| = 0$, the other edge spectrum $AB$ should be gapless: $g_{1,2}(k_1) = 0$, occurring at $|r^*_{L, 1} r_{R, 1}| = 1$ or $|\Delta P_1| = 1$.}

 \begin{figure}[t]
\centering
\includegraphics[width=0.7\columnwidth]{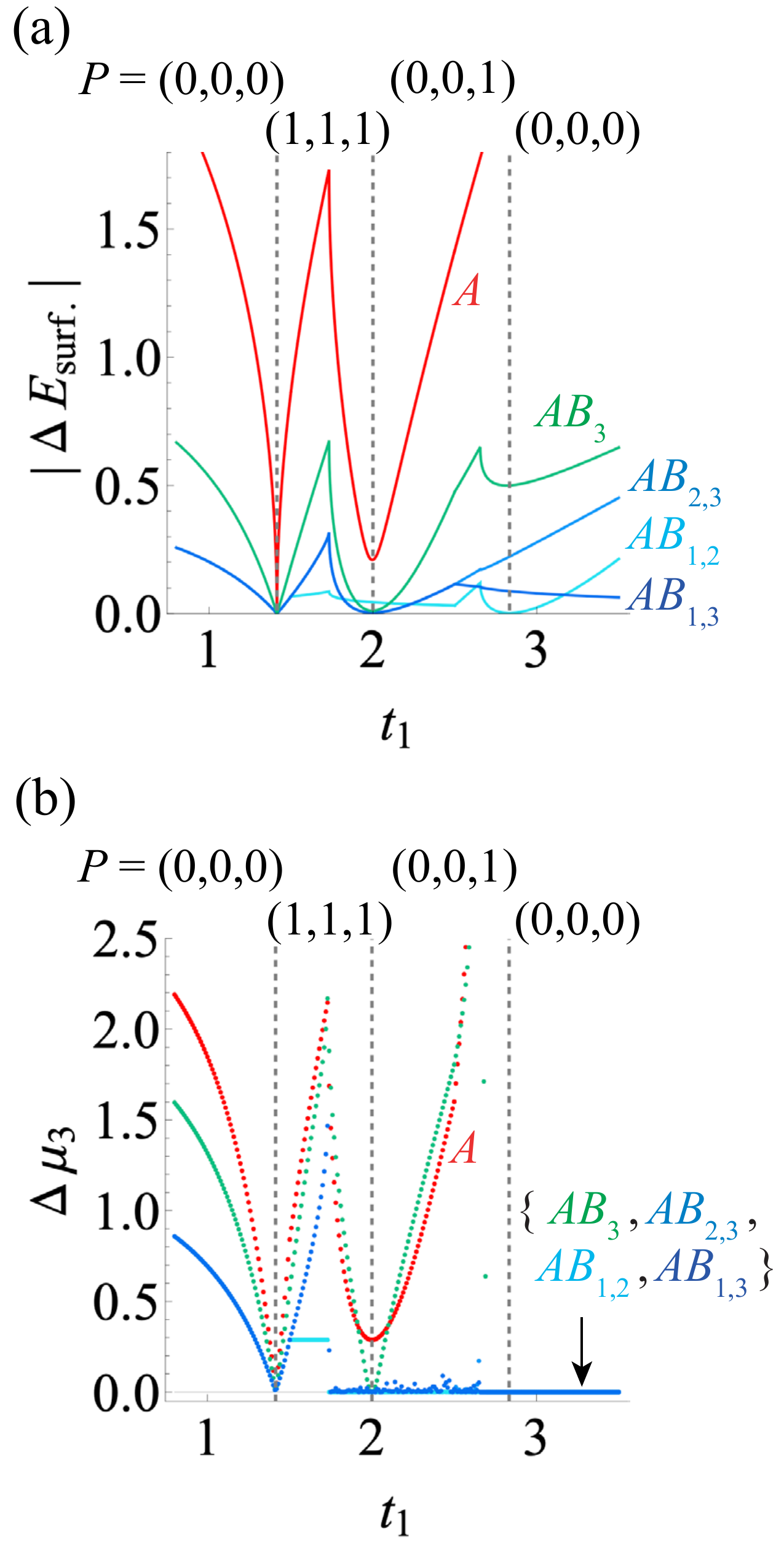}
\caption{NH cubic model. Surface gap closings signified by (a) polarization jump vs (b) separation gap closings captured by the absence of amoeba hole  regarding $O(N^n)$ skin modes  (marked by the motifs they occupy $AB_{l_1,l_2, \dots, l_n}$). Across topological phase transition lines of edge modes,  $\Delta P_{\text{tot}} = 3,2,1$ according to $|r^*_{L,l}r_{R,l}|=1$.  We choose $N_{1,2,3} = 26$ and keep $t_{2,4,6}=1$, $\gamma_{1,3,5}= \sqrt{3}$, $\gamma_{2,4,6}=0$. The path varies with $t_1$: $ t_1 = t_{3,5}$ for $t_1 \in [0.8, 1.5]$; $t_3 = t_1, t_5= 1.5$ for $t_1 \in (1.5, 2.5]$; $t_3 = 2.5, t_5= 1.5t_1-2.25$ for  $t_1 \in (2.5, 3.5]$. At $t_1=1.8$, the spectra of different skin modes projected on the complex-energy plane is shown  in \Cref{fig:rk}(d) where the corner mode $A$ (red) and the edge mode $AB_3$ (green) form a finite separation gap from the bulk (gray). To enhance visibility, the other two edge modes $AB_1$ and $AB_2$ are not shown in both plots.}
\label{fig:trans}
\end{figure}

 \textcolor{black}{In $d=3$, the NH cubic lattice support more types of boundary modes, residing on the corner $(A)$, the edge $(AB_{l_1})$ and the face $(AB_{l_1,l_2})$.
From \Cref{fig:trans}(a), our diagnostic using biorthogonal polarization in \Cref{eq:ptot} accurately predicts their surface gap closings with the bulk: the corner mode $A$  at $|\Delta P_1| = |\Delta P_2| =  |\Delta P_3| = 1$, the edge mode $AB_3$ at $|\Delta P_1| = |\Delta P_2|  = 1$ and the surface mode $AB_{2,3}$ at $|\Delta P_1| = 1$, etc.} 

One further observes the total polarization jump imposes a constraint on the type of $O(N^n)$ skin modes that can enter the $d$-dimensional bulk spectrum across the transition lines in \Cref{eq:gc}:
 \begin{gather} 
  n \ge d - \Delta P_{\text{tot}}, \label{eq:dc}
 \end{gather}
 where $\Delta P_{\text{tot}} = \sum_l |\Delta P_l|$.
It naturally follows from the fact that $\Delta P_{\text{tot}}$ registers the number of zeros in the $g$ functions, thus the level of collapse from the bulk to the boundary spectrum. In particular, when the total polarization change reaches the maximum $|\Delta P_{\text{tot}}| = d$, zero solution exists in all $g$ functions, enforcing every type of surface gap to close, e.g. when $\vec{P}$ jumps from $(0,0,0)$ to $(1,1,1)$ in \Cref{fig:trans}(a). Consistent with the constraint of  \Cref{eq:dc}, $n$ is allowed to take any integer value between $0$ and $d-1$. It also entails that each  boundary and bulk dispersive band $(\alpha = \pm)$ now supports one zero-energy mode, crossing the same energy of the corner mode.

\subsection{Interplay with higher-order NHSE}
\label{sec:int}
 \begin{figure}[t]
\centering
\includegraphics[width=1\columnwidth]{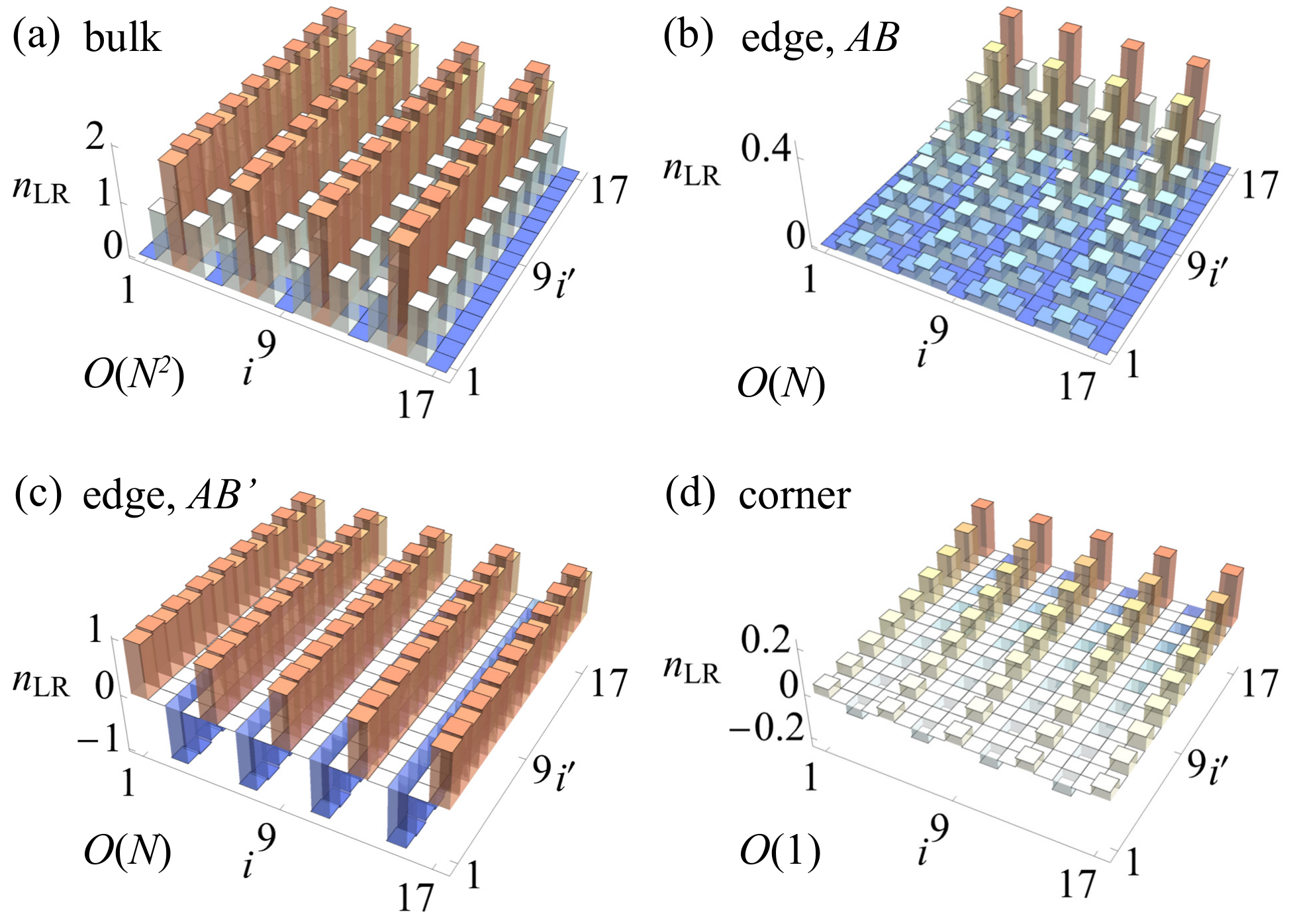}
\caption{\textcolor{black}{Total biorthogonal density $n_{LR}$ of the NH Lieb model at the polarization transition: $t_{1} = \sqrt{17}/2$ in \Cref{fig:gap}(a). We take a larger system size $N_{1,2} = 9$. As $\vec{P}$ changes from $(1,0)$ to $(0,0)$ accompanied by a surface gap closing with the bulk, the edge $AB'$ skin mode of panel (c)  loses the localization feature along the $x_1$ direction since $|r^*_{L,1} r_{R,1}| =1$. Simultaneously, the edge $AB$ skin mode of panel (b) shares zero energy with the corner mode of panel (d), both now exponentially localized towards the upper edge $AB$ of the Lieb lattice due to $|r^*_{L,2} r_{R,2}| > 1$. It manifests that when the polarization jumps according to $\Delta P_1 = 1$ and $\Delta P_2 = 0$, the edge $AB'$ skin mode enters the bulk while the corner mode leaks into the edge $AB$, in clear distinction from the case without polarization jump (compare \Cref{fig:loc}, lower panel).}}
\label{fig:lsg}
\end{figure}
\textcolor{black}{In the end of this section, we reveal that biorthogonal polarization predicts the real-space diffusion among bulk and higher-order skin modes.}

\textcolor{black}{It turns out that when the polarization vector jumps, different orders of $O(N^n)$ skin modes displaying a closed  surface gap  share the same localization lengths. It can be seen from the GSBZ of \Cref{eq:gsbz}, by the definition $r_l = (r_{R,l}/r^*_{L,l})^{1/2}$, the hybrid imaginary momentum shifts become identical once
  \begin{gather}
    |r_l| = |r_{R,l}|    \Longleftrightarrow |r^*_{L, l} r_{R, l}| = 1.
  \end{gather}
In terms of normal densities in \Cref{eq:den},  associated skin modes are not only localized at the same corner but characterized by the same set of localization parameters.
 Taking the NH Lieb lattice as an example, when $\vec{P}$ jumps from $(1,0)$ to $(0,0)$ in \Cref{fig:gap}(a), one verifies that at the transition  $|r^*_{L,1} r_{R,1}| =1$,  the measurement of $n_{RR}$ in \Cref{eq:loc1} gives
     \begin{align}
       n_{RR}^{\text{bulk}
      }, \  n_{RR}^{\text{edge}, AB'
      }  &\propto |r_1|^{2j}|r_2|^{2j'}, \\
       n_{RR}^{\text{edge}, AB
      },  \ n_{RR}^{\text{corner}
      } &\propto |r_1|^{2j}|r_{R,2}|^{2j'}. \notag
    \end{align}}

\textcolor{black}{Whereas from biorthogonal densities,  the contrast is made more clear for higher-order skin modes start to lose localization along the $x_l$ direction. We find that across the same polarization jump, the biorthogonal densities in \Cref{eq:loc2} exhibit the scaling:
    \begin{align}
       n_{LR}^{\text{bulk}}, \  n_{LR}^{\text{edge}, AB'} &\propto 1,  \\
       n_{LR}^{\text{edge}, AB},  \   n_{LR}^{\text{corner}} &\propto (r_{L,2}^*r_{R,2})^{j'}. \notag
    \end{align}
Figure~\ref{fig:lsg} shows consequences on the higher-order NHSE: the edge mode $AB'$ enters into the  bulk and the corner mode into the edge $AB$. When the diffusion happens, the surface gaps between two merged skin modes remain closed.} 

\textcolor{black}{Hence, biorthongal polarization proves to be an efficient tool to detect the diffusion among skin modes of different orders. In the extreme case $|\Delta P_{\text{tot}}| = d$, for instance  when $\vec{P}$ jumps from $(0,0,0)$ to $(1,1,1)$ in the three-dimensional (3D) cubic lattice of \Cref{fig:trans}(a), all boundary modes become delocalized in biorthogonal densities and share the same localization lengths with bulk modes.} 

\section{The amoeba}
\label{sec:af}
In this section, we apply the amoeba formulation to our exactly solvable models in higher dimensions. We aim to use the amoeba as a probe of the separation gap between boundary and bulk modes, in comparison with the surface gap studied by biorthogonal polarization in \Cref{sec:bp}. For the universal spectrum, another important quantity obtained from the amoeba, we examine the influence of higher-order skin modes with a focus on finite-size lattices, a more realistic setting for generic open-boundary NH systems. 

\subsection{Amoeba hole, Ronkin function and relation to GSBZ}

First, we briefly introduce the amoeba formulation, recently proposed in Ref.~\cite{wang2022} for the study of the non-Bloch bands beyond 1D.

Given a real-space NH Hamiltonian $\mathcal{H}$ on a $d$-dimensional lattice, the amoeba is defined as a logarithmic map of solutions to the  characteristic equation with respect to its Bloch Hamiltonian:
 \begin{gather}
     \mathcal{A}_h = \{ \vec{\mu} \  \text{with} \   \mu_l = \log |\beta_l|: h(\vec{\beta}) = 0 \},
 \end{gather}
where $l = 1,2, \dots, d$ and
  \begin{gather}
      h (\vec{\beta}) = \det [H(\vec{\beta}) - E]/E^{d-1}. \label{eq:hbeta}
  \end{gather}
In our models, $H(\vec{\beta})$ denotes the Bloch Hamiltonian in \Cref{eq:hbloch}  with a replacement  $e^{ik_l} \to \beta_l$ and $E$ stands for the reference energy. To remove trivial solutions coming from $(d-1)$ zero-energy bulk flat bands, we slightly modify the initial definition in Ref.~\cite{wang2022} by dividing a factor $E^{d-1}$ in the characteristic equation.

\begin{figure}[t]
\centering
\includegraphics[width=1\columnwidth]{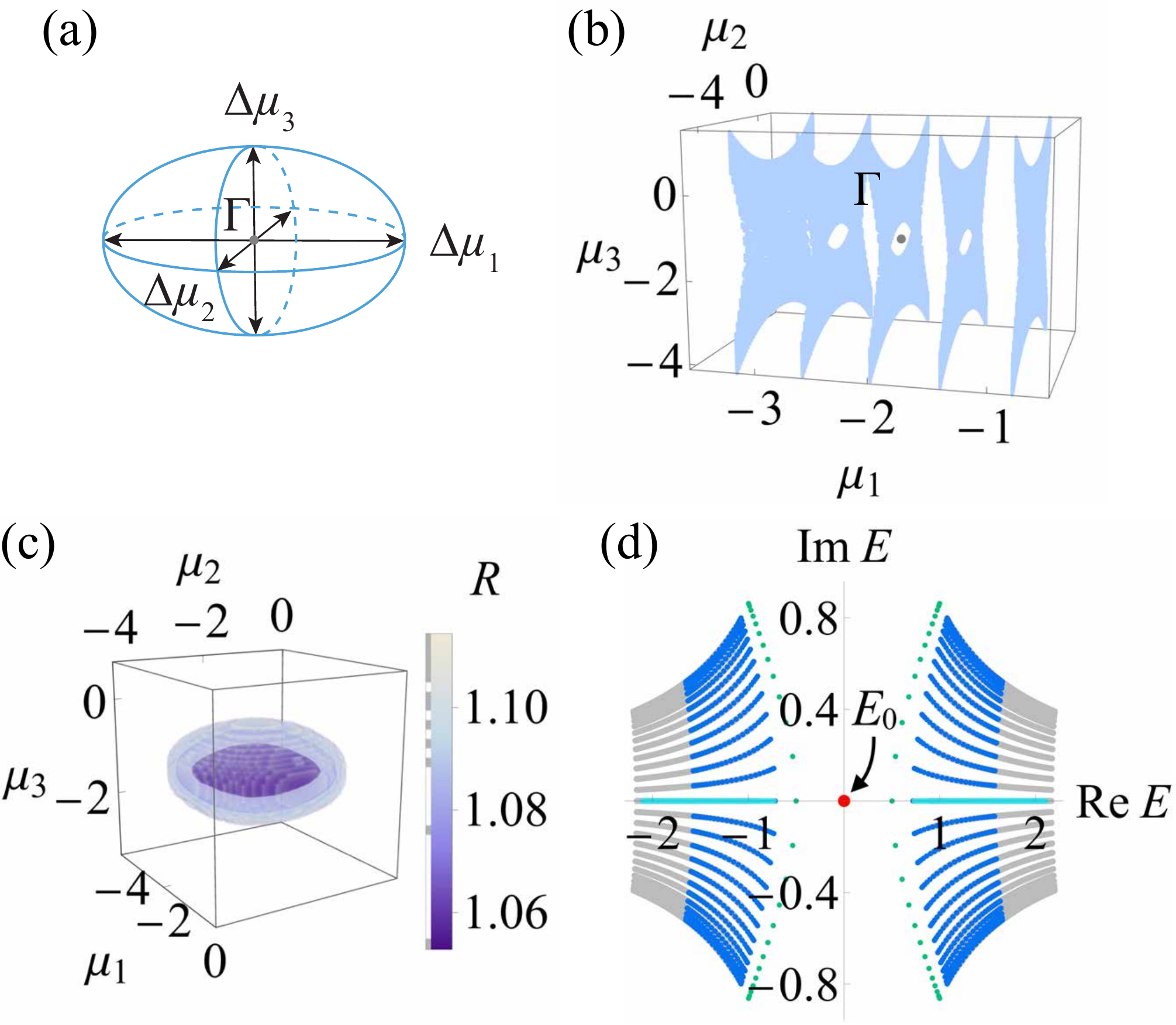}
\caption{\textcolor{black}{Illustration of 3D amoeba hole (a) in the NH cubic model, (b) in terms of ``vacuole" inside amoeba body, and (c) of Ronkin minimum. From the complete spectrum on the complex-energy plane (d), we choose the reference energy at eigen energy of the corner mode (red)  $E = E_0 = 0$, which exhibits a distinct separation gap with  bulk modes (gray)  at $t_1=1.8$ of \Cref{fig:trans} and with a system size $N_{1,2,3} = 26$. The energies of surface and edge modes are denoted by the same colors as  \Cref{fig:trans}. In (c), to evaluate the Ronkin function integral, we take a grid size $M_{1,2,3} = 20$.}}
\label{fig:rk}
\end{figure}

The key information the topology of amoeba can offer is that if $E$ belongs (does not belong) to the OBC bulk spectrum of $\mathcal{H}$, the amoeba forms a continuous body (has a hole) in $d$-dimensional space \cite{wang2022}. Or equivalently,
\begin{gather}
     \mathcal{V}_{\text{hole}} \ne 0
     \quad \text{if} \quad  E \notin \left. E^{\text{OBC}}_{\text{bulk}}\right|_{N \to \infty}, \label{eq:ath}
 \end{gather}
where $\mathcal{V}_{\text{hole}}$ denotes the volume of amoeba hole. Figure \ref{fig:rk}(a) depicts a generic 3D amoeba hole in the shape of an ellipsoid with its three principal axes extending with lengths $\Delta \mu_l$. In \Cref{eq:ath},
the thermodynamic limit $N \to \infty$ is imposed  as the amoeba addresses properties of the universal spectrum. 
Let us look at the NH cubic model, with its OBC spectra of both bulk and boundary modes shown in \Cref{fig:rk}(d). In the considered parameter region, the zero-energy corner mode (red) is isolated from the bulk spectrum (gray).
If we choose it as a reference point $E = E_0 = 0$, the amoeba coming from the solutions to the characteristic polynomial would hold a hole, as confirmed by  \Cref{fig:rk}(b). The existence of hole also infers that if one enlarges current finite system size to the thermodynamic limit, the corner mode still remains outside of the bulk spectrum. 

The prediction of the amoeba in \Cref{eq:ath} can be proved in our models using exact solutions. It is useful to introduce the
convex Ronkin function \cite{ronkin1974,forsberg2000,passare2004},  which reaches its minimum
in the amoeba hole [as shown in Figs.~\ref{fig:rk}(b) and \ref{fig:rk}(c)]:
 \begin{gather}
     R_h(\vec{\mu}) = \int_{T^d}  \frac{\prod_{l=1}^{d} dq_l}{(2\pi)^d}  \log |h(e^{\vec{\mu} + i \vec{q}})|. \label{eq:dron}
 \end{gather}
Here, $h(e^{\vec{\mu} + i \vec{q}}) = h(\vec{\beta})$ following a separation of  modulus and phase: $\beta_l \to e^{\mu_l + i q_l}$.
The integral is taken over the phase space, i.e. a $d$-dimensional torus  $T^d = [-\pi, \pi]^d$. 

We show if the amoeba hole exists,  the Ronkin function reaches the minimum at the 
central hole $\Gamma$, which is independent of the reference energy and determined by localization parameters of bulk modes ($k_l \ne 0 \ \forall \ l=1,2,\dots, d$) in the GSBZ in \Cref{eq:gsbz}: 
\begin{gather}
 \Gamma = \vec{\mu}_c = (\log |r_1|, \log |r_2|,\dots, \log |r_d| ).
\end{gather}
Let us take a gradient of the real Ronkin function along the $x_l$ direction, 
 \begin{align}
   \partial_{\mu_l} R_h(\vec{\mu}) &= \int_{T^{d-1}} \frac{\prod_{i \ne l} dq_i}{(2\pi)^{d-1}} \omega_{l}(\vec{\mu}), \notag \\
   \omega_l(\vec{\mu}) &= \frac{1}{2\pi i } \int_{-\pi}^{\pi} d q_l \partial_{q_l}
 \log |h(e^{\vec{\mu} + i\vec{q}})|, \label{eq:drk}
 \end{align}
where the relation $i\partial_{\mu_l}h(e^{\vec{\mu} + i\vec{q}}) = \partial_{q_l} h(e^{\vec{\mu} + i\vec{q}})$ is applied. With the amoeba defined on two dispersive bulk bands in \Cref{eq:hbeta},
 \begin{align}
   h(e^{\vec{\mu} + i\vec{q}})
   = &\sum_{l=1}^{d} t_{2l-1}^+t_{2l-1}^- + t_{2l}^+t_{2l
}^- + t_{2l-1}^-t_{2l}^- e^{-\mu_l - iq_l}  \notag \\
 &+ t_{2l-1}^+t_{2l}^+e^{\mu_l + iq_l} - E^2, 
 \end{align}
 the generalized winding number  $\omega_l$ vanishes at $\Gamma$,
 \begin{align}
   \omega_l(\vec{\mu}_c) &=  \frac{1}{2\pi i } \int_{-\pi}^{\pi} d q_l  \frac{\partial_{q_l} h(e^{\vec{\mu}_c + i\vec{q}})}{h(e^{\vec{\mu}_c + i\vec{q}})} \notag \\
 &=  \sqrt{t_{2l-1}^+t_{2l}^+} \sqrt{t_{2l-1}^-t_{2l
}^-}  \int_{-\pi}^{\pi} \frac{d q_l}{2\pi}   \frac{e^{iq_l}-e^{-iq_l}}{[E^{\text{OBC}}_{d,\pm} (\vec{q})]^2 -E^2 } \notag \\
   &=0. \label{eq:wz}
 \end{align}
Recalling the complete integration domain in \Cref{eq:drk}, $[E^{\text{OBC}}_{d,\pm} (\vec{q})]^2 = \sum_{l'=1}^{d} g_{2l'-1,2l'}(q_{l'})$ is defined at a given phase $\vec{q}$ on the torus $T^d$ and encodes the $g$ functions in \Cref{eq:gf}. Then, 
to reach the last step of \Cref{eq:wz}, we perform $q_l \to -q_l$ in the second half of the integral.  It cancels the first half due to spectral mirror symmetry: $E^{\text{OBC}}_{\pm}(q_l) = E^{\text{OBC}}_{\pm}(-q_l)$. We thus arrive at 
    \begin{gather}
   \forall \ E:   \left.\partial_{\mu_l} R_h(\vec{\mu})\right|_{\vec{\mu} = \vec{\mu}_c} = 0 \  \forall  \  l. \label{eq:rmin0}
  \end{gather}
As this gradient vanishes for arbitrary reference energy $E$, when the amoeba hole does not exist, the Ronkin function exhibits one single minimum at $\Gamma$.

We are now ready to demonstrate \Cref{eq:ath}.
Considering the convexity of the Ronkin function, the analytical structure of its minimum can be resolved at $\Gamma$:
    \begin{align}
     \min_{\vec{\mu}} R_h (\vec{\mu}) &= R_h(\vec{\mu}_c) \label{eq:rkm} \\
     &= \int_{T^d}  \frac{\prod_{l=1}^{d} dq_l}{(2\pi)^d}  \log |h(e^{\vec{\mu}_c + i\vec{q}})| \notag \\
     &= \int_{T^d}  \frac{\prod_{l=1}^{d} dq_l}{(2\pi)^d} 
   \sum_{\alpha=\pm} \log | E_{d, \alpha}^{\text{OBC}}(\vec{q}) - E|. \notag
    \end{align}
If $E_{d,\alpha}^{\text{OBC}}(\vec{q}) \ne E$ for any $\vec{q}$ and $\alpha$,  $h(e^{\vec{\mu}_c + i\vec{q}}) \ne 0$, indicating there is a hole at $\Gamma$ inside the amoeba: $\mathcal{V}_{\text{hole}} \ne 0$.
The limit $N \to \infty$ corresponds to  $\vec{q}$ taking continuous value on the torus. 
Equation~(\ref{eq:rmin0}) further implies that by varying $E$, the hole always encloses $\Gamma$ and changes shape continuously in its vicinity. Once $E$ enters $E_{\text{bulk}}^{\text{OBC}}$, there exists $\vec{q}$ and $\alpha$ such that $h(e^{\vec{\mu}_c + i\vec{q}}) = 0$, inferring in this opposite scenario that the amoeba hole must close at $\Gamma$.

\subsection{Hole-free amoeba and separation gap closing}
\textcolor{black}{Next, we apply the amoeba formulation to locate the separation gap between  boundary and bulk modes. 
Based on its definition in \Cref{eq:spg},
for the $O(N^n)$ skin modes with exact spectra solved in \Cref{eq:gf},  the separation gap from the bulk shares the form
\begin{align}
|\Delta E_{\text{Sep.}} | = &\min_{\forall \vec{p},\vec{q},\vec{k}_{n}, \alpha} \left|   \sqrt{\sum_{l \in \bar{l}_n} g_{2l-1,2l}(p_l) + \sum_{l' \notin \bar{l}_n} g_{2l'-1,2l'}(q_{l'})}  \right. \notag \\
    &- \alpha  \left. \sqrt{\sum_{l \in \bar{l}_n} g_{2l-1,2l}(k_l)} \right|, 
\end{align}
where $\bar{l}_n = 
\{l_1, \dots, l_n \}$ denotes  occupied $B_l$ motifs supporting nonzero surface momentum $(k_l \ne 0)$ of boundary modes  and $\alpha = \pm 1$ distinguishes a pair of dispersive bands at opposite energies. 
When $|\Delta E_{\text{Sep.}}| = 0$, the spectra of $O (N^n)$ skin modes are inseparable from the bulk on the complex-energy plane at normally different nonzero surface momenta: $p_l \ne k_l$ [see \Cref{fig:gap}(b), upper panel]. By fixing $p_l = k_l$, one retrieves the definition of the surface gap in \Cref{eq:surg}, which often displays intrinsically different behaviors for higher-dimensional skin modes ($n \geq 1$) [compare with \Cref{fig:gap}(b), lower panel].  In our models, the corner mode in \Cref{eq:cor} with fully suppressed occupancy on all $B$ motifs  ($\bar{l}_0 = \varnothing$) is the only type of boundary modes for which the two gaps become identical: $\Delta E_{\text{Sep.}} \equiv \Delta E_{\text{Surf.}}$.}

\textcolor{black}{Notably, from the statement of \Cref{eq:ath}, the separation gap closings are captured by the absence of an amoeba hole:
\begin{gather}
    \left. \Delta E_{\text{Sep.}} \right|_{N \to \infty} = 0 \quad \text{if} \quad \left. \mathcal{V}_{\text{hole}} \right|_{E = E^{\text{OBC}}_{n,\text{min}}} = 0, \label{eq:amoE}
 \end{gather}
where $E^{\text{OBC}}_{n,\text{min}}$ denotes the energy of the $O(N^n)$ skin modes that minimizes the separation gap. Shown in \Cref{fig:gap}(a), in the NH Lieb model, the separation gap (black) of the edge $AB'$ mode closes simultaneously with its amoeba hole (red), signaled by the vanishing length of one principal axis $\Delta \mu_2$ [see illustration in \Cref{fig:rk}(a)].} 

\textcolor{black}{Moreover, we compare in  \Cref{fig:trans}(b) the analytical tools of polarization and amoeba in 3D. Recalling our diagnostics in Eqs.~(\ref{eq:ptot}) and (\ref{eq:amoE}),
the absence of amoeba hole does not correlate with the surface gap closings of edge ($AB_3$, green) and surface  ($AB_{1,2}$, $AB_{2,3}$, $AB_{1,3}$, blue) modes since  $\Delta E_{\text{Surf.}} \ne \Delta E_{\text{Sep.}}$ along the major part of the parameter path. The two gaps only strictly match at the corner mode ($A$, red). As a consequence, the closing of the amoeba hole can predict the diffusion of the corner mode into the bulk, yet
is not sensitive to the diffusion of other higher-dimensional boundary modes (with a codimension $D = n \ge 1$). 
At large $t_1$ in \Cref{fig:trans}, for instance, although one observes  closed separation gaps of edge and surface modes from the absence of amoeba hole, in real space these boundary modes cannot enter the bulk without a surface gap closing.}

  \textcolor{black}{In fact, since the topology of our model is characterized by the non-Bloch winding number of higher-order edge modes in \Cref{eq:wn}, its equivalent,  the biorthogonal polarization vector is related to 1D amoebae defined on edge Bloch Hamiltonians:
\begin{gather}
  \mathcal{A}_{{h},\text{edge}}^{l} = \{  {\mu}_{\text{edge}}^l = \log |\beta_l|: 
  {h}({\beta}_l) = 0 \}, 
  \end{gather}
  with $l = 1,2, \dots, d$, and
    \begin{gather}
       {h}({\beta}_l) = \det [H_{\text{edge}, AB_l}({\beta}_l) -  E], \\
 H_{\text{edge}, AB_l}({\beta}_l)=  \begin{pmatrix}
	  		0 & t_{2l-1}^+ + t_{2l}^- {\beta}_l^{-1} \\
			t_{2l-1}^- + t_{2l}^+{\beta}_l & 0 
		    \end{pmatrix}. \notag
  \end{gather}
  From \Cref{eq:gc}, the jump of $\vec{P}$ is in sync with the closure of these 1D amoeba holes if  the reference energy is chosen at gapless lines  $E = 0$:
  \begin{gather} 
    |\Delta P_l| = 1 \quad \text{if} \quad   \Delta 
 {\mu}_{\text{edge}}^l = 0. 
   \end{gather}}

\subsection{Spectral consequence of higher-order skin modes}
 \begin{figure}[t]
\centering
\includegraphics[width=0.95\columnwidth]{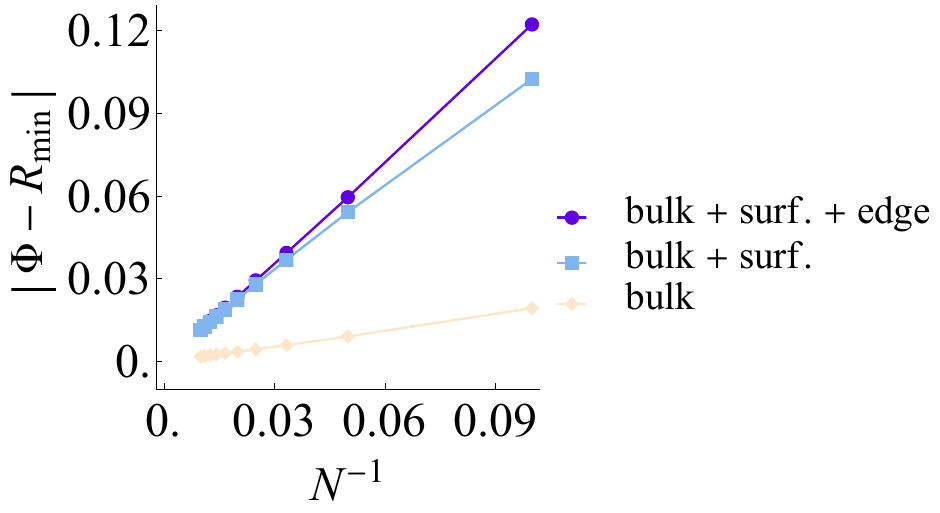}
\caption{Difference between Coulomb potential and Ronkin minimum as a function of linear system size $N = N_{1,2,3}$ in the NH cubic model. We choose the reference energy  $E = 0$ and fix the parameter region to $t_1 = 0.8$ of \Cref{fig:trans}. To calculate $R_{\min}$, we use $\vec{\mu}_{\min} = \vec{\mu}_c$ and take an integral grid with size $M_{1,2,3}= 100$.}
\label{fig:spec}
\end{figure}
Finally, we address the effects of higher-order skin modes on the universal spectrum. Let us focus on the isolated spectrum in \Cref{fig:trans}(b) at small $t_1$ $< 1$ with separation gaps of boundary modes all open. It is known  that the Coulomb potential \cite{lee2022d} constructed from a finite-size OBC spectrum approaches Ronkin function minimum in the limit $N \to \infty$, and their difference scales according to \cite{wang2022}  
\begin{gather}
  \Phi(E) = R_{\text{min}}(E) + {O(N^{-1})}. \label{eq:rmin}
\end{gather}
In our models, the Coulomb potential with contributions coming from $O(N^n)$ skin modes   can be resolved through a total number of $M$ eigenenergies ($M \leq N^d$):
    \begin{gather}
        \Phi(E) = \frac{1}{M} \sum_m \log |E_m - E|,
    \end{gather}
where each $E_m$ is obtained in our exact solutions of \Cref{eq:gf}.
While $\Phi(E)$ depends on the linear system size $N$, $R_{\text{min}}(E)$ given by \Cref{eq:rkm} becomes a universal quantity.  Figure~\ref{fig:spec} shows that, when $N$ increases,
the convergence to the universal spectrum goes slower when the potential profile includes more types of boundary modes. 
The DOS of the universal spectrum is determined directly from Ronkin function \cite{wang2022}:
  $\rho(E) = \left. \frac{1}{2\pi}\Delta \Phi(E) \right|_{N \to \infty}  =  \frac{1}{2\pi}\Delta  R_{\text{min}}(E)$, with $\Delta = \frac{\partial^2}{\partial (\re E)^2} + \frac{\partial^2}{\partial (\im E)^2} $; it also excludes any contribution from boundary modes.

\section{Generalizations}
\label{sec:gen}
In this section, we address our NH hypercubic models under hybrid boundary conditions ($d_c < d$): $\text{OBC}_{d_c} + \text{PBC}_{d-d_c}$. By constructing an enlarged motif $\tilde{A}$ which includes the $A$ motif and all $B$ motifs along the PBC directions as its internal degrees of freedom, the solvability of the model class remains intact. We then generalize the concepts of biorthogonal polarization and the amoeba to hybrid boundaries. A minimal model in this context is given by NH Lieb lattice on a cylinder geometry where we enrich the physics by attaching $\pi$ magnetic flux to each plaquette (see \Cref{fig:lattice}). 

\subsection{Hybrid boundaries}
\subsubsection{Solvability and GSBZ}
As a first step, it is useful to clarify notations related to dimensions given hybrid boundaries. Our hypercubic lattice 
under $\text{OBC}_{d_c} + \text{PBC}_{d-d_c}$ lives in  spatial dimension $d$  with OBC in $d_c$ directions and PBC in $(d - d_c)$ directions. It yields skin modes of codimension $D = 0,1,\ldots,d_c$, the total number of which is proportional to $O(N^n)$ where $n = d-d_c +D$ and $  d-d_c \le n \le d$.
For bulk (boundary) modes, $D = d_c$, $n = d$ ($D < d_c$, $n < d$). 

One discerns that the full NH lattice Hamiltonian can be separated to two parts: 
$\mathcal{H} = \mathcal{H}_\perp + \mathcal{H}_\parallel$, where $ \mathcal{H}_\perp$ ($\mathcal{H}_\parallel$) is subjected to OBC (PBC). We denote all $B$ motifs in the OBC (PBC) directions  by the symbol $B_\perp$ ($B_{\parallel}$) consisting of  elements $B_l$ with $l \in \bar{l}_\perp$ ($\bar{l}_\parallel$).
For convenience, we designate $\bar{l}_d = \{1, 2, \dots, d \}$,  $\bar{l}_\parallel = \{l_{m_1}, l_{m_2}\dots, l_{m_{d-d_c}} \}$
and $\bar{l}_\perp = \bar{l}_d - \bar{l}_\parallel$.
By Fourier transform in all PBC directions, 
\begin{gather}
  c_{\vec{j}_\perp, \lambda} (\vec{j}_\parallel)=\frac{1}{\sqrt{\prod_{l \in \bar{l}_\parallel} N_l}} \sum_{\vec{k}_\parallel}  e^{i \vec{k}_\parallel \cdot \vec{j}_\parallel} c_{\vec{j}_\perp, \lambda} (\vec{k}_\parallel), \label{eq:ftp}
\end{gather}
with the motif index $\lambda \in A  \cup B_\perp \cup B_\parallel$, 
it is natural to 
create a new $\tilde{A}$ motif from $(d-d_c+1)$ independent eigenmodes of $\mathcal{H}_\parallel (\vec{k}_\parallel)$, such that the original $AB_\parallel$ motifs are incorporated into its internal independent degrees of freedom. We leave the details of this construction to the appendix.

On the new $\tilde{A} B_{\perp} = \{ \tilde{A}_+ ,\tilde{A}_- \}  \cup B_\perp$ motifs, we arrive at  a generalized NH hypercubic model: ${\mathcal{H}'} = {\mathcal{H}'}_\perp + {\mathcal{H}'}_\parallel = \sum_{\vec{k}_\parallel} {\mathcal{H}'}(\vec{k}_\parallel)$. At each $\vec{k}_\parallel$, ${\mathcal{H}'}(\vec{k}_\parallel)$ extends along all OBC directions:
  \begin{gather}
    {\mathcal{H}'} (\vec{k}_\parallel) = \sum_{\alpha = \pm} \sum_{\vec{j}_\perp} \sum_{l \in \bar{l}_\perp}  \epsilon_{\tilde{A}_\alpha}(\vec{k}_\parallel)  c_{\vec{j}_\perp, \tilde{A}_\alpha}^\dagger ( \vec{k}_\parallel) c_{\vec{j}_\perp, \tilde{A}_\alpha}  (\vec{k}_\parallel) \notag \\
    + \ {t}_{2l-1, \alpha}^+  c^\dagger_{\vec{j}_\perp, \tilde{A}_\alpha} (\vec{k}_\parallel) c_{\vec{j}_\perp, B_l}(\vec{k}_\parallel) \notag \\
    + \  t_{2l-1,\alpha}^-c^\dagger_{\vec{j}_\perp, B_l} (\vec{k}_\parallel)c_{\vec{j}_\perp,\tilde{A}_\alpha} (\vec{k}_\parallel)   \notag \\
    + \ t_{2l, \alpha}^+ c^\dagger_{\vec{j}_\perp, B_l} (\vec{k}_\parallel) c_{\vec{j}_\perp+\vec{e}_l, \tilde{A}_\alpha}(\vec{k}_\parallel)   \notag \\
   +\ t_{2l,\alpha}^-  c^\dagger_{\vec{j}_\perp + \vec{e}_l, \tilde{A}_\alpha}(\vec{k}_\parallel)  c_{ \vec{j}_\perp, B_l}(\vec{k}_\parallel), \label{eq:gnh}
  \end{gather}
where 
  \begin{gather}
       \epsilon_{\tilde{A}_\pm} (\vec{k}_\parallel) = \pm \sqrt{\sum_{l \in \bar{l}_\parallel} f_{2l-1,2l}(\vec{k}_\parallel)}, \label{eq:hzero}
  \end{gather}
  and the $f$ functions take the analytical form in \Cref{eq:gf}. From the appendix, it turns out that only two dispersive eigenmodes $\tilde{A}_\pm$ of $\mathcal{H}_\parallel (\vec{k}_\parallel)$  are elevated to skin modes through renormalized nonreciprocal couplings $t_{l,\alpha}^\pm$ to  the $B_\perp$ motifs [the explicit forms can be found in \Cref{eq:rcou}].  By contrast, the remaining $(d - d_c -1)$ zero-energy eigenmodes on the flat bands are decoupled from $B_\perp$, turning into nonlocalized modes.  Their contributions do not enter ${\mathcal{H}'}$. 

In spite of the fact that the nonzero effective mass term $\epsilon_{\tilde{A}_\alpha}(\vec{k}_\parallel)$ at $\alpha = \pm$ breaks generalized chiral symmetry in ${\mathcal{H}'}(\vec{k}_\parallel)$, our models are still solvable with spectral mirror symmetry respected along all OBC directions [see \Cref{eq:mirror}]. Equally important, from Eqs.~(\ref{eq:rcou}), (\ref{eq:rRL}) and (\ref{eq:rl}),  a unique localization parameter $r_{l}^\alpha$ is shared by 
 two independent particles $\tilde{A}_{\alpha}$, which keeps the same value as the case under complete OBC: 
 \begin{gather}
   r_{R,l}^\alpha = -\frac{t_{2l-1,\alpha}^-}{t_{2l,\alpha}^+}  = r_{R,l}, \notag \\   r_{L,l}^{\alpha *} = -\frac{t_{2l-1,\alpha}^+}{t_{2l,\alpha}^-} =  r^*_{L,l},  \notag \\
   r_{l}^\alpha = \sqrt{\frac{r_{R,l}^\alpha}{r_{L,l}^{\alpha *}}}  = r_l.\label{eq:0rl}
 \end{gather}
As before, in search of exact solutions, nonunitary gauge transforms $S$ and $U$ can be established for  ${\mathcal{H}'}(\vec{k}_\parallel)$ by replacing in Eqs.~(\ref{eq:gauge_s}), (\ref{eq:gauge_s1})
 and (\ref{eq:map2_0}), (\ref{eq:map2_1}):
   \begin{gather}
  \bar{l}_d \to \bar{l}_\perp ,  \quad \bar{l}_n \to \bar{l}_D, \quad c_{\vec{j}, \lambda} \to c_{\vec{j}_\perp, \lambda}(\vec{k}_\parallel), \label{eq:gtg}
   \end{gather}
where  $\bar{l}_{D} = \{l_1, \dots, l_{D} \} \in \bar{l}_\perp$ denotes the motifs $\lambda \in  \tilde{A}_\alpha \cup B_{\perp}^D =  \{ \tilde{A}_\alpha, B_{l_1}, \dots, B_{l_D}\}$ covered by skin modes of codimension $D$. Key localization parameters remain the same from \Cref{eq:0rl}.
These nonunitary gauge transforms also lead to the exact GSBZ for the right $O(N^n)$ skin modes under hybrid boundary conditions ($n = d - d_c + D$), with a generic function present in \Cref{eq:gsbz}.

Moreover, the solvability of our models can be further relaxed to different 
$r_{R/L,l}^\alpha$ in \Cref{eq:0rl}, which corresponds to distinct localization factors $\tilde{r}_l^\alpha = r^\alpha_{R,l}/r_l$ in $U$, since the contributions from the occupied  $\tilde{A}_\alpha$ sites on the unoccupied $B_\perp$ motifs can be canceled independently [see  \Cref{fig:rs}(a) and also the example of Lieb lattice at $\pi$ flux in \Cref{sec:cyl}]. However, the gauge transform $S$ acts on the shared $B_\perp$ motifs, requiring a unique 
$r_{l}$ factor [see \Cref{fig:rs}(b)].

\subsubsection{Exact spectrum, skin and nonlocalized modes}
Next, we illuminate the structures of the spectrum and the eigenmodes including skin and nonlocalized modes entailed by hybrid boundaries.

Along all OBC directions, one performs a second
 Fourier transform:
\begin{gather}
 c_{\vec{j}_\perp, \lambda} (\vec{k}_\parallel) =\frac{1}{\sqrt{\prod_{l \in \bar{l}_\perp} N_l}} \sum_{\vec{k}_\perp}  e^{i \vec{k}_\perp \cdot \vec{j}_\perp} c_{\lambda} (\vec{k}_\perp; \vec{k}_\parallel).
\end{gather}
As eigen energies do not change with basis, we can  choose to diagonalize ${\mathcal{H}'}$ in the new motifs  $\lambda \in \tilde{A} \cup B_{\perp}^D$ or $\mathcal{H}$ in the original motifs  $\lambda \in A \cup B_\parallel \cup B_{\perp}^D$. It turns out that while the former is convenient for the identification of corner modes ($D = 0$), the latter is a more natural choice to get access to skin modes of higher codimension ($D > 0$). 
In the original basis $\u{\psi} (\vec{k}) = (c_{A} (\vec{k}), c_{B_\parallel} (\vec{k}), c_{B_\perp^D} (\vec{k}))^T$ with $\vec{k} = (\vec{k}_\perp; \vec{k}_\parallel) = (\vec{k}_D, \vec{0};\vec{k}_\parallel)$, the associated non-Bloch Hamiltonian shares the structure:
    \begin{gather}
      H^{\text{non-Bloch}}_{D}(\vec{k}) 
      = \begin{pmatrix}
	  		0 & Y_{\parallel,-} & Y_{\perp,-}\\
      Y_{\parallel,+}^T &  0 & 0 \\
      Y_{\perp,+}^T & 0 & 0
		    \end{pmatrix}. \label{eq:hnb_h}
    \end{gather}
While $Y_{\parallel, \pm} =
       ( t_{2m_{1}-1}^\mp + t_{2m_1}^\pm e^{\pm ik_{m_1}}, \ \dots,\  t_{2m_{d-d_c}-1}^\mp  + t_{2m_{d-d_c}}^\pm e^{\pm ik_{m_{d-d_c}}})$ in the normal BZ, 
     $Y_{\perp, \pm} = (t_{2l_1-1}^\mp + t_{2l_1}^\pm (r_{l_1})^{\pm 1} e^{\pm ik_{l_1}},\  \dots,\  t_{2l_D-1}^\mp + t_{2l_D}^\pm (r_{l_D})^{\pm 1}  e^{\pm ik_{l_D}})$ following momentum shifts in the GSBZ.
The exact spectrum ($d_c < d$) follows 
 \begin{align}
   E_{D, \pm}(\vec{k}) &= \pm \sqrt{\sum_{l \in \bar{l}_\parallel} f_{2l-1,2l}(k_l)+ \sum_{l \in \bar{l}_D}g_{2l-1,2l}(k_l)}, \notag \\
   E_{D, 0 }(\vec{k}) &= 0 \quad  (D>0), \notag \\
   E_{\text{non-local}}(\vec{k}) &= 0 \quad (d-d_c-1 > 0), \label{eq:spec_h}
 \end{align}
with the $f$ and $g$ functions in \Cref{eq:gf}.

When $D = 0$ and $\bar{l}_D = \varnothing$, there arise in $E_{D, \pm}(\vec{k})$ two corner modes that are endowed with opposite energies equal to \Cref{eq:hzero}. As eigenmodes of ${\mathcal{H}'}(\vec{k}_\parallel)$ in \Cref{eq:gnh}, they become localized on $\tilde{A}_+$ and $\tilde{A}_-$ motifs:
\begin{align}
 E_{D=0, \pm}(\vec{k}) &= \epsilon_{\tilde{A}_\pm} (\vec{k}_\parallel), \label{eq:cor_h} \\
 |{\psi}_{R/L, 0,  \pm} (\vec{k}) \rangle 
 &= \mathcal{N}_{R/L, \pm}  \sum_{\vec{j}_\perp} \prod_{l \in \bar{l}_\perp} (r_{R/L,l})^{j_{\perp,l}} c^\dagger_{ \vec{j}_\perp, \tilde{A}_\pm} (\vec{k}_\parallel)|0\rangle. \notag
\end{align}

Whereas, for codimension $D > 0$, by taking $\bar{l}_\parallel = \varnothing$ in $E_{D, \pm}(\vec{k})$, the two dispersive bands under hybrid boundary conditions ($d_c < d$) are connected to the ones under the complete OBC  ($d_c =d$) in \Cref{eq:gf}.
At the same time, in addition to $(d-d_c - 1)$ zero-energy nonlocalized modes on $B_\parallel$ motifs captured by the normal BZ along the PBCs, there emerge $D$ zero-energy  skin modes  exponentially localized on the $B_\perp$  motifs characterized by the GSBZ. In total, we retrieve  $(1 + d-d_c + D)$ eigenmodes for the non-Bloch Hamiltonian.

The exact GSBZ in \Cref{eq:gsbz} also predicts
the hybrid localization behaviors of skin modes of codimension $D$ in terms of non-Bloch waves:
    \begin{align}
       &\u{\psi}_{R, D, (\alpha, \vec{k})}(\vec{j}) \\
       &\propto   \prod_{l \in \bar{l}_D} (r_{l})^{j_{\perp, l}}  \prod_{l' \in (\bar{l}_\perp - \bar{l}_D)} (r_{R,l'})^{j_{\perp,l'}}  e^{i\vec{k} \cdot \vec{j}}  \u{u}_{R,D,\alpha}(\vec{k} ), \notag \\
       &\u{\psi}_{L,D,(\alpha, \vec{k})}(\vec{j}) \notag \\
       &\propto  \prod_{l \in \bar{l}_D} (r_{l}^*)^{-j_{\perp, l}}  \prod_{l' \in (\bar{l}_\perp - \bar{l}_D)} (r_{L,l'})^{j_{\perp,l'}}  e^{i\vec{k} \cdot \vec{j}}  \u{u}_{L,D,\alpha}(\vec{k}), \notag
    \end{align}
where $\u{u}_{D,R/L, \alpha}(\vec{k})$ with $\alpha = \pm, 0$ represents the biorthogonal eigenvector of the non-Bloch Hamiltonian in \Cref{eq:hnb_h}. In comparison, the nonlocalized modes are composed of normal Bloch waves:
 \begin{gather}
   \u{\psi}_{R/L,\vec{k}}^{\text{non-local}}(\vec{j}) \propto e^{i\vec{k}\cdot\vec{j}} \u{u}_{R/L}^{\text{non-local}}(\vec{k}).
 \end{gather}
Being also an eigenvector of the non-Bloch Hamiltonian,  $\u{u}_{R/L}^{\text{non-local}}(\vec{k})$ vanishes on all $B_\perp^D$ motifs and is immune to the momentum shifts.

\subsubsection{Generalized biorthogonal polarization and amoeba formulation}
Now, we extend the analytical tools of biorthogonal polarization and the amoeba to hybrid boundaries, such that they can still be used to detect surface and separation gap closings between boundary and bulk modes. 

On one hand, with chiral symmetry broken in ${\mathcal{H}'}(\vec{k}_\parallel)$ of \Cref{eq:gnh}, the non-Bloch winding number in \Cref{eq:wn} is no longer a good invariant. In order to  capture surface gap closings, one can apply the more robust biorthogonal polarization based on two corner modes in \Cref{eq:cor_h}. 
Given $\vec{k}_\parallel$, 
let us define a generalized polarization vector $\vec{P}(\vec{k}_\parallel)$ with its components in each OBC direction $x_l$ ($l \in \bar{l}_\perp$) given by \begin{widetext}
    \begin{gather}
      P_l(\vec{k}_\parallel)  = 2  
     -\lim_{N_l \to \infty} \frac{1}{N_l} \sum_{\alpha = \pm}\left| \langle \psi_{L0,\alpha}(\vec{k}_\parallel)| \sum_{\vec{j}_\perp} j_{\perp,l} \Pi_{\vec{j}_\perp} (\vec{k}_\parallel) | \psi_{R0, \alpha}(\vec{k}_\parallel)\rangle \right|. \label{eq:gpol} 
     \end{gather}
     \end{widetext}
The density operator acts on the $\vec{j}_\perp$th unit cell in OBC directions: $\Pi_{\vec{j}_\perp} (\vec{k}_\parallel) = \sum_{\lambda} |e_{\vec{j}_\perp, \lambda}(\vec{k}_\parallel) \rangle \langle e_{\vec{j}_\perp,\lambda}(\vec{k}_\parallel)|$ with  $|e_{\vec{j}, \lambda} (\vec{k}_\parallel) \rangle = c^\dagger_{\vec{j}_\perp,\lambda} (\vec{k}_\parallel)|0\rangle$ and $\lambda \in \tilde{A}_+ \cup \tilde{A}_- \cup B_\perp $.
Since the localization factors of two corner modes $r_{R/L,l}$ in \Cref{eq:0rl} do not depend on $\vec{k}_\parallel$, $P_l$ does not vary with $\vec{k}_\parallel$. Taking into account  $\langle {\psi}_{L0, \alpha } (\vec{k}_\parallel) |{\psi}_{R0, \alpha} (\vec{k}_\parallel) \rangle = 1$, we obtain quantized polarization components: $P_l = 2 \ (0)$ when $|r^*_{L,l}r_{R,l}| < 1 \ (> 1)$. $\vec{P}$ can signal the surface gap closings for higher-order dispersive skin modes ($\alpha = \pm$) in \Cref{eq:spec_h}:
\begin{align}
   & | \Delta E_{\text{Surf.}} (\vec{k}_\parallel)|  \label{eq:surf-h} \\
   & \phantom{=} = \min_{\forall \vec{q},\vec{k}_D,\alpha} \{ |E_{d_c,+} (\vec{k}_D, \vec{q};\vec{k}_\parallel ) - \alpha E_{D,+} (\vec{k}_D, \vec{0};\vec{k}_\parallel)| \}. \notag
\end{align}
More precisely,
\begin{gather}
 \left. \Delta E_{\text{Surf.}} (\vec{k}_\parallel) \right|_{N \to \infty} = 0 \quad  \text{if} \quad |\Delta P_{l'}| = 1 \  \forall \ l' \in \bar{l}_\perp - \bar{l}_D. 
 \end{gather}

On the other hand, the separation gaps with bulk modes of codimension $d_c$ can be captured by the generalized amoeba. At each $\vec{k}_\parallel$, we define
    \begin{gather}
       \mathcal{A}_{f} (\vec{k}_\parallel)= \{ \vec{\mu}\ \text{with} \  \mu_l  = \log |{\beta}_l|: h(\vec{\beta}, \vec{k}_\parallel) = 0 \}, 
    \end{gather}  
    where 
    \begin{gather}
       h(\vec{\beta}, \vec{k}_\parallel) = \det [H(\vec{\beta}, \vec{k}_\parallel) - E]/E^{d-1}, \label{eq:hb_h}
    \end{gather}
 By analogy to the complete OBC, here $H(\vec{\beta}, \vec{k}_\parallel)$ comes from the bulk Bloch Hamiltonian with a replacement $\beta_l = e^{ik_{l}}$ for any $l \in \bar{l}_\perp = \{ l_1 , \dots, l_{d_c} \}$ in the OBC directions,
    \begin{gather} 
      H(\vec{\beta}, \vec{k}_\parallel) 
      = \begin{pmatrix}
	  		0 & Y_{\parallel,-}(\vec{k}_\parallel) & Y_{\perp,-}(\vec{\beta})\\
      Y_{\parallel,+}^T (\vec{k}_\parallel) &  0 & 0 \\
      Y_{\perp,+}^T(\vec{\beta}) & 0 & 0
		    \end{pmatrix}. \label{eq:gamo}
    \end{gather}
$Y_{\parallel,\pm} (\vec{k}_\parallel)$ is given below \Cref{eq:hnb_h} and $ Y_{\perp, \pm}(\beta) = (t_{2l_1-1}^\mp + t_{2l_1}^\pm \beta_{l_1}^\pm,\  \dots,\  t_{2l_{d_c}-1}^\mp + t_{2l_{d_c}}^\pm \beta_{l_{d_c}}^\pm)$.
Trivial solutions from a total of $(d - 1)$ zero-energy flat bands are excluded by the denominator in \Cref{eq:hb_h}. Under hybrid boundary conditions, one evaluates the separation gap between boundary and bulk dispersive modes ($\alpha = \pm$) according to 
\begin{align}
   &| \Delta E_{\text{Sep.}} (\vec{k}_\parallel)| \label{eq:sep-h} \\
   &\phantom{=} = \min_{\forall \vec{p},\vec{q},\vec{k}_D,\alpha} \{ |E_{d_c,+} (\vec{p}, \vec{q};\vec{k}_\parallel ) - \alpha E_{D,+} (\vec{k}_D, \vec{0};\vec{k}_\parallel)| \},  \notag
\end{align}
where $d_c$ ($D$) denotes the codimension of bulk (boundary) modes.
A hole-free amoeba indicates a separation gap closing: 
\begin{gather}
     \Delta E_{\text{Sep.}}(\vec{k}_\parallel) |_{N \to \infty} = 0 \quad   \text{if} \quad   \mathcal{V}_{\text{hole}} (\vec{k}_\parallel)|_{E = E_{D,\text{min}}} = 0.
 \end{gather}
Without loss of generality, we take from the boundary spectrum a reference energy at $E_{D,\text{min}}$ which minimizes the separation gap.

   \begin{figure*}[t]
\centering
\includegraphics[width=1\linewidth]{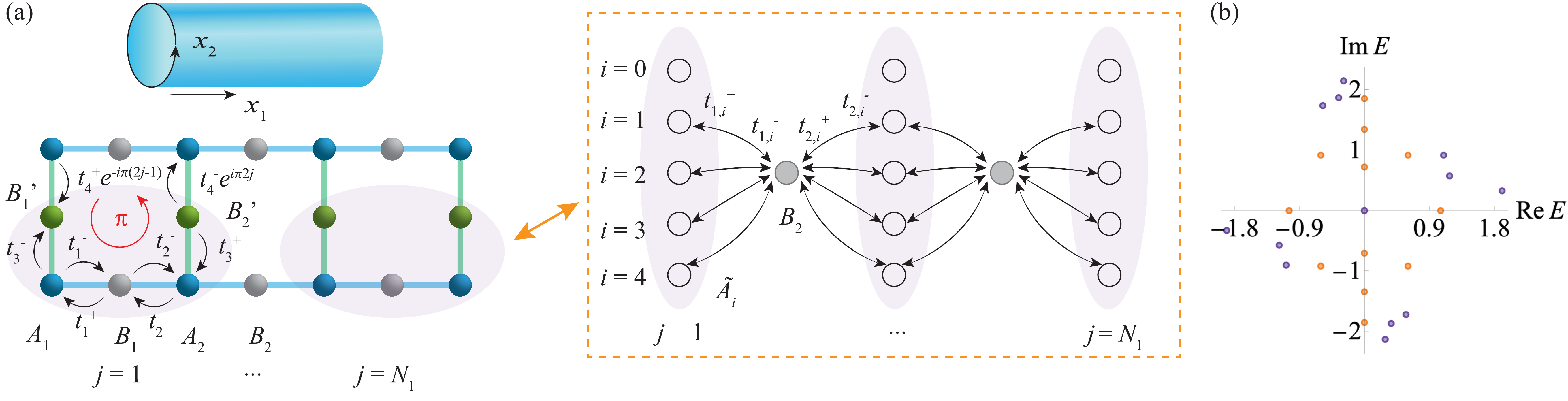}
\caption{(a) Cylinder geometry for the NH Lieb lattice of size $(4N_1-1)\times(2N_2-1)$ at $\pi$ flux with a mapping to a generalized NH SSH model along the $x_1$ (OBC) direction. The four zero-energy corner modes on the chain form two chiral edge pairs on the cylinder. We fix the gauge of $\pi$ flux per plaquette by assigning a phase factor $e^{\mp i\pi(2j-1)}$ ($e^{\mp i\pi 2j}$) to the $t_4^\pm$ bonds connected to the $B'_1$ ($B'_2$) motif. (b) Comparison of complex eigenvalues between the numerical (dark dots) and analytical (light dots) results for a finite-size cylinder with $N_{1} =3$ at given momenta $k_2 = \pi/10$ (purple) and $k_2 = \pi$ (orange). Distinct values are taken for the whole set of hopping parameters: $\{t_1, t_2, t_3, t_4 \} = \{0.8, 1, 0.9, 1.2 \}$, $\{\gamma_1, \gamma_2, \gamma_3, \gamma_4 \} = \{\sqrt{3}, 0.1, \sqrt{2}, 0.2 \}$.}
\label{fig:cyl}
\end{figure*}  
\subsection{Example of cylinder geometry: Non-Hermitian Lieb lattice at $\pi$ flux}
\label{sec:cyl}

For completeness, we give an example of our solvable NH hypercubic models under hybrid boundary conditions, the NH Lieb lattice on a cylinder geometry ($d_c = 1 < d$, $\text{OBC}_1 + \text{PBC}_1$) shown in Figs.~\ref{fig:lattice} and \ref{fig:cyl}(a). An external magnetic field is introduced at $\pi$ flux per plaquette, bringing an enriched  phase diagram in \Cref{fig:cyl_pol}(a). We derive the exact spectra for bulk modes and four chiral edge modes. Similar to the corner mode under complete OBC,  these edge modes on the cylinder share zero codimension and display identical surface and separation gaps with the bulk. 
It enables us to apply both biorthogonal polarization and amoeba formulation  to locate complex gap closings in the system in Figs.~\ref{fig:cyl_pol} and \ref{fig:cyl_spec}, through which biorthogonal bulk-boundary correspondence is manifested.

\subsubsection{Spectrum and chiral edge modes}
First, as illustrated in \Cref{fig:cyl}(a), we introduce the NH Lieb model on a cylinder geometry with OBC in  $x_1$ direction and PBC in  $x_2$ direction, such that spectral mirror symmetry is present along the OBC direction: $E(k_1, k_2) = E(-k_1, k_2)$. In presence of an external magnetic field,  each plaquette is attached to $\pi$ flux, which enlarges 
the unit cell to include six sites. Guided by spectral mirror symmetry that requires both edges of the cylinder terminate on the same motif, we establish motifs $A \cup B_2$, where the motif $A = \{ A_1, B_1, B_1', A_2, B_2'\}$ contains five internal degrees of freedom (marked by pink). In our convention,  the new unit cell is located by $\vec{j} = (j,j')$  with $j = 1, \dots, N_1$, $j' = 1, \dots, N_2$. The gauge can now be fixed by assigning a phase factor $e^{\mp i\pi(2j-1)}$ ($e^{\mp i\pi 2j}$) to the $t_4^\pm$ bonds connected to the $B'_1$ ($B'_2$) motif. In this manner, the flux per plaquette or unit cell recovers $\Phi_{\tot} = \pi$.

Along the lines of the generic approach for hybrid boundaries in \Cref{eq:gnh}, one can map the current model to a generalized NH SSH chain on the new $\tilde{A}\cup B_2$ motifs: ${\mathcal{H}'} = \sum_{k_2} {\mathcal{H}'}(k_2)$, with
  \begin{align}
  &{\mathcal{H}'}(k_2) = \sum_{j=1}^{N_1}  \sum_{i=1}^{4} \epsilon_i(k_2) c^\dagger_{j, \tilde{A}_i}(k_2) c_{j, \tilde{A}_i} (k_2) \label{eq:th-cyl} \\
  &+ t_{1,i}^+  c^\dagger_{j, \tilde{A}_i}(k_2) c_{j, B_2}(k_2)  + t_{1,i}^-c^\dagger_{j, B_2}(k_2) c_{j,\tilde{A}_i}(k_2)  \notag \\ 
  &+ t_{2,i}^+  c^\dagger_{j, B_2} (k_2) c_{j+1, \tilde{A}_i}(k_2)   + t_{2,i}^- c^\dagger_{ j + 1, \tilde{A}_i}(k_2) c_{j, B_2}(k_2), \notag
  \end{align}
by diagonalizing $\mathcal{H}_\parallel$ inside the $A$ motif (see more details in the appendix). Shown in \Cref{fig:cyl}(a) (right panel), this mapping establishes a NH SSH chain
with an enlarged  $\tilde{A}$ motif: $\tilde{A} = \{\tilde{A}_1, \dots, \tilde{A}_4 \}$. Each $\tilde{A}_i$ holds an effective mass: 
\begin{gather}
 \epsilon_i(k_2) = \pm \sqrt{\tau_0 \pm \sqrt{\tau(k_2)}},
\end{gather}
where
\begin{align}
  \tau_0 &= t_1^+ t_1^- + t_2^+ t_2^- + 2(t_3^+ t_3^- + t_4^+ t_4^-), \notag \\
  \tau(k_2) &= [t_1^+t_1^- - t_2^+t_2^- -2 (t_3^+t_4^+e^{ik_2}+t_3^-t_4^-e^{-ik_2})]^2 \notag \\
  &\phantom{=} + 4t_1^+ t_1^- t_2^+ t_2^-, \label{eq:taui}
\end{align}
and displays asymmetric hopping terms $\{t_{1,i}^\pm, t_{2,i}^\pm\}$ identified in Eqs.~(\ref{eq:dis}) and (\ref{eq:cyl-c}).
The index $i=1,2,3,4$ refers to  four modes taking $(+,+),(-,+),(+,-),(-,-)$ signs in $\epsilon_i(k_2)$.
In addition, there is one zero-energy eigenmode $\tilde{A}_0$  decoupled from the rest, and it becomes the nonlocalized mode occupying only $B_1B_1'B_2'$ motifs in the PBC direction. 

Next, we solve the analytical spectra of bulk and edge modes on the cylinder geometry. The total number of edge modes is proportional to $O(N_2)$, equal to the degrees of freedom in $k_2$ and consistent with $n=d-d_c + D = 1$. For each $k_2$, there arise in ${\mathcal{H}'}$ four chiral edge modes with a dispersive energy equal to the effective mass, each exponentially localized on the $\tilde{A}_i$ motif:
\begin{align}
  E_{\text{edge},i} (k_2) &= \epsilon_i(k_2),  \label{eq:c2} \\ 
 |{\psi}_{R/L, i}^{\text{edge}} (k_2) \rangle &= \mathcal{N}_{R/L, i}(k_2) \sum_{j=1}^{N_1}  r_{R/L,i}^j (k_2) c^\dagger_{j, \tilde{A}_i} (k_2)|0\rangle. \notag
\end{align}
From the biorthogonal relation $\langle {\psi}_{L, i}^{\text{edge}} (k_2) |{\psi}_{R, i}^{\text{edge}} (k_2) \rangle =1 $, the normalization factors read
\begin{align}
    &\mathcal{N}^*_{L,i}(k_2) \mathcal{N}_{R,i} (k_2) \notag \\
     &= \frac{[r^*_{L,i}(k_2)r_{R,i}(k_2)]^{-1}[r^*_{L,i}(k_2)r_{R,i}(k_2)-1]}{[r^*_{L,i}(k_2)r_{R,i}(k_2)]^N-1}. 
\end{align}
In contrast to the scenario without a magnetic field in \Cref{eq:0rl}, the localization factors  vary with $k_2$ at $\pi$ flux:
 \begin{align}
      r_{R,i}(k_2) &= - \frac{t_{1,i}^-}{t_{2,i}^+} = \left( \frac{t_1^-}{t_2^+} \right)^2 [1-\eta_i(k_2)], \notag \\   
      r^*_{L,i}(k_2) &=  - \frac{t_{1,i}^+}{t_{2,i}^-}  = \left( \frac{t_1^+}{t_2^-} \right)^2 [1-\eta_i(k_2)],
 \end{align}
where we introduce the auxiliary $\eta$ functions:
    \begin{gather}
     \eta_i(k_2) = \frac{1}{t_1^+t_1^-} [\epsilon_i^2 -  (t_3^+ - t_4^- e^{- ik_2})(t_3^- -t_4^+ e^{ ik_2})]. \label{eq:eta}
     \end{gather}
Since $\eta_i$ is an even function of $\epsilon_i$, the four edge modes form two chiral edge pairs (CEPs). Each pair is characterized by the same localization length but opposite energies, 
\begin{align}
r_{R/L,1}(k_2) &= r_{R/L,2}(k_2), \quad \epsilon_1(k_2) = - \epsilon_2(k_2), \notag \\
r_{R/L,3}(k_2) &= r_{R/L,4}(k_2), \quad \epsilon_3(k_2) = - \epsilon_4(k_2). \label{eq:cep}
\end{align}

The localization factor of bulk modes on the cylinder is also unique: 
 \begin{gather}
   r = \sqrt{\frac{r_{R,i}(k_2)}{r^*_{L,i}(k_2)}} = \frac{t_1^-t_2^-}{t_1^+t_2^+} = r_1^2, \label{eq:rcyl}
 \end{gather}
where we make links with $r_1$, hosted by the NH Lieb model at zero flux in \Cref{eq:liebr} under complete OBC. Intuitively, $\pi$ flux makes the unit cell twice in size, thus doubling the localization factor in the exponential: $\log|r| = 2\log|r_1|$.

In the OBC direction, the uniqueness of the localization factor $r$  together with spectral mirror symmetry contribute to an exact GSBZ, established through gauge transforms of \Cref{eq:gtg}. Similar to \Cref{eq:hnb_h}, the bulk spectrum can be obtained conveniently through an imaginary momentum shift  in the Bloch Hamiltonian in the original basis $\u{\psi}(\vec{k}) = (c_{A_1}(\vec{k}), c_{B_1}(\vec{k}), c_{B'_1}(\vec{k}), c_{A_2}(\vec{k}), c_{B'_2}(\vec{k}),  c_{B_2}(\vec{k}))^T$:
  \begin{gather}
    H(k_1, k_2) =     
     \begin{pmatrix}
              0 & t_1^+ & p^+  & 0 & 0 & t_2^- e^{-ik_1} \\
        t_1^- & 0 & 0 & t_2^+ & 0  & 0 \\
        p^- & 0 & 0 & 0 & 0 & 0 \\
        0 & t_2^- & 0 & 0 & q^+ & t_1^+ \\
        0 & 0 & 0 & q^- & 0  & 0 \\
        t_2^+ e^{ik_1} & 0 & 0 & t_1^- & 0 & 0
      \end{pmatrix}, \label{eq:cyl_b}
    \end{gather}
where $p^\pm = t_3^\pm - t_4^\mp e^{\mp ik_2},  q^\pm = t_3^\pm + t_4^\mp e^{\mp ik_2}$. A second Fourier transform has been performed along $x_1$ direction: $c_{j,\lambda} (k_2) = \frac{1}{\sqrt{N_1}} \sum_{k_1} e^{ik_1 \cdot j} c_{\lambda} (k_1,k_2)$ with  $k_1 = \frac{\pi \tilde{m}}{N_1} \in (0, \pi)$ and $\tilde{m} = 1, \dots, N_1 -1$. The exact bulk spectrum follows from $H(k_1 - i\ln r, k_2)$:
  \begin{align}
     E_{\text{bulk}, 0}(\vec{k}) &= 0, \notag \\
    E_{\text{bulk},(\alpha,\alpha')}(\vec{k}) &= \alpha \sqrt{\zeta_0 +\alpha' \sqrt{\zeta(\vec{k})}},  \label{eq:c1} 
  \end{align}
with $\alpha, \alpha' = \pm$ and
  \begin{align}
      \zeta_0 &= \sum_{l=1}^4 t_l^+ t_l^-,  \\
      \zeta(\vec{k}) &= 4 t_1^+ t_1^- t_2^+ t_2^- \cos^2(k_1/2)+(t_3^+t_4^+e^{ik_2}+t_3^-t_4^-e^{-ik_2})^2. \notag
  \end{align}
The $\pi$ flux doubles the number of dispersive bands to four and through the $1/2$ factor in $\zeta (\vec{k})$ establishes the  magnetic BZ along OBC direction: $k_1 /2 \in (0, \pi/2)$. Meanwhile, a second zero-energy flat band emerges in the GSBZ, which generates skin modes that occupy four $B$ motifs, exponentially localized on the $B_2$ motif while delocalized on $B_1B_1'B_2'$ motifs. 

  To sum up,  in Eqs.~(\ref{eq:c2}) and (\ref{eq:c1}), we obtain the complete spectrum of the NH Lieb lattice at $\pi$ flux on a cylinder geometry, which includes chiral edge pairs, nonlocalized and skin modes at zero energy, as well as bulk dispersive modes. Figure~\ref{fig:cyl}(b) verifies that our analytical solutions of the eigenenergies are consistent with numerical results at different $k_2$ on a finite-size lattice.

\subsubsection{Biorthogonal bulk-boundary correspondence}
\begin{figure}[t]
\centering
\includegraphics[width=1\linewidth]{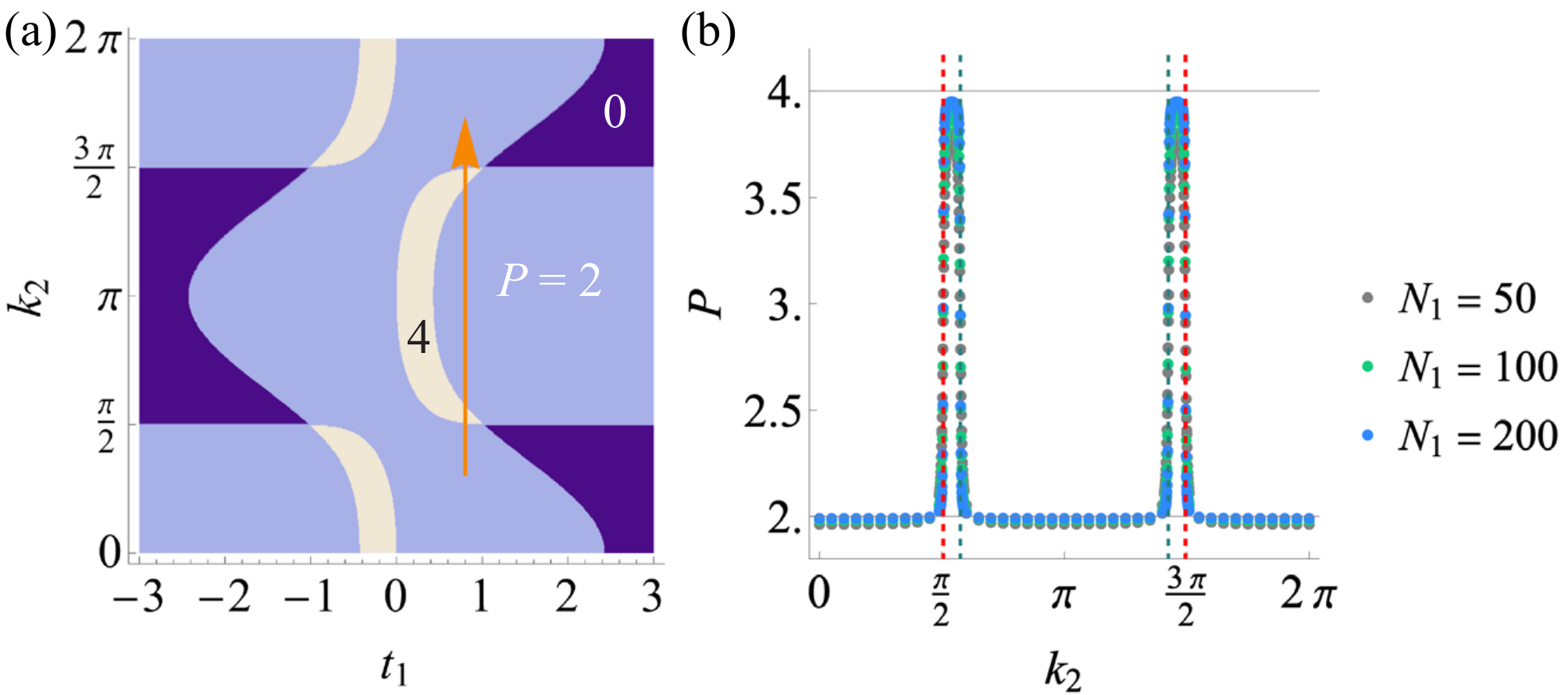}
\caption{(a) Phase diagram of the NH Lieb lattice on a cylinder geometry at $\pi$ flux characterized by biorthogonal polarization. The formation of four chiral edge modes into two pairs, each pair displaying same localization length, leads to $P \in \{ 0, 2, 4 \}$. We vary $k_2$, the momentum in the PBC direction of the cylinder and the hopping parameter $t_1 (= t_3) $ with fixed $t_{2,4}=1$, $\gamma_{1,3}= 0.2$ and $\gamma_{2,4}= 0$. The orange arrow signifies a parameter path as a function of $k_2$ at $t_1 = 0.8$. Following this path,  the quantization of polarization for different system sizes $N_1$ is shown in panel (b). $P$ jumps by two across the transitions  $|r^*_{L,1}(k_2)r_{R,1}(k_2)| = 1$ (red dashed lines) and $|r^*_{L,3}(k_2)r_{R,3}(k_2)| = 1$ (green dashed lines).}
\label{fig:cyl_pol}
\end{figure}

\begin{figure}[htb]
\centering
\includegraphics[width=1\linewidth]{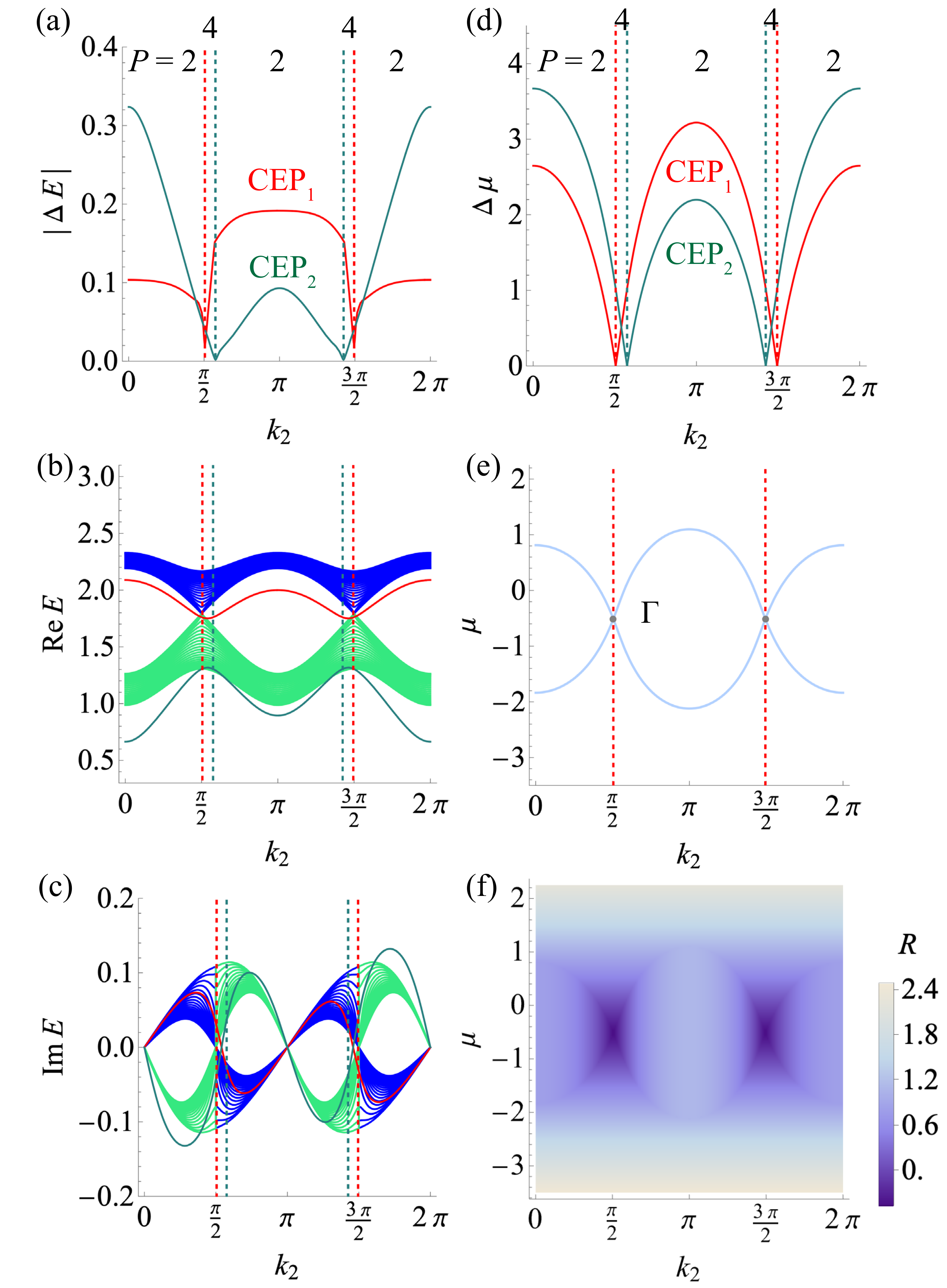}
\caption{NH Lieb lattice on a cylinder at $\pi$ flux. The parameter path follows the orange arrow in \Cref{fig:cyl_pol}(a). We use edge modes at $i=1, 3$ or equivalently $(+,+)$ and $(+,-)$ to represent CEP$_1$ (red) and CEP$_2$ (dark green). For each CEP, there are two solutions to $|r^*_{L,i}(k_2)r_{R,i}(k_2) | = 1$ over $k_2 \in [0, 2\pi)$ denoted by red and dark-green dashed lines. Across these transition lines, due to zero codimension of CEPs, the amoeba hole closes while the biorthogonal polarization jumps in panel (d). 
   Both predict that the CEP enters the bulk with a  complex gap closing in panel (a). Panels (b) and (c) further show real and imaginary energy spectra  of bulk modes [$(+,+)$ mode in blue and $(+,-)$ mode in light green] as well as CEPs. We take a system size $N_1 = 26$ ($N_2 = 200$) along the OBC (PBC) direction. Panels
(e) and (f) display the amoeba and Ronkin function at $E = 
\epsilon_1$ from CEP$_1$. A single Ronkin mininum is reached when the amoeba hole disappears at the central point $\Gamma$. The Ronkin integral takes a grid size $M_1 = 100$.
}
\label{fig:cyl_spec}
\end{figure}

To restore biorthogonal bulk-boundary correspondence between bulk and edge modes on the cylinder, we introduce a complex energy gap at given $k_2$:
 \begin{gather}
  |\Delta E (k_2)| = \min_{\forall k_1} \{ |E_{\text{bulk}}(k_1,k_2) - E_\text{edge} (k_2)| \}. \label{eq:ceg}
 \end{gather}
Since edge modes have zero codimension, by setting $\vec{k}_D = \varnothing$ in Eqs.~(\ref{eq:surf-h}) and (\ref{eq:sep-h}), $|\Delta E (k_2)|$ is equivalent to both the surface gap  and the separation gap. Consequently, complex energy gap closings can be captured by biorthogonal polarization as well as the generalized amoeba formulation (see \Cref{fig:cyl_spec}). 

First, let us look at the approach of generalized biorthogonal polarization. 
As chiral symmetry is broken by  effective mass terms in ${\mathcal{H}'}(k_2)$,  
the non-Bloch winding number is no longer a good invariant  at $\pi$ flux.  
Yet, based on four chiral edge modes constructed in \Cref{eq:c2},  one can still define a quantized polarization at each $k_2$ by analogy to \Cref{eq:gpol}: 
  \begin{align}
    &P(k_2) \\
    &= 4 - \lim_{N_1 \to \infty} \frac{1}{N_1} \sum_{i=1}^{4} \left|\langle \psi^{\text{edge}}_{L,i}(k_2)|\sum_{j=1}^{N_1} j \Pi_j (k_2)  | \psi^{\text{edge}}_{R,i}(k_2)\rangle \right|. \notag 
\end{align}
The density operator  $
 \Pi_j (k_2) = \sum_{\lambda} |e_{j, \lambda} (k_2) \rangle \langle e_{j,\lambda} (k_2)|$
with $|e_{j, \lambda} (k_2) \rangle = c^\dagger_{j,\lambda} (k_2)|0\rangle$ is measured on motifs $\lambda \in \tilde{A} \cup B_2$. Due to $\pi$ flux, the polarization varies with $k_2$ in contrast with being independent of $\vec{k}_\parallel$ at zero flux. Meanwhile, 
as shown in \Cref{fig:cyl_pol}(b), the quantization of $P(k_2)$ remains well defined for different system sizes. By further varying one of the hopping parameters, \Cref{fig:cyl_pol}(a) depicts the change of polarization over the parameter space $(t_1, k_2)$. The boundaries of the polarization phase diagram are captured by $|r^*_{L,i}(k_2)r_{R,i}(k_2)| = 1$, where $i=1, 3$ from \Cref{eq:cep}. The fact that the four edge modes form two CEPs, each sharing the same localization length, also indicates that $P(k_2)$ jumps by two every time it crosses the transition (or four when two transition lines overlap). Figures~\ref{fig:cyl_spec}(a)-\ref{fig:cyl_spec}(c) further shows that although the real and imaginary parts of bulk and edge spectra overlap at different regions, their complex energy gap defined in \Cref{eq:ceg}  closes precisely at the polarization transition lines $|r_{R,i}(k_2) r^*_{L,i}(k_2)| = 1$, where one CEP enters the bulk.

Next, we extend the amoeba formulation to the cylinder. It turns out that the hole disappears in response to complex energy gap closings. Let us define
    \begin{align}
        \mathcal{A}_{f} (k_2)&= \{ \mu (k_2) = \log |\beta (k_2)|: f(\beta, k_2) = 0 \}, \notag \\
      f(\beta, k_2) &= \det [H(\beta, k_2) - E]/E^2, 
    \end{align}
 where the reference energy $E = \epsilon_{i}(k_2)$ and  $H(\beta, k_2)$ is given by the Bloch Hamiltonian in \Cref{eq:cyl_b} with the replacement $e^{ik_1/2} \to \beta $, consistent with the magnetic BZ: $k_1/2 \in (0, \pi/2)$.  Two zero-energy flat bands are excluded by the denominator. The amoeba is solvable in 1D. From $f(\beta, k_2) = 0$, one obtains $\beta_{\pm} (k_2) = (-b \pm \sqrt{b^2 - 4ac})$/(2a) with $a = (t_1^+t_2^+)^2$, $b= - [(E^2-\sum_{l=1}^4 t_l^+t_l^-)^2 - 2t_1^+t_1^-t_2^+t_2^- - (t_3^+t_4^+e^{ik_2}+t_3^-t_4^-e^{-ik_2})^2 ]$ and $c =  (t_1^- t_2^-)^2$.
Given $k_2$,  the line segment between two solutions $\beta_{\pm} (k_2)$ comprises the 1D amoeba hole depicted in \Cref{fig:cyl_spec}(e), which is also captured by the Ronkin minimum in \Cref{fig:cyl_spec}(f).
When the two solutions become degenerate ($b^2 - 4ac = 0$), the hole is absent.  Comparing  \Cref{fig:cyl_spec}(d) with \Cref{fig:cyl_spec}(a), the amoeba hole disappears simultaneously with complex energy gap closings: $\Delta \mu = |\mu_+ - \mu_-| = 0$, at $|r^*_{L,i}(k_{2})r_{R,i}(k_{2})|=1$. Moreover, the closing point, or the center $\Gamma$, can be linked to the same localization parameter which gives the imaginary momentum shift in the GSBZ in \Cref{eq:rcyl}:
\begin{gather}
 \beta_{+} (k_2) = \beta_{-} (k_2)  = -r.  
\end{gather}
Figure~\ref{fig:cyl_spec}(f) also shows that the Ronkin function descends quickly from a line of minima to a single minimum at $\Gamma$. 

As expected, the transition lines $|r^*_{L,i}(k_{2})r_{R,i} (k_{2})|=1$ along which chiral edge modes enter the bulk, are accompanied by both a jump of polarization and an absence of amoeba hole on a cylinder. It demonstrates the equivalence of these two approaches for restoring biorthogonal bulk-boundary correspondence between boundary modes of codimension $D=0$ and bulk modes ($D = d_c$). Yet, to characterize higher-dimensional boundary modes ($D > 0$) from our earlier analysis (cf.~\Cref{fig:gap,fig:trans,fig:lsg}),  biorthogonal polarization plays a more effective role.

\section{Discussion}
In this work, we have exactly solved a class of NH models in any spatial dimension $d$, and with open boundary conditions in $d_c \leq d$ directions. 
Our exact solutions make it explicit that no previously suggested approach is able to fully account for the interplay of bulk and boundary modes in these systems due to the presence of higher-dimensional NHSEs. However, we here successfully remedied these problems by combining and extending several previously distinct approaches. Specifically, we showed that the very recently proposed amoeba theory \cite{wang2022} fully accounts for the $O(N^d)$ bulk modes. While the amoeba approach does not account for any of the $O(N^{d_c})$ boundary modes, we managed to fully describe those by generalizing the GBZ \cite{yao2018} and biorthogonal polarization \cite{flore2018} approaches to higher dimensions.   

While full analytical solvability of our models facilitates a direct confirmation of the aforementioned approaches, this is in general an exceedingly challenging task relying essentially on numerical tests. Some insights can however be obtained analytically by suitably extending the models described here. First of all, a prominent feature of our models is a generalized chiral symmetry.  
Insights can be gleaned from breaking chiral symmetry in one dimension where the non-Bloch winding numbers and the biorthogonal polarization still provide key information about phase transitions despite the winding number (but not the polarization) losing its quantization \cite{Zelenayova2024,Mandal2024}. Second, the $A$ motif in our unit cell can be enlarged, allowing  multiple $r_{R,l}$'s, as long as the GSBZ still has a unique localization parameter $r_l$ in one direction.
These two points are demonstrated in the general case with hybrid boundary conditions $d_c < d$, showing the solvability of 
our models when distinct mass is added to each motif with broken chiral symmetry in every direction, and when the $A$ motif includes internal degrees of freedom. In the example of NH Lieb lattice on a cylinder at $\pi$ flux, two emergent zero-energy flat bands, one producing nonlocalized modes and the other one skin modes, are reminiscent of the recently studied  NH flat-band topology \cite{banerjee2023non,flore2024}.
We also note that our models can be generalized to open quantum systems where the non-Hermiticity described by engineered Lindblad dissipators gives rise to the Liouvillian skin effect \cite{fei2019,ueda2021,yang2022,kohei2023,ekman2024,yoshida2023,brighi2024nonreciprocal}. A dynamical distillation \cite{emil2023,emil2023n,meng2023} of higher-order skin modes in the full master equation framework will be featured in a future work.

\textcolor{black}{For nonsolvable NH models in higher dimensions, the universal bulk spectrum together with the GBZ can be numerically obtained via the amoeba and Ronkin function minimum \cite{wang2022}, despite missing information on boundary modes. Vice versa, from desired OBC spectra of NH systems, it is possible to reconstruct parent Hamiltonians through electrostatics \cite{lee2022d}. Although efficient in 1D, the extension of the approach to higher dimensions requires the formation of an array of 1D chains, and the complexity escalates with a multiband Hamiltonian as the target. Compared with these two methods, our class of NH hypercubic models offers insights into higher-order skin modes in generic NH systems beyond 1D, in the sense that the approach of dimensional reduction to $O(N^n)$ subgroups and the emergent  GSBZ description might be generalized to capture generic boundary modes in a systematic way. }

In conclusion, our present work marks a significant step towards a general quantitative understanding of the NHSE and its interplay with boundary modes in higher dimensions. 


\section*{Acknowledgements}
	We thank Rodrigo Arouca, Kohei Kawabata, J. Lukas K. K{\"o}nig, Paolo Molignini, and Kang Yang for discussions. This work was supported by the Swedish Research Council (grant 2018-00313), the Knut and Alice Wallenberg Foundation (KAW) via the Wallenberg Academy Fellows program (2018.0460) and the project Dynamic Quantum Matter (2019.0068) as well as the G\"oran Gustafsson Foundation for Research in Natural Sciences and Medicine.

\renewcommand{\theequation}{A\arabic{equation}}

\appendix

 \section*{\uppercase{Appendix: Derivation of non-Hermitian Hamiltonians for hybrid boundaries}}
\label{app:dia}
In the appendix, we focus on our NH hypercubic models in \Cref{fig:lattice} under hybrid boundary conditions: $\text{OBC}_{d_c} + \text{PBC}_{d-d_c}$. In particular, we present details on the construction of NH Hamiltonian ${\mathcal{H}'}$ on the new motifs $\tilde{A}B_\perp$ by diagonalizing $\mathcal{H}_\parallel$ in PBC directions. Among its eigenmodes, the ones coupled nonreciprocally to the $B_\perp$ motifs in OBC directions give rise to skin modes, while the ones decoupled contribute to the emergence of nonlocalized modes. In addition to hybrid boundaries, we address two flux conditions: generic hypercubic lattice at zero flux ($d_c < d$) and Lieb lattice at $\pi$ flux ($d_c = d - 1 = 1$).

\subsection{Zero flux for $d_c < d$}
At zero flux, for our model in arbitrary spatial dimension $d$ with $d_c$ open boundaries, 
we first obtain the Bloch Hamiltonian of $\mathcal{H}_\parallel$ by Fourier transform in all PBC directions according to \Cref{eq:ftp}.
 In the BZ subjected to PBC, the momentum related to the $x_l$ direction takes discrete values: 
  $k_{\parallel,l} = \frac{2\pi \tilde{n}}{N_l} \in [0, 2\pi)$  with $\tilde{n} = 0, 1, \dots, N_l -1 $.  The subscript $\vec{j}_\perp$ can be ignored for the moment.  Given each $\vec{k}_\parallel$, in the basis $\u{\varphi} (\vec{k}_\parallel) = (c_{A} (\vec{k}_\parallel), c_{ B_\parallel} (\vec{k}_\parallel))^T$, the Bloch Hamiltonian of $\mathcal{H}_\parallel$ shares the form
  \begin{gather}
    H_{\parallel} (\vec{k}_\parallel) = \begin{pmatrix}
	  		0 & Y_{\parallel,-}\\
     Y^T_{\parallel,+} & 0
		    \end{pmatrix},  \label{eq:hpa}
\end{gather} 
with  $Y_{\parallel, \pm} =
       ( t_{2m_{1}-1}^\mp + t_{2m_1}^\pm e^{\pm ik_{m_1}}, \ \dots,\  t_{2m_{d-d_c}-1}^\mp  + t_{2m_{d-d_c}}^\pm e^{\pm ik_{m_{d-d_c}}})$. 
After diagonalization, there emerge two dispersive bands and $(d-d_c-1)$ zero-energy flat bands:
  \begin{gather}
    H_{\parallel} (\vec{k}_\parallel) = \sum_{\alpha} \epsilon_{\tilde{A}_\alpha} (\vec{k}_\parallel) | \varphi_{R, \alpha} (\vec{k}_\parallel) \rangle \langle \varphi_{L, \alpha}(\vec{k}_\parallel)|, \label{eq:hpe} \\
    \begin{split}
   \quad \epsilon_{\tilde{A}_\pm} (\vec{k}_\parallel) &= \pm \sqrt{\sum_{l \in \bar{l}_\parallel} f_{2l-1,2l}(\vec{k}_\parallel)}, \notag \\
   \epsilon_{\tilde{A}_0} (\vec{k}_\parallel)  &= 0 \ (d-d_c > 1), \notag
   \end{split}
  \end{gather}
where the $f$ functions are given by \Cref{eq:gf}. Identifying each eigenmode of $H_{\parallel} (\vec{k}_\parallel)$ as a particle on the new $\tilde{A}$ motif,  
\begin{gather}
 | \varphi_{R, i} (\vec{k}_\parallel) \rangle = c_{\tilde{A}_i}^\dagger (\vec{k}_\parallel) | 0 \rangle, \quad  \langle \varphi_{L, i}(\vec{k}_\parallel)| = \langle 0 |c_{\tilde{A}_i} (\vec{k}_\parallel), \label{eq:ta}
 \end{gather}
we arrive at 
 \begin{gather}
 H_{\parallel} (\vec{k}_\parallel) = \sum_i \epsilon_{\tilde{A}_i}(\vec{k}_\parallel)  c_{\tilde{A}_i}^\dagger ( \vec{k}_\parallel) c_{\tilde{A}_i}  (\vec{k}_\parallel). \label{eq:hpa}
 \end{gather}

It is interesting to see the coexistence of nonlocalized modes and skin modes under hybrid boundary conditions. 
Particles on the $\tilde{A}$ motif are coupled to the $B_\perp$ motifs through the original $A$ motif according to
 \begin{align}
 c_{A}^\dagger (\vec{k}_\parallel) &= \sum_i \varphi_{L, i}^* (\vec{k}_\parallel,1)  c_{ \tilde{A}_i }^\dagger (\vec{k}_\parallel), \notag \\
 c_{A} (\vec{k}_\parallel) &= \sum_i \varphi_{R, i} (\vec{k}_\parallel,1)  c_{ \tilde{A}_i} (\vec{k}_\parallel). \label{eq:mta}
 \end{align}
In the above, we apply to \Cref{eq:ta} the eigenvalue solutions
 $ | \varphi_{R/L, i} (\vec{k}_\parallel) \rangle = \sum_{\lambda \in A \cup B_\parallel } \varphi_{R/L, i} (\vec{k}_\parallel, \lambda) c_{\lambda}^\dagger (\vec{k}_\parallel) |0 \rangle $ and the identity $\mathbbm{1}_\lambda = \sum_i  | \varphi_{R, i} (\vec{k}_\parallel, \lambda) \rangle  \langle \varphi_{L, i}(\vec{k}_\parallel,\lambda)|$.
One further notices that in \Cref{eq:hpe}, the wave functions of zero-energy eigenmodes $\tilde{A}_0$ on the flat bands vanish completely on the $A$ motif: 
    \begin{gather}
        \varphi_{L, 0}^* (\vec{k}_\parallel,1) = \varphi_{R, 0} (\vec{k}_\parallel,1) =0.
    \end{gather}  
It leads to a decoupling from all $B_\perp$ motifs and 
preserves $(d-d_c-1)$  nonlocalized modes at zero energy.

In contrast, through couplings to the $B_\perp$ motifs the two dispersive eigenmodes $\tilde{A}_\pm$ of $H_{\parallel} (\vec{k}_\parallel)$ evolve into skin modes. By restoring the subscript $\vec{j}_\perp$ and expressing $\mathcal{H}_\perp$ in the subspace of $\{ \tilde{A}_+ ,\tilde{A}_- \}  \cup B_\perp$ using \Cref{eq:mta},  one builds  ${\mathcal{H}'}$ in \Cref{eq:gnh} of the main text, where the
 renormalized asymmetric hopping parameters read ($\alpha = \pm$)
  \begin{align}
    t_{2l-1,\alpha}^+ &= t_{2l-1}^+ \varphi_{L,\alpha}^*(\vec{k}_\parallel, 1), \quad 
    t_{2l-1,\alpha}^- = t_{2l-1}^- \varphi_{R,\alpha}(\vec{k}_\parallel, 1), \notag \\
     t_{2l,\alpha}^+ &= t_{2l}^+ \varphi_{R,\alpha}(\vec{k}_\parallel, 1), \quad
     t_{2l,\alpha}^- = t_{2l}^- \varphi^*_{L,\alpha}(\vec{k}_\parallel, 1). \label{eq:rcou}
  \end{align}

\subsection{$\pi$ flux for $d_c = d - 1 =1$}
At $\pi$ flux, we give the example of NH Lieb lattice on a cylinder geometry. In \Cref{fig:cyl}(a), $\mathcal{H}_\parallel$ can be diagonalized 
 explicitly by a Fourier transform along the $x_2$ (PBC) direction:
$c_{j, \lambda} (j') = \frac{1}{\sqrt{N_2}} \sum_{k_2} e^{ik_2 \cdot j'} c_{j, \lambda} (k_2)$ where $k_2 = \frac{2\pi \tilde{n}}{N_2} \in [0, 2\pi)$ and  $\tilde{n} = 0, 1, \dots, N_2 -1$. $\vec{j} = (j,j')$ locates the unit cell with $j = 1, \dots, N_1$, $j' = 1, \dots, N_2$ and the motif index $\lambda \in A \cup B_2$. As before, we ignore the subscript $j$ for the moment. 
In the basis $\u{\varphi}(k_2) = (c_{A_1}(k_2), c_{B_1}(k_2), c_{B'_1}(k_2), c_{A_2}(k_2), c_{B'_2}(k_2))^T$, the Bloch Hamiltonian of $\mathcal{H}_\parallel$ reads: $\mathcal{H}_{\parallel} = \sum_{k_2} \u{\varphi}^\dagger(k_2) H_{\parallel}(k_2)\u{\varphi} (k_2)$, 
  \begin{gather}
    H_{\parallel}(k_2) =     
     \begin{pmatrix}
              0 & t_1^+ & p^+  & 0 & 0 \\
        t_1^- & 0 & 0 & t_2^+ & 0 \\
        p^- & 0 & 0 & 0 & 0 \\
        0 & t_2^- & 0 & 0 & q^+ \\
        0 & 0 & 0 & q^- & 0  
      \end{pmatrix},  \\
      p^\pm = t_3^\pm - t_4^\mp e^{\mp ik_2}, \quad q^\pm = t_3^\pm + t_4^\mp e^{\mp ik_2}. \notag
    \end{gather}
The sign change in front of $t_4^\mp$ reproduces $\pi$ flux per plaquette. A direct diagonalization leads to 
 \begin{gather}
   H_{\parallel}(k_2) = \sum_{i=0}^4 \epsilon_{i} (k_2) |\varphi_{R,i} (k_2) \rangle \langle \varphi_{L,i} (k_2)|, \notag \\  \epsilon_0 = 0, \quad
  \epsilon_i(k_2) = \pm \sqrt{\tau_0 \pm \sqrt{\tau(k_2)}}, \label{eq:c0}
\end{gather}
with $\tau$ functions given in \Cref{eq:taui}.
For simplicity, we assign the subscript $i=0$ to denote one zero-energy mode and $i=1,2,3,4$ four dispersive modes on $\epsilon_i(k_2)$ bands, taking signs $(+,+),(-,+),(+,-),(-,-)$ respectively. 

It can be checked that the zero-energy mode has no occupancy on $A_1$ and $A_2$ motifs, thus decoupled from the $B_2$ motif in the OBC direction [see \Cref{fig:cyl}(a)]
 \begin{align}
   |\varphi_{R,0, \text{bulk}} (k_2) \rangle &\sim 
   (0,  \ p^+, \ -t_1^+, \ 0,\  -t_2^-p^+/q^+)^T, \notag \\
    \langle \varphi_{L,0,  \text{bulk}} (k_2) | &\sim
    (0, \  p^-, \ -t_1^-, \ 0, \ -t_2^+p^-/q^-).
 \end{align}
It resides on $B_1 B'_1 B'_2$ motifs only and becomes a nonlocalized mode, immune from the NHSE.

By contrast, four dispersive eigenmodes take finite occupancy on $A_1$ and $A_2$ motifs: \begin{widetext}
  \begin{align}
    |\varphi_{R,i}(k_2) \rangle &=  c_{\tilde{A}_i}^\dagger (k_2) | 0 \rangle
 \sim (\epsilon_i, \ t_1^- \eta_i, \ p^-, \ -\frac{t_1^-}{t_2^+}(1-\eta_i)\epsilon_i, \ -\frac{t_1^-}{t_2^+}(1-\eta_i)q^-)^T, \notag \\
    \langle \varphi_{L,i}(k_2)| &=\langle 0 |  c_{\tilde{A}_i} (k_2) \sim (\epsilon_i, \ t_1^+ \eta_i, \ p^+, \ -\frac{t_1^+}{t_2^-}(1-\eta_i)\epsilon_i, \ -\frac{t_1^+}{t_2^-}(1-\eta_i)q^+), \label{eq:dis} 
 \end{align}
with $\eta$ functions given in \Cref{eq:eta}.

It allows us to use them to construct a  new $\tilde{A}$ motif: $\tilde{A} = \{ \tilde{A}_1, \tilde{A}_2, \tilde{A}_3, \tilde{A}_4 \}$, which is coupled nonreciprocally to 
the $B_2$ motif as shown in \Cref{fig:cyl}(a).
The coupling strengths can be read from a change of basis associated with the original $A_1$ and $A_2$ motifs:
 \begin{gather}
   c_{A_1}^\dagger(k_2) = \sum_{i=1}^4  \varphi_{L,i}^*(1) c_{\tilde{A}_i}^\dagger(k_2), \quad
   c_{A_1}(k_2) = \sum_{i=1}^4  \varphi_{R,i}(1) c_{\tilde{A}_i}(k_2), \notag \\
   c_{A_2}^\dagger(k_2) = \sum_{i=1}^4  \varphi_{L,i}^*(4) c_{\tilde{A}_i}^\dagger(k_2), \quad
   c_{A_2}(k_2) = \sum_{i=1}^4  \varphi_{R,i}(4) c_{\tilde{A}_i}(k_2).  \label{eq:map_cyl}
 \end{gather}
 It leads to 
   \begin{align}
    t_{1,i}^+ &= t_1^+ \varphi_{L,i}^*(4),\quad
    t_{1,i}^- = t_1^- \varphi_{R,i}(4), \notag \\ t_{2,i}^+ &= t_2^+ \varphi_{R,i}(1), \quad t_{2,i}^- = t_2^- \varphi_{L,i}^*(1).
    \label{eq:cyl-c}  
  \end{align}
After restoring the subscript $j$, one maps the original Hamiltonian to a generalized NH SSH chain along the $x_1$ (OBC) direction of the cylinder in \Cref{eq:th-cyl}. 

  \end{widetext}

\bibliography{sample}

\end{document}